\documentclass[10pt,cleanfoot,cleanhead]{asme2ej}

\usepackage{epsfig} %% for loading postscript figures
\usepackage{subfigure}
\usepackage{graphicx,color}
\usepackage{amsmath,amssymb}
\usepackage{lipsum}
\usepackage{cleveref}
\usepackage{appendix}

\def\eeq{\relax}
\def\beq#1#2\eeq{\begin{equation}\label{#1}#2\end{equation}}
\def\bal#1#2\eal{\begin{align}\label{#1}#2\end{align}}
\def\bse#1#2\ese{\begin{subequations}\label{#1}#2\end{subequations}}
\def\ba{\begin{aligned}}   \def\ea{\end{aligned}}

\title{Non-reciprocal wave transmission in a bilinear spring-mass system}

\author{Zhaocheng Lu
   \affiliation{
   Department of Mechanical and Aerospace Engineering\\
   Rutgers University\\
   Piscataway, NJ 08854-8058\\
   e-mail: zhaocheng.lu@rutgers.edu
   }
}

\author{Andrew N. Norris%\thanks{Corresponding author.}
    \affiliation{
   	Department of Mechanical and Aerospace Engineering\\
   Rutgers University\\
   Piscataway, NJ 08854-8058\\
   e-mail: norris@rutgers.edu
   }
}

\begin{document}

\maketitle    

%%%%%%%%%%%%%%%%%%%%%%%%%%%%%%%%%%%%%%%%%%%%%%%%%%%%%%%%%%%%%%%%%%%%%%
\begin{abstract}
{\it 
Significant amplitude-independent and passive non-reciprocal wave motion can be  achieved in a  one dimensional (1D) discrete chain of masses and  springs with bilinear elastic stiffness. Some fundamental asymmetric spatial modulations of the bilinear spring stiffness are first examined for their non-reciprocal properties. These are combined as building blocks into more complex configurations with the objective of maximizing non-reciprocal wave behavior.  The non-reciprocal  property is demonstrated by the significant difference between the transmitted pulse displacement amplitudes and energies for incidence in  opposite directions.  Extreme non-reciprocity is realized when almost-zero transmission is achieved for the propagation from one direction with a noticeable transmitted pulse for incidence from the other. 
These models provide the basis  for  a class of simple 1D non-reciprocal   designs and can serve as the building blocks for  more complex and higher dimensional non-reciprocal wave systems.
}
\end{abstract}

%\begin{keywords}
%non-reciprocity \sep bilinear spring \sep zero transmission \sep one-way propagation
%\end{keywords}

%%%%%%%%%%%%%%%%%%%%%%%%%%%%%%%%%%%%%%%%%%%%%%%%%%%%%%%%%%%%%%%%%%%%%%
\section{Introduction}
	
Reciprocity is a fundamental physical principle of wave motion which requires   symmetry in wave transmission between any two points. 
The same incident wave traveling in   opposite directions should result in the same transmitted wave. Recent advances  have shown that the principle of reciprocity can be violated under special conditions in electromagnetism \cite{Caloz2018}, acoustics \cite{Wang2015,Nassar2017a,Nassar2018a,Liang2009,Liang2010,Fronk2019,Moore2018,Bunyan2018,Luo2018} and other physical systems supporting wave propagation \cite{Wallen2019}. 
The ability to violate  reciprocity in a controlled and passive manner opens the possibility of extreme  wave dynamics and control mechanisms such as one-way propagation, acoustic diodes, etc, and provide revolutionary solutions to   existing problems and useful tools for promising applications.   The focus of the present work is breaking   reciprocity of elastic waves   using   bilinear material properties. 
	
There are several  ways to break dynamic reciprocity. One is to remove time-reversal symmetry. For instance,  gyroscopic inertial effects are used to break the time-reversal symmetry in a one-way phononic waveguide \cite{Wang2015}. 
Spatiotemporal modulation of density and elastic properties provides another way to violate reciprocity. The direct modulation of elastic moduli and mass density simultaneously in both space and time introduces non-reciprocity due to the tilting of dispersion bands \cite{Nassar2017a,Nassar2018a}. Non-reciprocal elastic wave propagation  can be achieved via modulated stiffness  realized by applying a wavelike deformation that alters the effective on-site stiffness \cite{Wallen2019}. 
A third way is using nonlinearity, which unlike the previous active examples, provides a passive method to achieve non-reciprocity. Usually, the non-reciprocal phenomenon can be obtained by combining the nonlinearity with other assistant properties. In the acoustical domain, an acoustic diode can be achieved by utilizing the second-harmonic generation property of the nonlinear medium and the frequency selectivity of the sonic crystal \cite{Liang2009,Liang2010}. Nonlinear acoustic non-reciprocity is also reported theoretically and experimentally in lattice structures incorporating strong stiffness nonlinearities, internal scale hierarchy and asymmetry in their unit cell designs \cite{Fronk2019,Moore2018,Bunyan2018}. Weak nonlinearity can be used in  band gap manipulation which in turn leads to non-reciprocal behavior \cite{Luo2018}.
	
Bilinear springs present a unique case of nonlinearity,   consisting of two different linear load-deformation relations.  Unlike other nonlinearities, such as cubic \cite{Narisetti2010a,Narisetti2012}, the bilinear relation is amplitude-independent; the nonlinearity enters only through the sign of the displacement.  The analogous phenomenon in continuum mechanics occurs in materials with bilinear (also known as heteromodular or bimodular) constitutive  elastic behavior, which have been proposed as nonlinear models for studying contact forces \cite{Shaw1983a},  elastic solids containing cracks \cite{Scalerandi2003} and for the dynamics of geophysical systems, including  granular media \cite{Kuznetsova2017}. 
The discontinuity of the piecewise linear relation gives rise to a strong nonlinearity, for which it is difficult to find   analytical solutions for simple wave problems.  
Wave motion in bimodular media has been studied extensively \cite{Benveniste1980,Maslov1985,Nazarov90,Ostrovsky1991,AbeyaratneKnowles1992,GavrilovHerman2012,Nazarov2015,Rudenko2016,Rudenko2016a,Nazarov2016}. 
Even a small difference between the moduli in tension and compression immediately causes  the appearance of shock waves \cite{GavrilovHerman2012}; however linear viscosity eliminates the shocks. A good review of the literature of wave motion in continuous bimodular media, particularly the considerable work done by Russian researchers, can be found in  \cite{GavrilovHerman2012}, while \cite[p.\ 32]{NO1998} provides an earlier review. There are far fewer studies of wave motion in discrete spring-mass chains with bilinear spring forces.  Of particular interest is  the study \cite{Kuznetsova2017} which analyzed impulse  harmonic wave propagation in a 1D system of bilinear oscillators.  Although they did not emphasize non-reciprocal effects, the authors noticed that sign inversion of a signal can be obtained, from tension to compression {that} can lead to pulse spreading or shortening and possible shock formation. However, none of these prior studies of either continuous or discrete systems considered spatial inhomogeneity and asymmetry, which are necessary for producing non-reciprocal wave motion in the presence of material nonlinearity. 
	
Here we leverage the bilinear property to break wave reciprocity in a simple mechanical structure. Specifically, we consider  a 1D bilinear spring-mass chain system in which  the spatial modulation of bilinear stiffness is carefully investigated and designed. Incident  pulses, generated by an external harmonic loading, are used to study transmitted pulse amplitudes for incidence from opposite directions in the modulated chain. The non-reciprocal property of the system is demonstrated by the significant difference between the transmitted wave amplitudes. Zero transmission can be approximately achieved when the absolute amplitude ratio of the transmitted and incident pulse is small enough. One-way propagation can be realized if almost-zero transmission is achieved for the propagation from one direction with  a noticeable transmitted pulse from the other. The final non-reciprocal structure is obtained after careful design of the inhomogeneous and asymmetric spatial modulation of the stiffness, and will be shown to display significant amplitude-independent non-reciprocity in a passive system. This  simple 1D design could  be the building block for complex, multiple dimensional non-reciprocal wave systems. 
	
The paper proceeds as follows: A 1D bilinear spring mass chain system is introduced and the governing equations are discussed in Sect.\ \ref{sec2}. Some fundamental spatial modulations of the bilinear stiffness are presented and investigated in Sect.\ \ref{sec3}. Two non-reciprocal models are designed and demonstrated in Sect.\ \ref{sec4}. 
%{which includes a discussion of the  effect of input frequency on the non-reciprocal wave behavior in Sect.\ \ref{sec4.2}.} 
Section \ref{sec6} concludes the paper.

%%%%%%%%%%%%%%%%%%%%%%%%%%%%%%%%%%%%%%%%%%%%%%%%%%%%%%%%%%%%%%%%%%%%%%
\section{1D Bilinear Chain Overview}  \label{sec2}

\begin{figure}[h] %%%%%%%%%%%%%%%%%%%%%%%%%%%%%%%%%%%%%%%%%%%%%%%%%%%%%%%%%%%%%%%%%%%%%%
\centerline{\psfig{figure=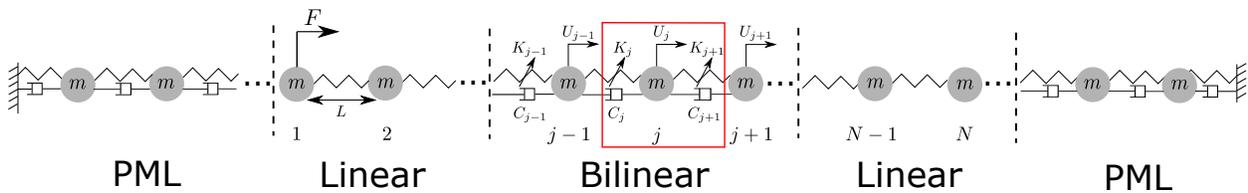,width=6.5in}}
\caption{Schematic view of 1D test chain system of identical masses with uniform linear springs and spatially varying bilinear springs. A perfectly matched layer (PML) is attached to each side of the test chain. An external force $F$ is applied on the first mass in the linear chain on the left. The length $L$, assumed uniform, denotes the  spacing between the equilibrium positions of two adjacent masses. Index $j$ denotes the mass number, with identical masses assumed. The box is the unit cell for equilibrium. $U_j$ denotes the displacement of the $j$th mass; $K_j$  and $C_j$ denotes the $j$th spring and damper between $j$th and $(j-1)$th mass respectively.}
\label{fig1}
\end{figure}

Consider a mass-spring chain system consisting of a linear part sandwiching a bilinear region as shown in Fig.\ \ref{fig1}. The bilinear section is a chain of identical masses, weak dampers and bilinear springs with possible spatially dependent stiffness. The linear section consists of two chains of identical masses and uniform linear springs attached to both ends of the bilinear chain. An external loading is applied to generate the incident pulse in the linear chain. A perfectly matched layer (PML) is added to each end of the test chain in order to eliminate reflections, and hence the bilinear chain lies between two  effectively semi-infinite linear chains. The PML is itself a chain of damped oscillators with  damping coefficients that are incrementally “ramped-up” to prevent internal reflections \cite{Narisetti2010}. 
%%%%%%%%%%%%%%%%%%%%%%%%%%%%%%%%%%%%%%%%%%%%%%%%%%%%%%%%%%%%%%%%%%%%%%

The governing equations of the test chain system can be found by concentrating on the $j$th mass and it nearest neighbors, Fig.\ \ref{fig1}. The index $j$, which starts from 1 at the beginning of the test chain, also represents the $j$th spring $K_j$ or damper $C_j$. Suppose that the number of masses in the linear part is $2 \, N_{l}$ and the number of masses in the bilinear part is $N_{bl}$. The total number of identical masses in this test chain is $N = 2 \, N_l+N_{bl}$. Therefore, the number of the linear springs is $2 \, (N_l + 1)$ and the number of the bilinear springs is $N_{bl} - 1$. The total number of all springs is $N+1$. The equation of motion for the $j$th mass follows as	
\beq{govern_eqn}
\ba
 M \, \ddot U_j - \big [ K_{j+1} \, ( U_{j+1} - U_j ) + K_{j} \, ( U_{j-1} - U_j ) \big ]
 - \big [ C_{j+1} \, ( \dot U_{j+1} - \dot U_j ) + C_{j} \, ( \dot U_{j-1} - \dot U_j ) \big ] = F \, \delta_{j1} \, , \quad{} j = 1, .. , N \, .
\ea
\eeq

\begin{figure} [hbt!]%%%%%%%%%%%%%%%%%%%%%%%%%%%%%%%%%%%%%%%%%%%%%%%%%%%%%%%%%%%%%%%%%%%%%%
\begin{center}
\subfigure[\ Compression-tension pulse ]{
\includegraphics[width=3.25in]{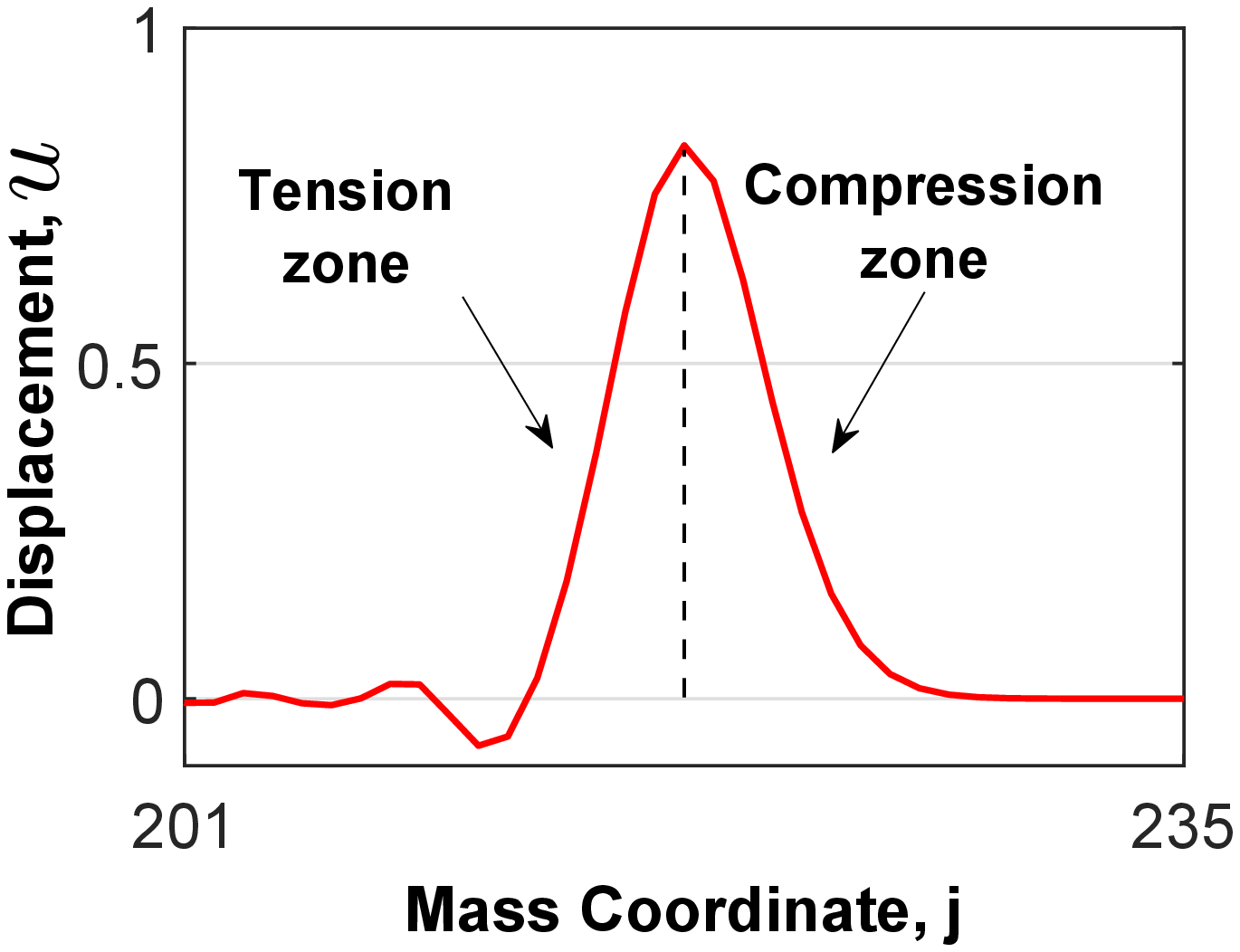}}
\subfigure[\ Tension-compression pulse ]{
\includegraphics[width=3.25in]{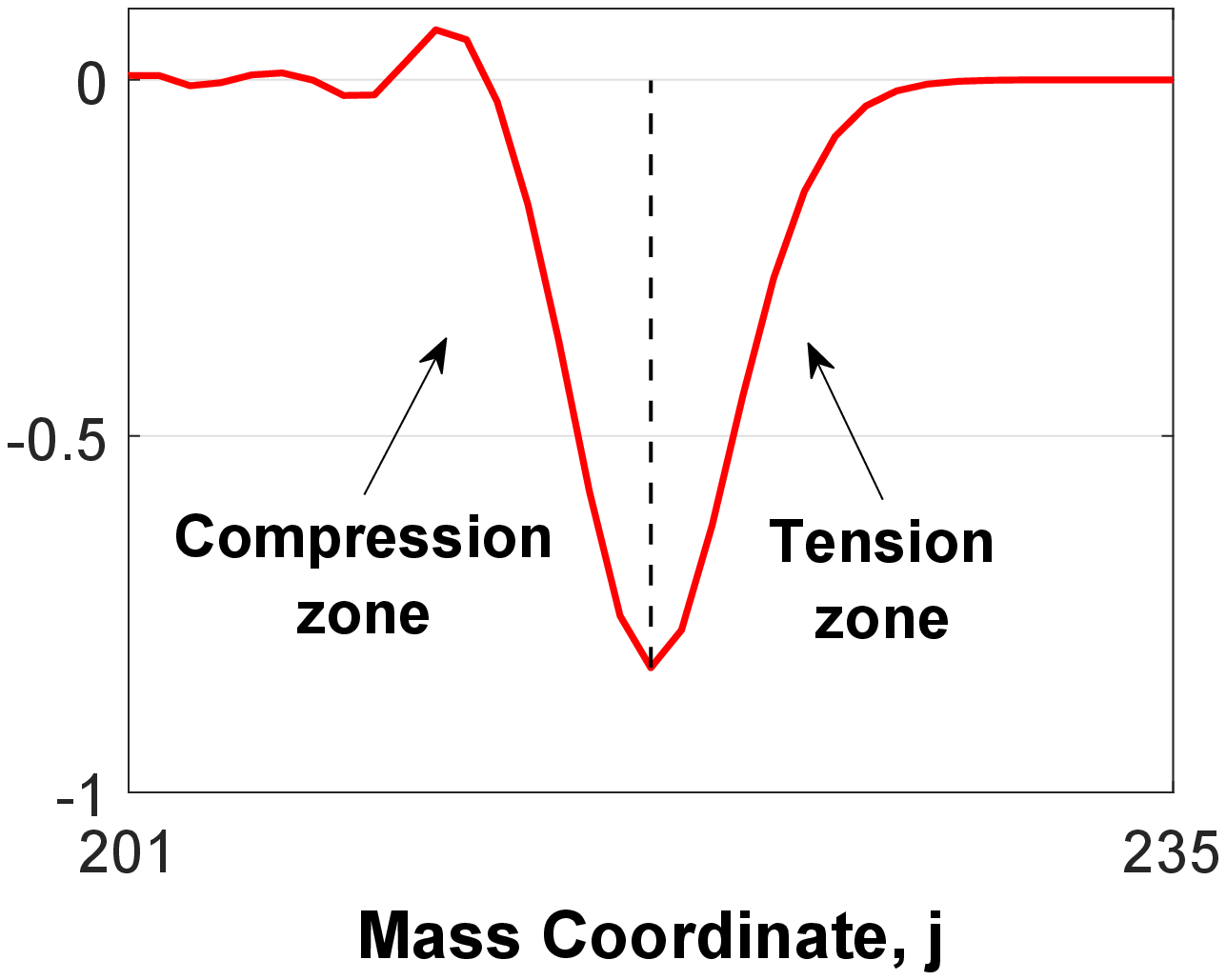}}
\caption{Compression-tension and tension-compression incident pulses generated by an external forcing propagate in the linear chain and form the incident wave on the bilinear section. The dashed line divides the pulse into two different parts each of which contains  springs in the same state, compression or tension.}
\label{fig2}
\end{center}
\end{figure}

A pulse is used for investigation of non-reciprocal properties \cite{KuznetsovaPasternakDyskin2017}. The specific pulse adopted here is generated  by an external harmonic loading of the form $ F = \pm \, F_0 \, \mathcal{H}(T) \, \mathcal{H}(\frac{2 \pi}{\Omega} - T) \, \sin \Omega T $, where $F_0 >0$ is the forcing amplitude and $\Omega$  the excitation frequency. The positive  sign  results in an incident pulse with compression followed by tension, called a compression-tension (CT) pulse, see Fig.\ \ref{fig2}.  Conversely, the negative sign produces a tension-compression (TC) type of pulse. 

We rewrite the equation of motion Eq.\ \eqref{govern_eqn} in terms of dimensionless displacement $u$ and time $t$,
\beq{dimensionless}
u =  \frac{U}{L} \, , \quad{} t = \Omega_0 \, T \, ,
\eeq
where $\Omega_0 = \sqrt{ \frac{K_0}{M} }$ is the basic frequency and $K_0$ is the linear stiffness. The dimensionless form of Eq.\ \eqref{govern_eqn} is
\beq{govern_eqn_dless}
  \ddot u_j - \big [ \kappa_{j+1} \, ( u_{j+1} - u_j ) + \kappa_{j} \, ( u_{j-1} - u_j ) \big ] \\
 - \big [ \zeta_{j+1} \, ( \dot u_{j+1} - \dot u_j ) + \zeta_{j} \, ( \dot u_{j-1} - \dot u_j ) \big ] = f \, \delta_{j1}  \, , \quad{} j = 1, .. , N \, ,
\eeq
where the dimensionless stiffness and damping coefficients are
\beq{dimensionless_mech}
\kappa_j = \frac{K_j}{K_0}	\, , \quad{} 
\zeta_j = \frac{C_j}{\sqrt{K_0 / M}}		\, ,
\eeq
and the dimensionless external forcing is
\beq{dimensionless_force}
f = \pm \, f_0 \, \mathcal{H}(t) \, \mathcal{H}(\frac{2 \pi}{\omega} - t) \, \sin \omega \, t \, ,
\eeq
with $f_0 = \frac{F_0}{L \, K_0}$ and $\omega = \frac{\Omega}{\Omega_0}$. Both the dimensionless stiffness $\kappa_j$ and the dimensionless damping coefficient $\zeta_j$ depend on their location, or the index $j$. In the linear sections the index takes the values $1 \leq j \leq N_l+1$ or $ N_l+N_{bl} < j \leq N+1$, and in the bilinear middle section $N_l+1 < j \leq N_l+N_{bl}$.
The stiffness is 
\beq{kappa}
\kappa_j = 
\begin{cases}
    	1 + \Delta_{j,c} \ \ \text{if} \ \ u_j - u_{j-1}<0 , & \text{bilinear section}  ,
			\\
			1 + \Delta_{j,t} \ \ \text{if} \ \ u_j - u_{j-1} > 0 , & \text{bilinear section}  ,
			\\
			  1 \, , & \text{linear sections} ,
\end{cases}
\eeq
%%%%%%%%%%%%%%%%%%%%%%%%%%%%%%%%%%%%%%%%%%%%%%%%%%%%%%%%%%%%%%%%%%%%%%
where  $\Delta_{j,c}$ and $\Delta _{j,t}$ are the dimensionless compressive and tensile stiffness increment, respectively. In terms of the bilinear stiffness,  $\Delta_j = \frac{\Delta K_j }{K_0}$  with $K_j$ assuming different values depending as the spring is in compression or tension. 
%To break the reciprocity, stiffness increment $\Delta_j$ need to be carefully designed. More details will be explored in the following sections. 
The damping coefficient  is taken as
\beq{zeta}
\zeta_j = 
\begin{cases}
      \zeta_0 \, , 		& \text{bilinear section} , \\
	0 \, , 		& \text{linear sections}  ,  
\end{cases}
\eeq
where $\zeta_0$ is the constant damping coefficient in the bilinear chain. Expressions for the stiffness and damping coefficient in the PML can be found in Appendix \ref{appA}. {Damping is not included in the linear sections so that we can more clearly see the incident and transmitted waves. The weak dampers in the bilinear section do not significantly effect the transmission. 
However, a small amount of damping has a strong smoothing effect in the presence of discontinuities, such as those that arise from shock-like structures that can arise in bilinear media. 
Details of  comparison with and without damping are  included in Appendix \ref{appA_1}.
In this paper we choose to include sufficient damping such that the transmitted wave forms are smooth.  This allows us to make quantitative comparisons between transmission amplitudes for incidence from opposite directions.  } 

The system of $N$ linear and bilinear ordinary differential equations, Eq.\ \eqref{govern_eqn_dless}, is solved numerically using ODE45 in Matlab with time step $\Delta t = 0.01$.

%%%%%%%%%%%%%%%%%%%%%%%%%%%%%%%%%%%%%%%%%%%%%%%%%%%%%%%%%%%%%%%%%%%%%%
\section{Spatial Modulation of the Bilinear Stiffness}  \label{sec3}

In this Section  we explore the dynamic properties of several underlying spatial modulations of the bilinear stiffness in the test chain. {The designs strictly follow the principle of spatial inhomogeneity and asymmetry which are critical for creating non-reciprocal propagation in the presence of nonlinearity.} 

A compression-tension pulse,  generated by setting a positive sign in the external forcing Eq.\ \eqref{dimensionless_force}, is used to test the different modulations. As Fig.\ \ref{fig2} depicts, the positive displacement corresponds to the movement of mass to the right and the negative to the left. Numerical results presented in this Section are obtained using the parameters listed in Table \ref{table:parameters}.
\begin{table}[ht]
\caption{Parameters for numerical experiments: $N_{pml},N_{l}$ and $N_{bl}$ are the numbers of masses in different parts of the chain system, see Eqs.\ \eqref{kappa} and \eqref{zeta}; $\zeta_0$ is the dimensionless damping coefficient in Eq.\ \eqref{zeta}; and $f_0$ is the dimensionless force amplitude in Eq.\ \eqref{dimensionless_force}.}
\begin{center}%%%%%%%%%%%%%%%%%%%%%%%%%%%%%%%%%%%%%%%%%%%%%%%%%%%%%%%%%%%%%%%%%%%%%%
\label{table:parameters}
\begin{tabular}{c l l l l l}
%& & & & & \\ % put some space after the caption
\hline \hline
$N_{pml}$ & $N_{l}$ & $N_{bl}$ & $\zeta_0$ & $f_0$ & $\omega$ \\
\hline
201 & 49 & 101 & 0.1 & 0.01 & 0.5 \\
\hline \hline
\end{tabular}
\end{center}
\end{table}%%%%%%%%%%%%%%%%%%%%%%%%%%%%%%%%%%%%%%%%%%%%%%%%%%%%%%%%%%%%%%%%%%%%%%

\begin{figure}%%%%%%%%%%%%%%%%%%%%%%%%%%%%%%%%%%%%%%%%%%%%%%%%%%%%%%%%%%%%%%%%%%%%%%
%[hbt!]
\begin{center}
\subfigure[$\,$ Linearly increasing stiffness.]{
\includegraphics[width=3.25in]{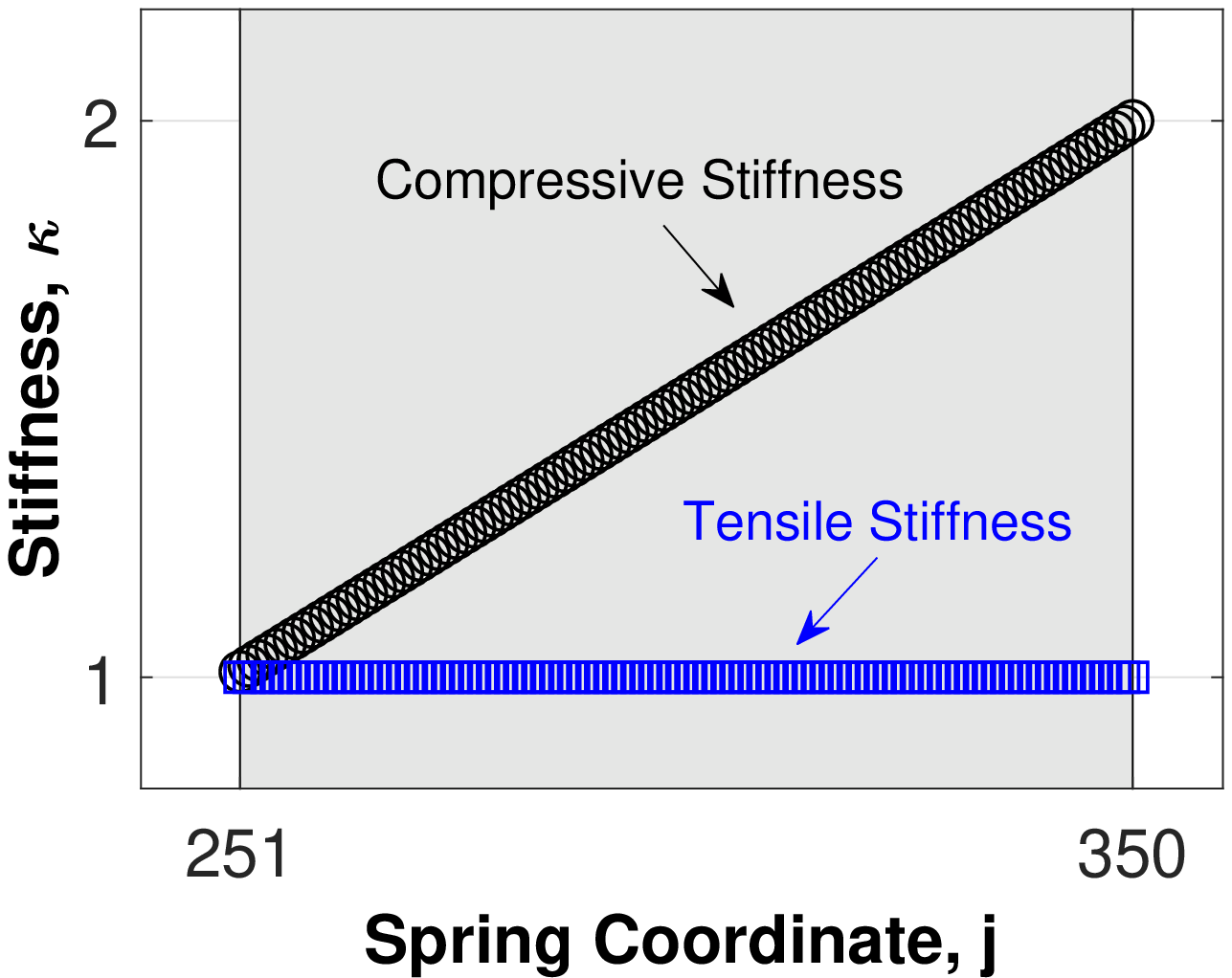}}
\subfigure[$\,$ Linearly decreasing stiffness.]{
\includegraphics[width=3.25in]{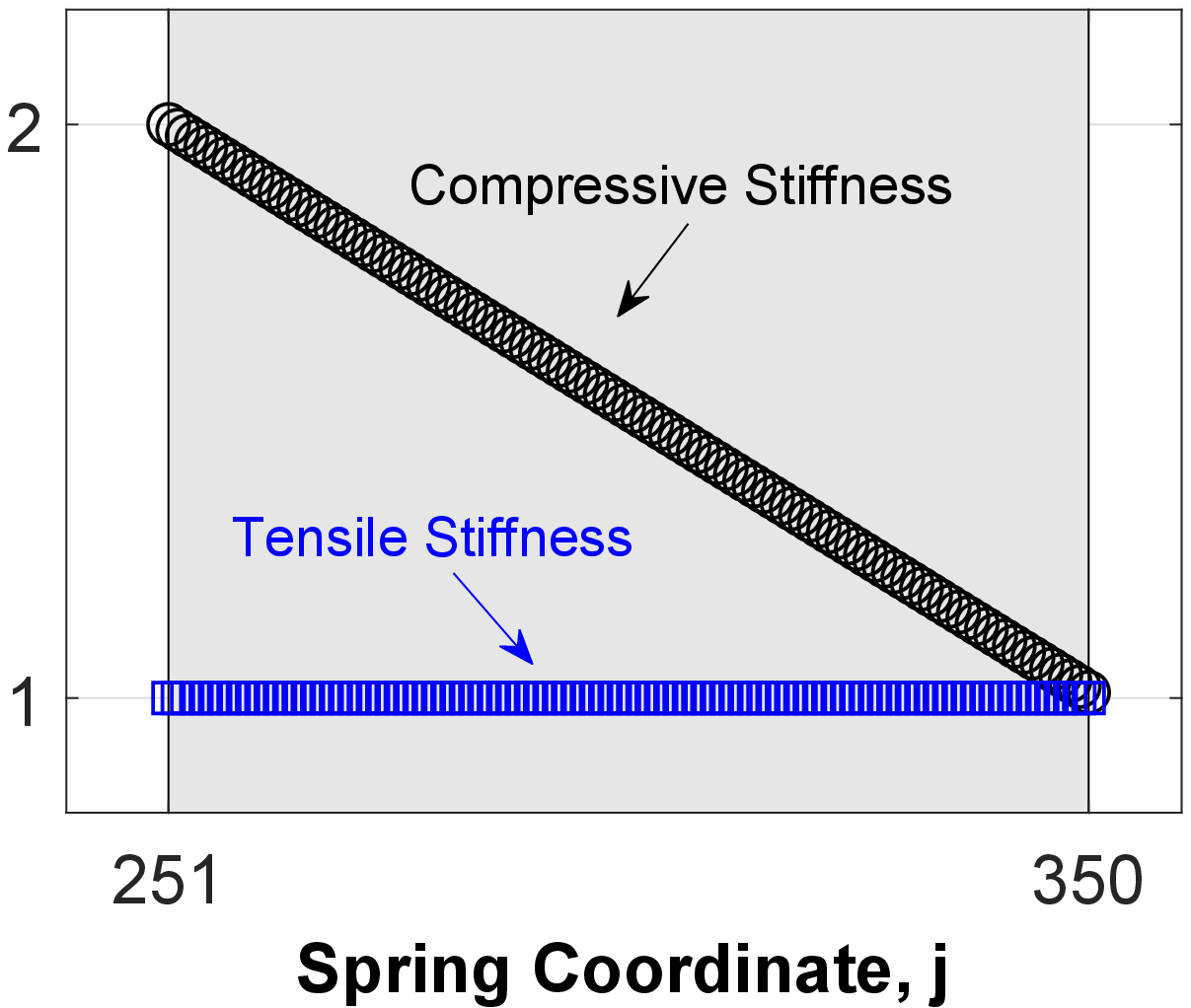}}
\subfigure[$\,$ Displacement   for modulation (a).]{
\includegraphics[width=3.25in]{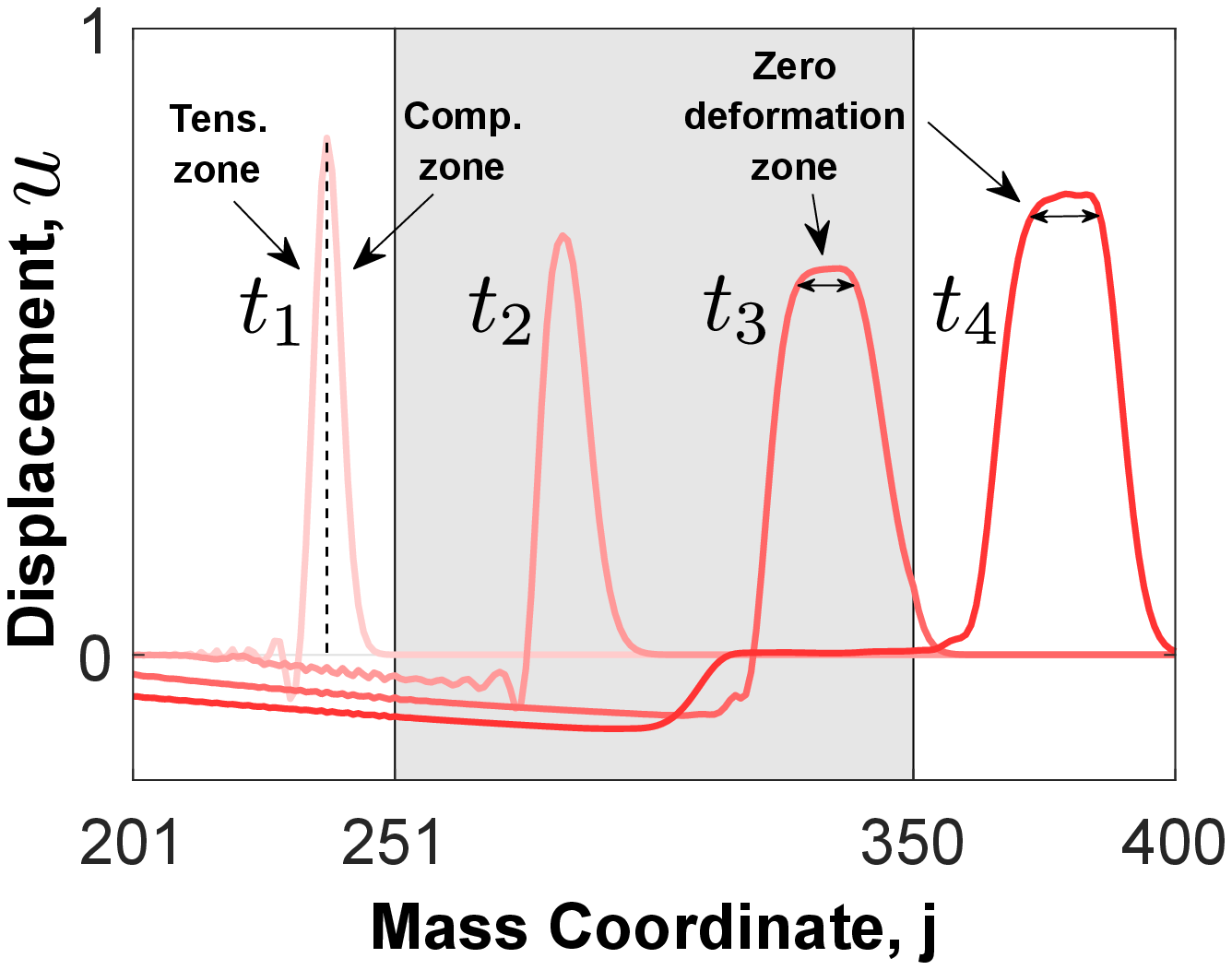}}
\subfigure[$\,$ Displacement   for modulation (b).]{
\includegraphics[width=3.25in]{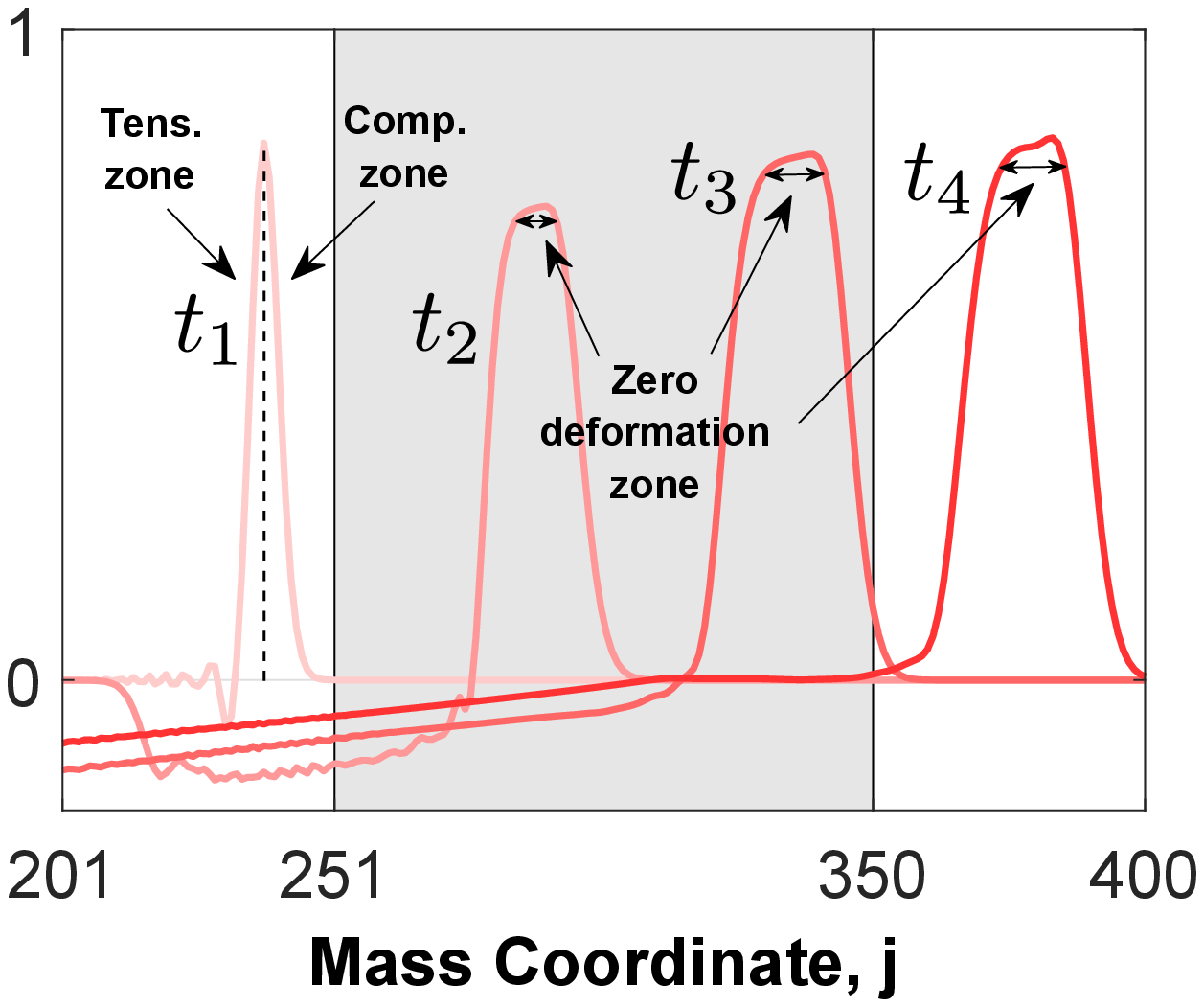}}
\caption{Dynamic properties of the bilinear spring chain with  compressive stiffness varying linearly with position. The shaded region is the bilinear part of the test chain. (a) and (b) are the  two basic bilinear stiffness spatial modulations. (c) and (d) shows displacement fields at different moments for modulations (a) and (b) respectively. }
\label{fig3}
\end{center}
\end{figure}

We first assume that the stiffness of the bilinear spring is greater in compression than in tension. For simplicity, we set the stiffness in tension equal to the linear stiffness. Therefore,   $\Delta_{j,c} > \Delta_{j,t} = 0$. Suppose that the increment $\Delta_{j,c}$ can be either linearly increasing or decreasing over location with non-dimensional stiffness ranging in values between $1$ and $2$. For the linearly increasing modulation, we have  for $N_l+1 < j \leq N_l+N_{bl}$ 
\beq{delta_incr}
\Delta_{j,c} = \frac{j - N_{l} - 1}{N_{bl} - 1}	\, ,
\eeq
and for the linearly decreasing modulation, 
\beq{delta_decr}
\Delta_{j,c} = \frac{N_{l}+N_{bl}+1-j}{N_{bl} - 1}		\, .
\eeq
Figures \ref{fig3}(a) and \ref{fig3}(b) depict the stiffness modulations of Eqs.\ \eqref{delta_incr} and \eqref{delta_decr}, respectively.

%%%%%%%%%%%%%%%%%%%%%%%%%%%%%%%%%%%%%%%%%%%%%%%%%%%%%%%%%%%%%%%%%%%%%%

Figures \ref{fig3}(c) and \ref{fig3}(d) shows the displacement fields along the test chain at four different moments $t_n = nt_1$ $(n=1,2,3,4)$ where $t_1 = 44$ for modulations \ref{fig3}(a) and \ref{fig3}(b), respectively. The incident pulse  propagates from left to   right. Since a compressive wave travels with a higher speed than a tensile one in the bilinear chain, we  expect an increase  with time of the distance between compressive and tensile zones for both modulations. A zero deformation zone, which is the horizontal regions with nearly constant positive displacement, clearly indicates the gap between compressive and tensile wave fronts is increasing with time. This phenomenon always happens when a faster wave is followed by a slower one.

{Note that  Figs.\ \ref{fig3}(a) and \ref{fig3}(b) consider a stiffness modulation slope of  1/100  (maximum value of $\Delta_{j,c}$ / number of bilinear spring) with the maximum $\Delta_{j,c}$ equal to  1. By modifying the slope of the stiffness curve (increasing or decreasing the maximum value of $\Delta_{j,c}$), we  find that the length of the zero deformation regime changes. See details in Appendix  \ref{appB}.}
	
It is clear from  the different modulations of the bilinear stiffness result in distinct propagation processes. The  decreasing modulation (Fig.\ \ref{fig3}(b) and Eq.\ \eqref{delta_decr}) leads to a more effective increase in the distance between two zones than the increasing modulation (Fig.\ \ref{fig3}(a) and Eq.\ \eqref{delta_incr}) because it is evident that an almost-zero deformation zone appears between times $t_1$ and $t_2$ in Fig.\ \ref{fig3}(d). By contrast, in Fig.\ \ref{fig3}(c)  a noticeable zero deformation zone appears between $t_2$ and $t_3$.

{The modulations of Figs.\ \ref{fig3}(a) and \ref{fig3}(b) induce impedance discontinuities, which in turn cause  reflections.  The discontinuity for \ref{fig3}(a) is seen as the incident wave from the left exits the bilinear chain, and conversely the discontinuity for \ref{fig3}(b) occurs as the wave enters the bilinear section.  Interestingly, the latter situation results in a larger transmitted displacement amplitude, and in both cases the transmitted displacement amplitudes are only slightly less than the incident. It is difficult to quantify the effect of impedance discontinuities for linear-to-bilinear springs since the latter do not have a single impedance. Any model of reflectivity at such interfaces would necessarily depend upon the pulse shape; however, we do not pursue that question in this paper.}

\begin{figure}%%%%%%%%%%%%%%%%%%%%%%%%%%%%%%%%%%%%%%%%%%%%%%%%%%%%%%%%%%%%%%%%%%%%%%
%[hbt!]
\begin{center}
\subfigure[$\,$ Linearly increasing stiffness.]{
	\includegraphics[width=3.25in]{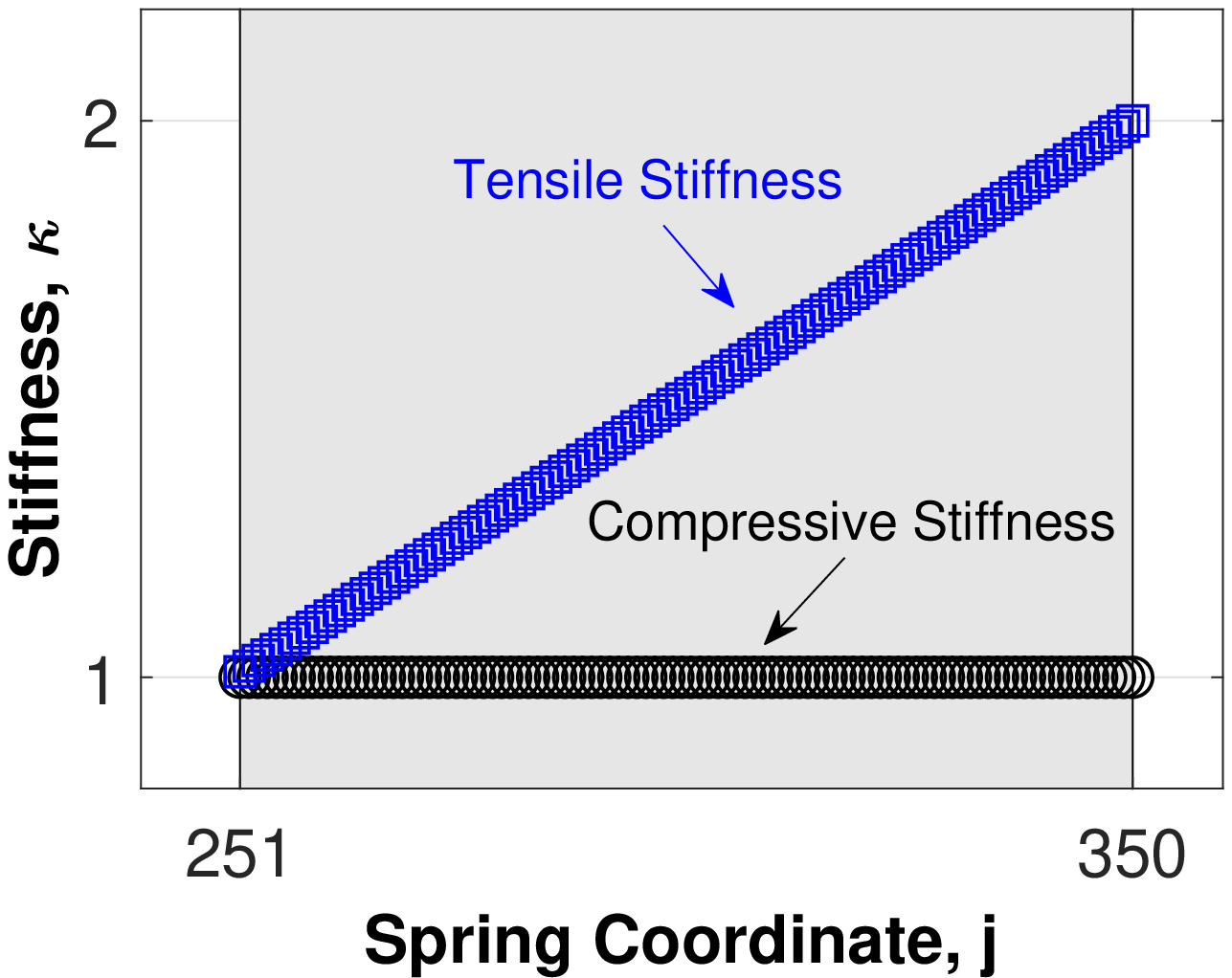}}
\subfigure[$\,$ Linearly decreasing stiffness.]{
	\includegraphics[width=3.25in]{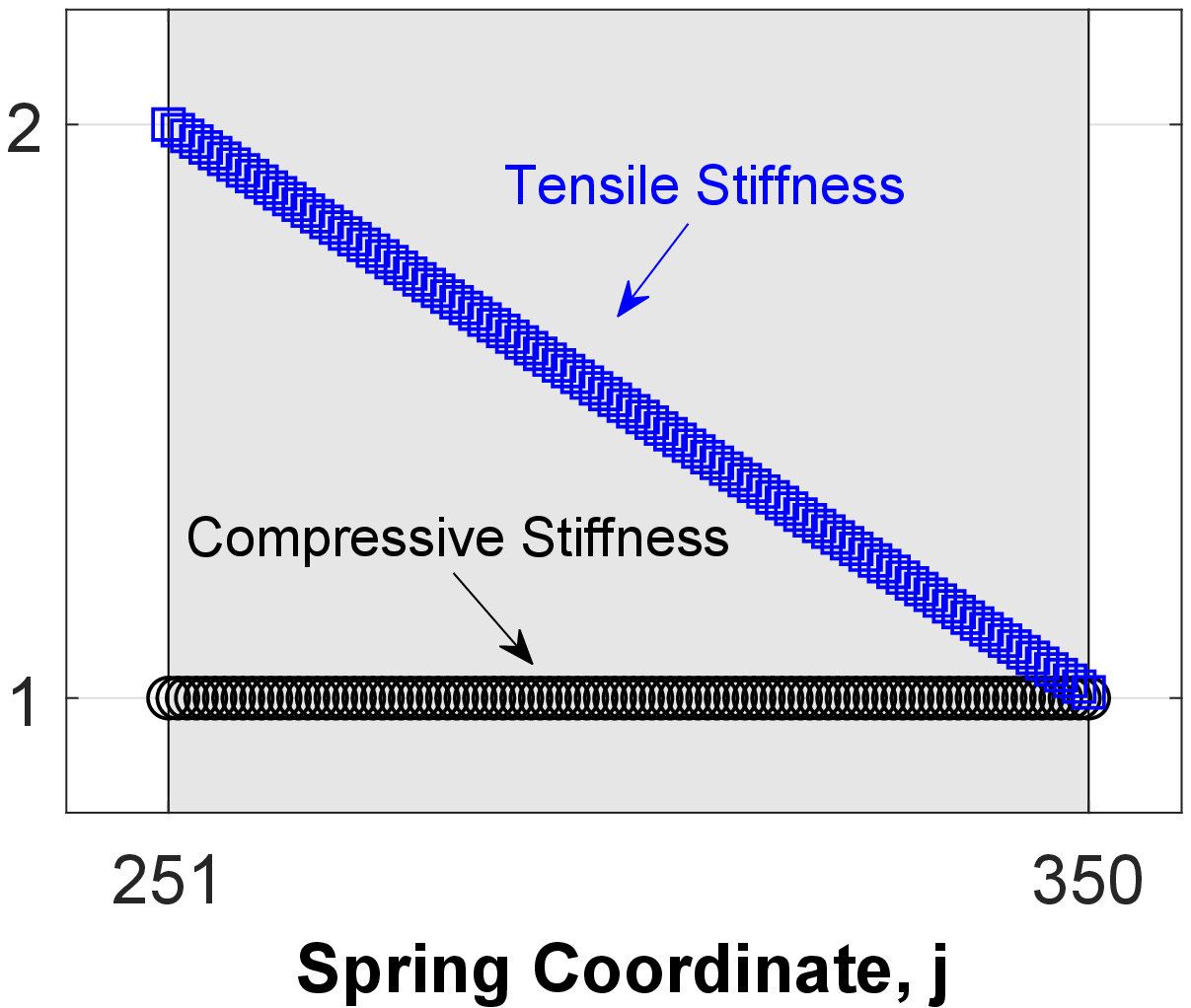}}
\subfigure[$\,$ Displacement   for modulation (a).]{
	\includegraphics[width=3.25in]{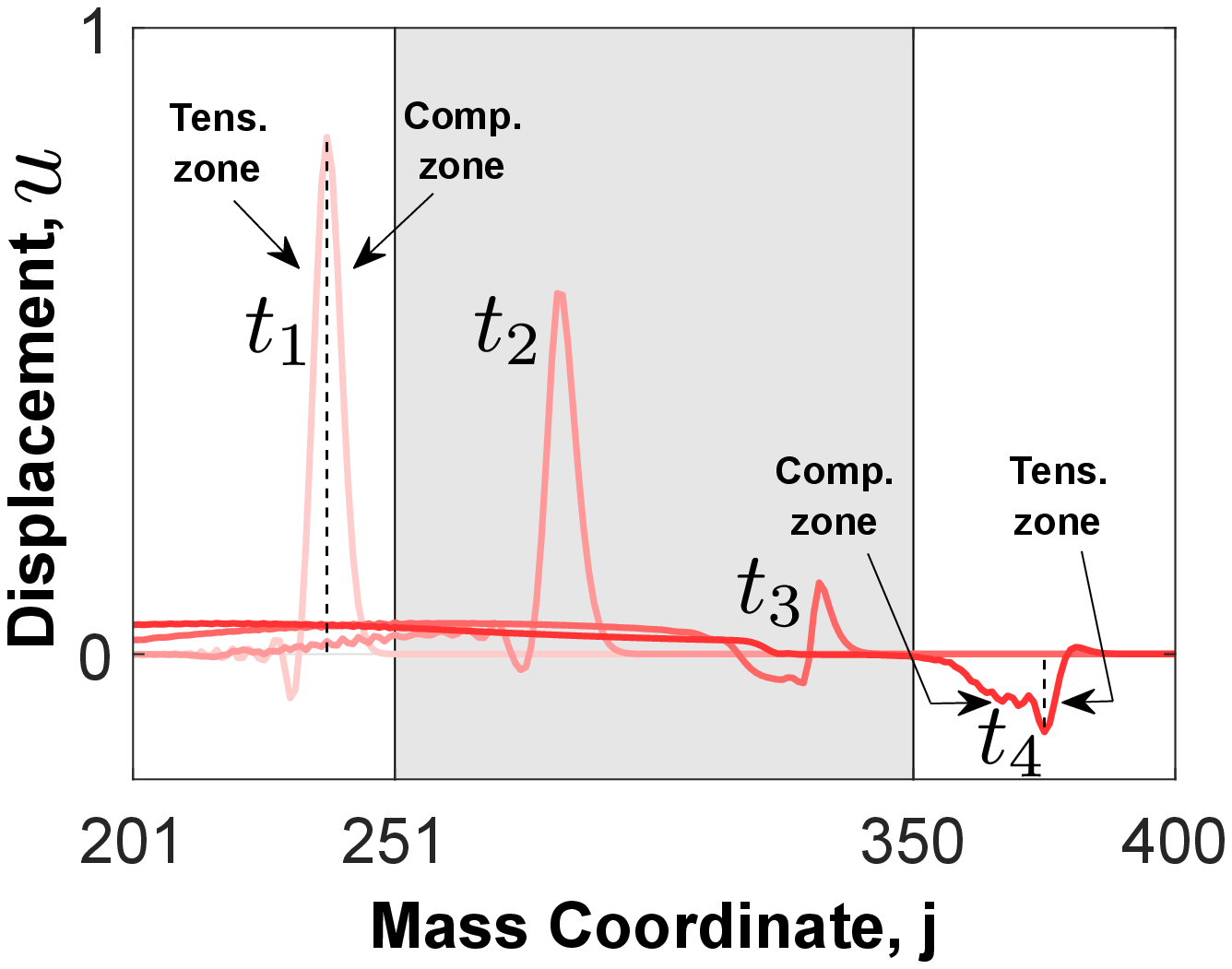}}
\subfigure[$\,$ Displacement   for modulation (b).]{
	\includegraphics[width=3.25in]{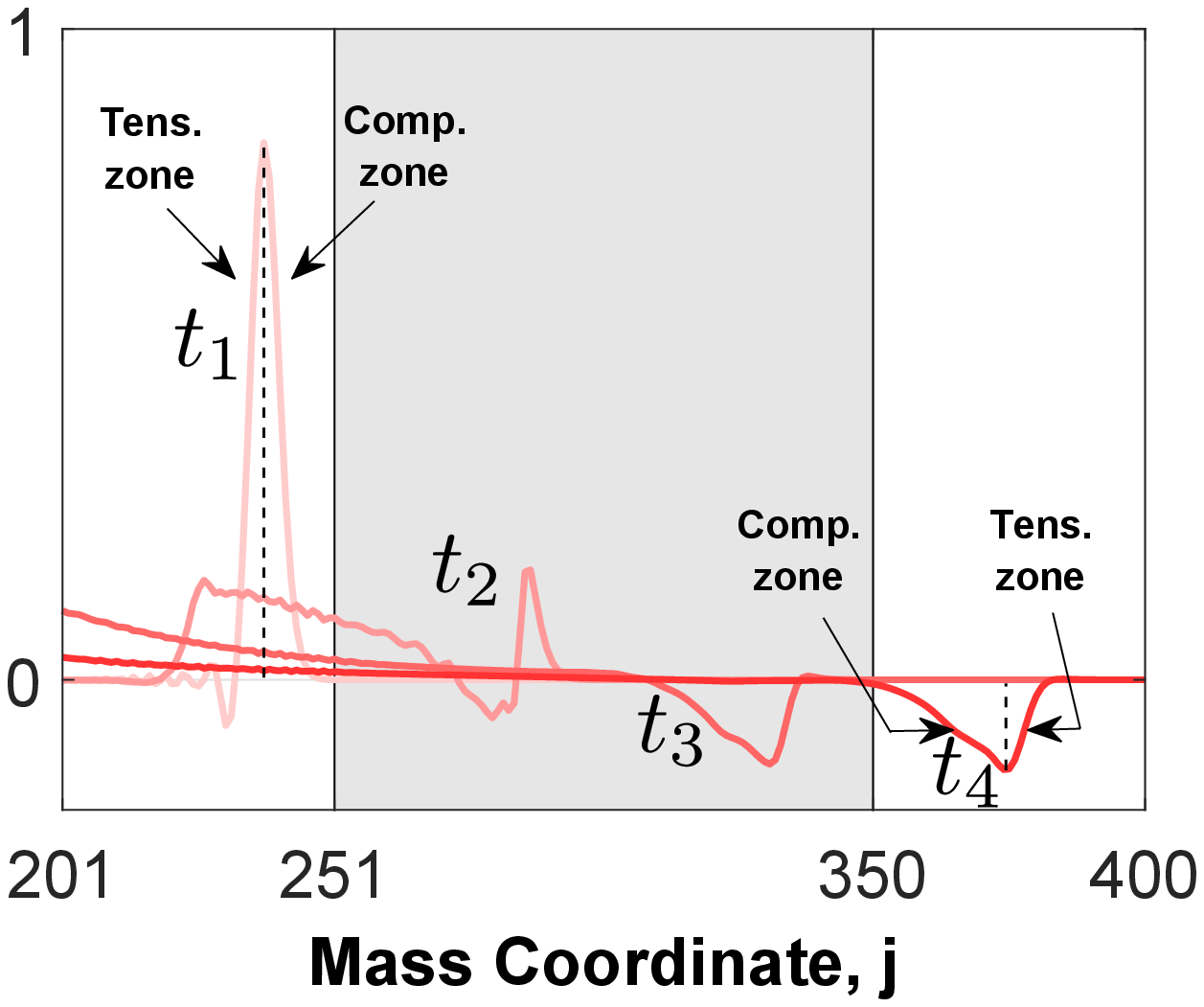}}
\caption{Dynamic properties of the bilinear spring chain with   linearly varying tensile stiffness. The two basic spatial modulations of bilinear stiffness are shown in (a) and (b) while (c) and (d) are the corresponding displacement fields of the test chain at different moments.}
\label{fig4}
\end{center}
\end{figure}

We now examine the case where the bilinear spring stiffness is greater in tension than in compression: $\Delta_{j,t} > \Delta_{j,c} = 0$. As before, two types of modulations are considered for  $N_l+1 < j \leq N_l+N_{bl}$: the linearly increasing modulation, 
\beq {delta_incr_2}
\Delta_{j,t} = \frac{j - N_{l} - 1}{N_{bl} - 1} \,  
\eeq
and  the linearly decreasing modulation, 
\beq{delta_decr_2}
\Delta_{j,t} = \frac{N_{l} + N_{bl} + 1 - j}{N_{bl} - 1} \, .
\eeq
Figures \ref{fig4}(a) and \ref{fig4}(b) show the modulations of Eqs.\ \eqref{delta_incr_2} and \eqref{delta_decr_2}, respectively.

%%%%%%%%%%%%%%%%%%%%%%%%%%%%%%%%%%%%%%%%%%%%%%%%%%%%%%%%%%%%%%%%%%%%%%
	
Figures \ref{fig4}(c) and \ref{fig4}(d) depicts the displacement fields for modulations \ref{fig4}(a) and \ref{fig4}(b), respectively. These modulations are of particular interest due to the fact that a slower wave speed is followed by a faster one in the bilinear chain. Consequently, the faster tensile wave front catches up with the slower compressive front, which makes a compression-tension pulse change to tension-compression one after transmission. This phenomenon of pulse type change always happens when a slower wave is followed by a faster one.
Similarly, we can observe  distinct propagation processes for the different modulations. The linearly decreasing modulation (Fig.\ \ref{fig4}(b) and Eq.\ \eqref{delta_decr_2}) results in an effective change of pulse type that happens between times $t_2$ and $t_3$ in Fig.\ \ref{fig4}(d). However, for the other modulation (Fig.\ \ref{fig4}(a) and Eq.\ \eqref{delta_incr_2}), the change in pulse type  takes place between $t_3$ and $t_4$ in Fig.\ \ref{fig4}(c).
	
Finally, we note the alternative tension-compression incident pulse, which is generated by setting the negative sign in Eq.\ \eqref{dimensionless_force}. Based on the previous findings  we   expect (i) a change of pulse type for the modulations with the greater compressive stiffness, and (ii)  an increase in the distance between two zones for the modulations with the greater tensile stiffness.  These expected dynamic properties are corroborated in Appendix  \ref{appB}.

To sum up,   the various spatial modulations of the bilinear stiffness investigated in this Section can be used as the building blocks for more complex bilinear chain models with the potential of significant violation of  wave reciprocity.  This possibility is explored in the next Section. 

%%%%%%%%%%%%%%%%%%%%%%%%%%%%%%%%%%%%%%%%%%%%%%%%%%%%%%%%%%%%%%%%%%%%%%
\section{Non-reciprocal Properties of the Bilinear Chain}  \label{sec4}

{\subsection{Two Non-reciprocal Models} \label{sec4.1}}

\begin{figure}[hbt!]%%%%%%%%%%%%%%%%%%%%%%%%%%%%%%%%%%%%%%%%%%%%%%%%%%%%%%%%%%%%%%%%%%%%%%
\begin{center}
\subfigure[Modulation Model I.]{
	\includegraphics[width=3.25in]{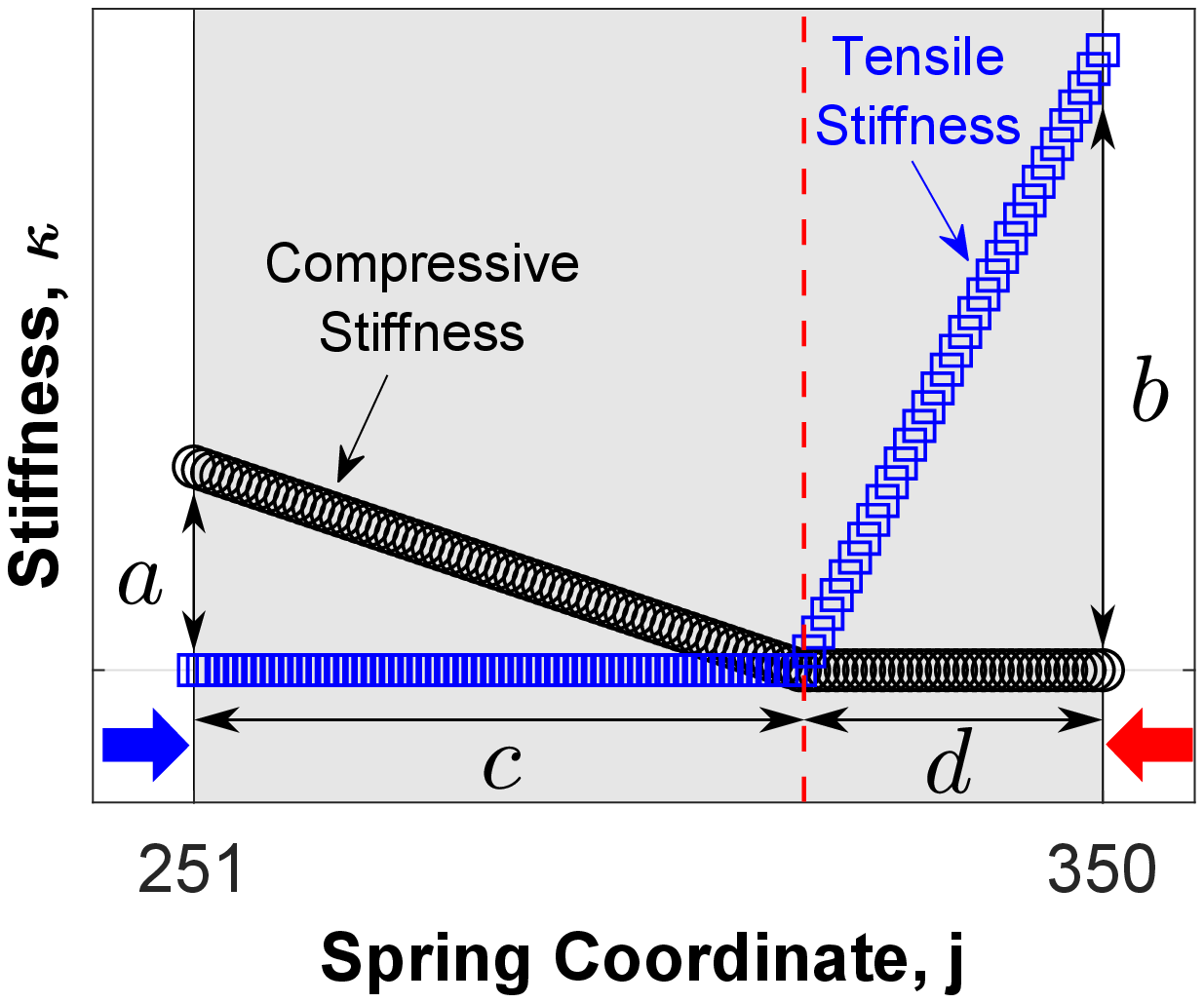}}
\subfigure[Modulation Model II.]{
	\includegraphics[width=3.25in]{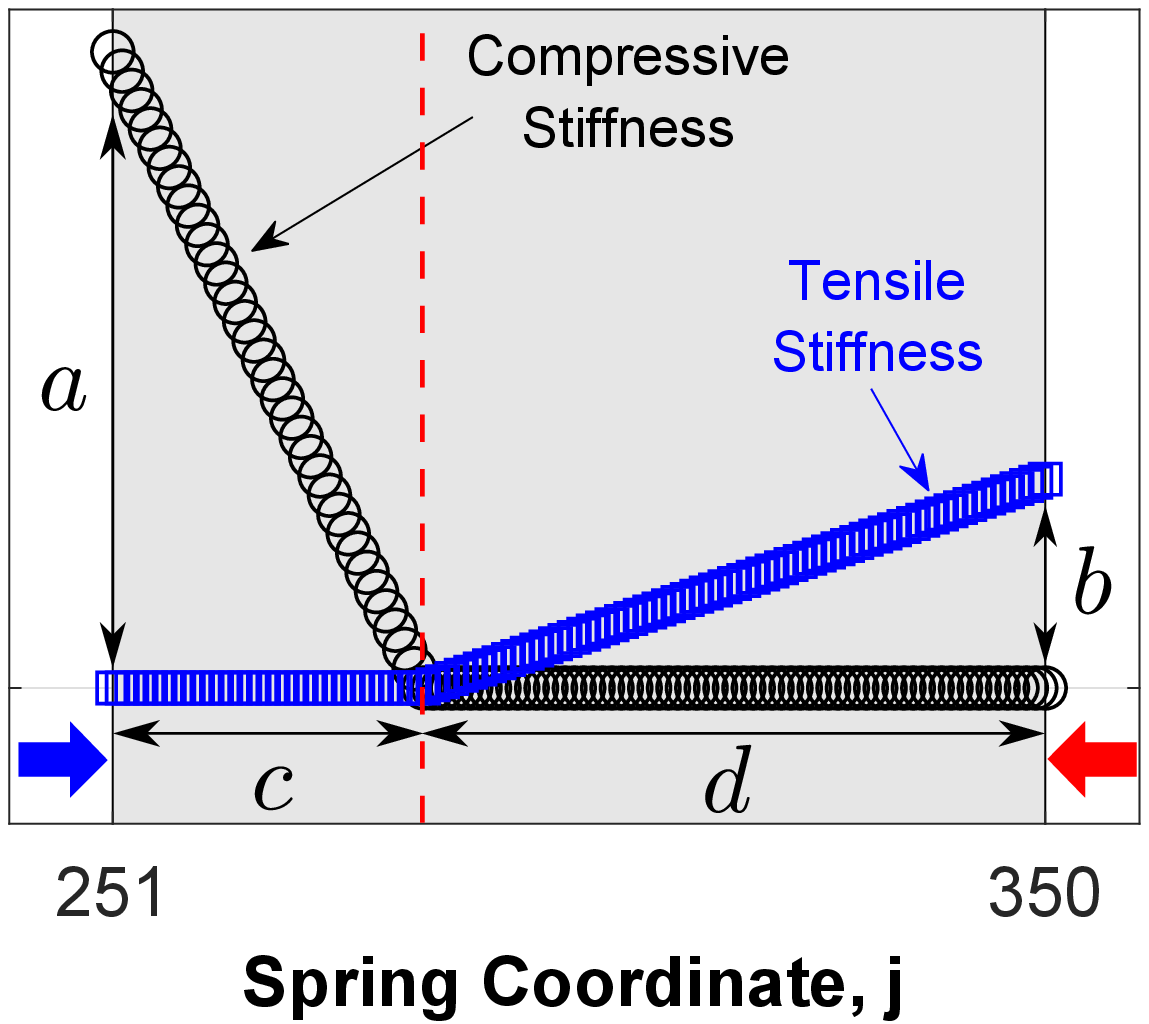}}
\caption{Asymmetric stiffness-location modulation models. The dashed line separates the bilinear chain into two parts. On the left-hand side $a$ is the maximum difference between the stiffness of the bilinear springs in compression and in tension, and $c$ is the   number of the bilinear springs. On the right-hand side, $b$ is the maximum stiffness difference, and $d$ is the   number of bilinear springs. In subsequent simulations the blue and red arrows are used to indicate the opposite pulse propagation directions: blue for incidence from the left, red from the right.}
\label{fig5}
\end{center}
\end{figure}

Based on the findings in Sect.\ \ref{sec3} we now design two spatially asymmetric and inhomogeneous  models to break wave reciprocity. The bilinear chain in either model of Fig.\ \ref{fig5} can be separated into two parts each of which is of the same form as one of the fundamental stiffness modulations discussed in Sect.\ \ref{sec3}. In the following discussions   blue and red arrows   indicate the opposite pulse propagation directions in these models; blue for incidence from the left, red from the right.
	
%%%%%%%%%%%%%%%%%%%%%%%%%%%%%%%%%%%%%%%%%%%%%%%%%%%%%%%%%%%%%%%%%%%%%%

We first define the models'  parameters. For a pulse incident from the left (blue arrow in Fig.\ \ref{fig5}), the section on the left-hand side of the dashed line is similar to the modulation in Fig.\ \ref{fig3}(b), but with different slope. The maximum difference between the compressive and tensile stiffness of the bilinear spring is $a$ and the total number of bilinear springs in this part is $c$. The section on the right is analogous to the modulation in Fig.\ \ref{fig4}(a) with the maximum stiffness difference $b$ and the total bilinear springs number $d$. The index takes the values $N_l +1 < j \leq N_l+\frac{c}{c+d} \, N_{bl}$ for the section on the left and $N_l+\frac{c}{c+d} \, N_{bl} < j \leq N_l+N_{bl}$ on the right. Hence, the dimensionless stiffness increments $\Delta_{j,c}$ and $\Delta_{j,t}$ can be expressed as
\beq{delta_1}
\Delta_{j,c} = 
\begin{cases}
a \, \, \frac{N_l \, + \, \frac{c}{c+d} \, N_{bl} \, + \, 1 \, - \, j}{\frac{c}{c+d} \, N_{bl} \, - \, 1} \, , & \, \text{left section} , \\
0 \, , & \, \text{right section} \, ,
\end{cases}
\qquad
\Delta_{j,t} = 
	\begin{cases}
		0 \, , & \, \text{left section} \, , \\
		b \, \, \frac{j \, - \, N_{l} \, - \, \frac{c}{c+d} \, N_{bl} \, - \, 1}{\frac{d}{c+d} \, N_{bl} \, 	- \, 1} \, , & \, \text{right section} \, .
    \end{cases}
\eeq
For pulse propagation from the right, the right-hand section    is analogous to the modulation in Fig.\ \ref{fig4}(b) and the section on the left to the modulation in Fig.\ \ref{fig3}(a). The dimensionless stiffness increments follow accordingly; see Appendix \ref{appC} for details. 
\begin{table}[h]%%%%%%%%%%%%%%%%%%%%%%%%%%%%%%%%%%%%%%%%%%%%%%%%
\caption{Parameters for the non-reciprocal Models of  Fig.\ \ref{fig5}, using  Eq.\ \eqref{delta_1}   to calculate the bilinear stiffness increments. For Model I, $a,c,d$ are fixed and only $b$ changes; for Model II, we set $b,c,d$ constant and vary only $a$.}
\begin{center}
\label{table2}
\begin{tabular}{c cccc}
%& & & & \\ % put some space after the caption
\hline\hline
\, & $a$ & $b$ & $c$ & \quad{} $d$ \\
\hline
Model I \quad{} & 1 & 1 $\sim$ 10 & 67 & \quad{} 33  \\
Model II \quad{} & 1 $\sim$ 10 & 1 & 33 & \quad{} 67  \\ 
\hline\hline
\end{tabular}
\end{center}
\end{table}%%%%%%%%%%%%%%%%%%%%%%%%%%%%%%%%%%%%%%%%%%%%%%%%

{Our objective is that a pulse incident from the left  produces a  transmitted pulse similar to the incident one and of comparable  amplitude. Therefore, a long enough distance between the compression and tension zone is necessary in the left section. We introduce the linearly decreasing modulation in this section. If we set $a$ small but $c$ large as Model I in Fig.\ \ref{fig5}(a) shows, the effect of the other section on the right is consequently weak because of the linearly increasing modulation  and the small value of $d$. For Model I, we have $a<b$ and $c>d$. Alternatively, we set $a$ large but $c$ small in the left section for Model II depicted in Fig.\ \ref{fig5}(b), and consequently the effect of the other section is weak because of the linearly increasing modulation  and the small value of $b$. We have $a>b$ and $c<d$ for Model II.} 
When a pulse is incident from the right we require that the catch-up phenomenon takes place sequentially in each section to gradually minimize the pulse amplitude. In order to ensure the change in pulse in the section on the right, we apply the linearly decreasing modulation in that region. We can either set $b$ large and $d$ small as Model I in Fig.\ \ref{fig5}(a) shows or $b$ small but $d$ large as Model II in Fig.\ \ref{fig5}(b) shows for this section. {The linearly increasing modulation  and the small value of $a$ in Model I ($a<b$ and $c>d$) and  increasing modulation plus small $c$ in Model II ($a>b$ and $c<d$), causes the catch-up process in the section on the left to be relative weak for both models.} 
\begin{figure}[h!]%%%%%%%%%%%%%%%%%%%%%%%%%%%%%%%%%%%%%%%%%%%%%%%%%%%%%%%%%%%%%%%%%%%%%%
%[t]
\begin{center}
\subfigure{
	\includegraphics[width=3.25in]{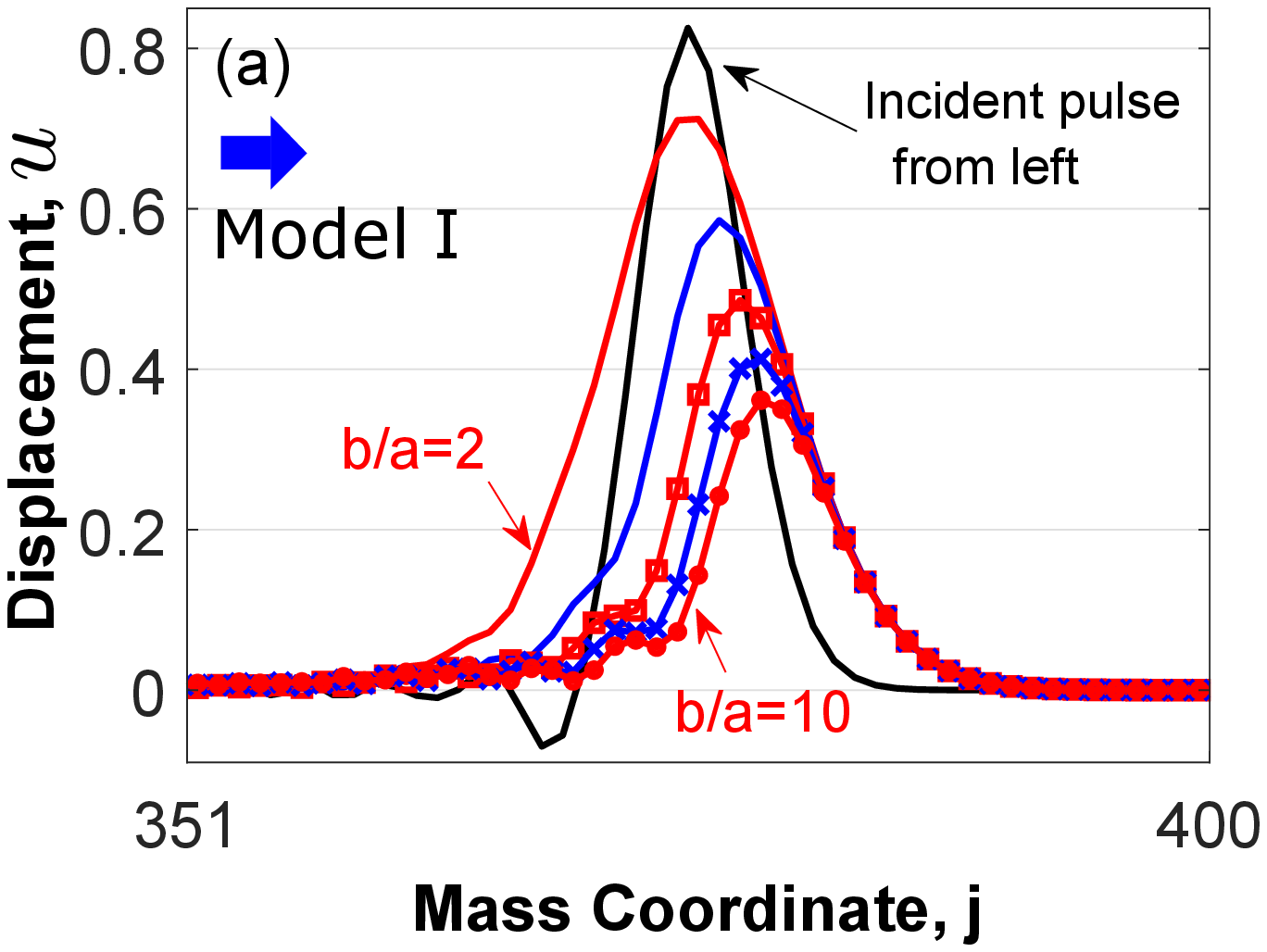}}
\subfigure{
	\includegraphics[width=3.25in]{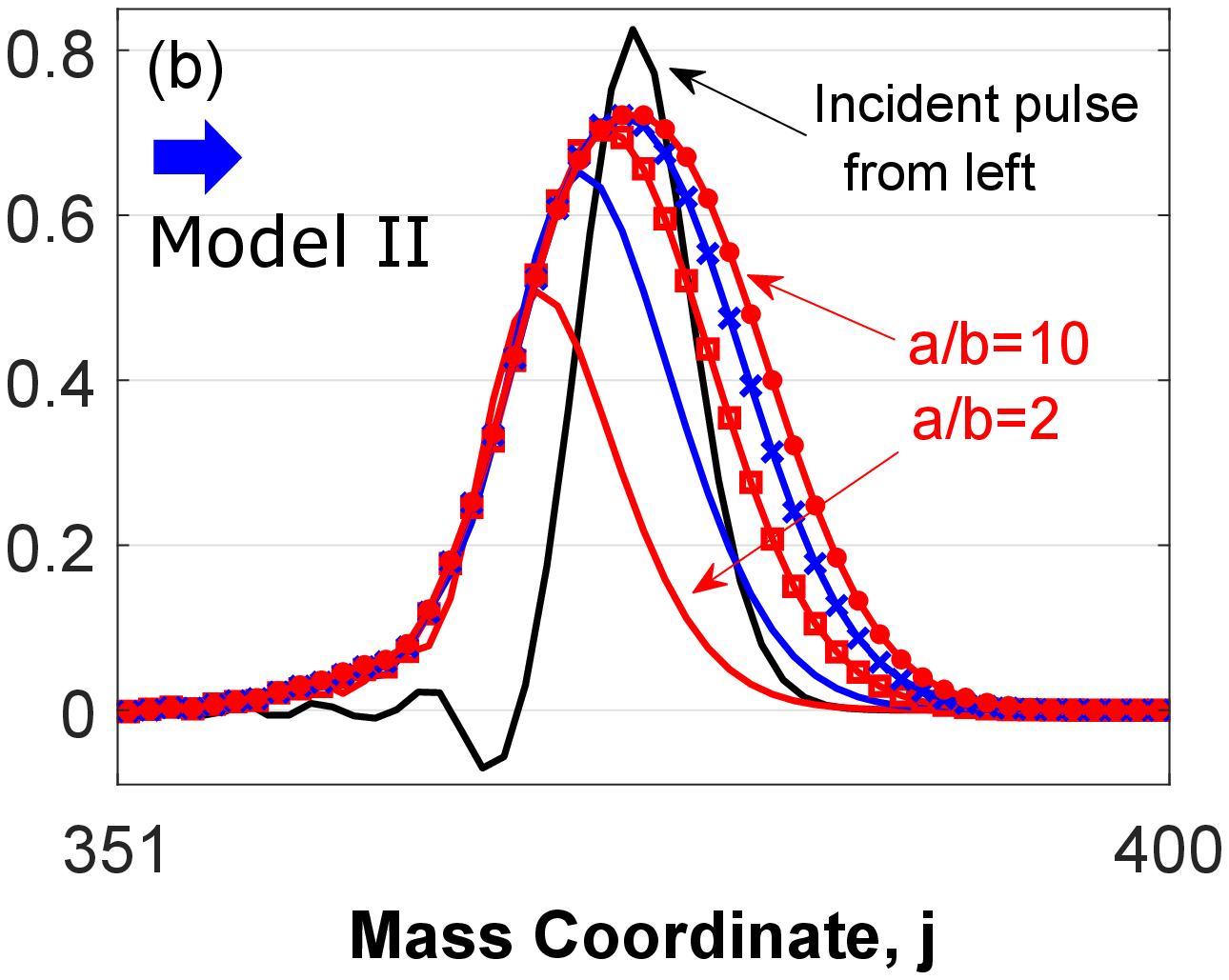}}
	%\vspace{-.5in}
\subfigure{
	\includegraphics[width=3.25in]{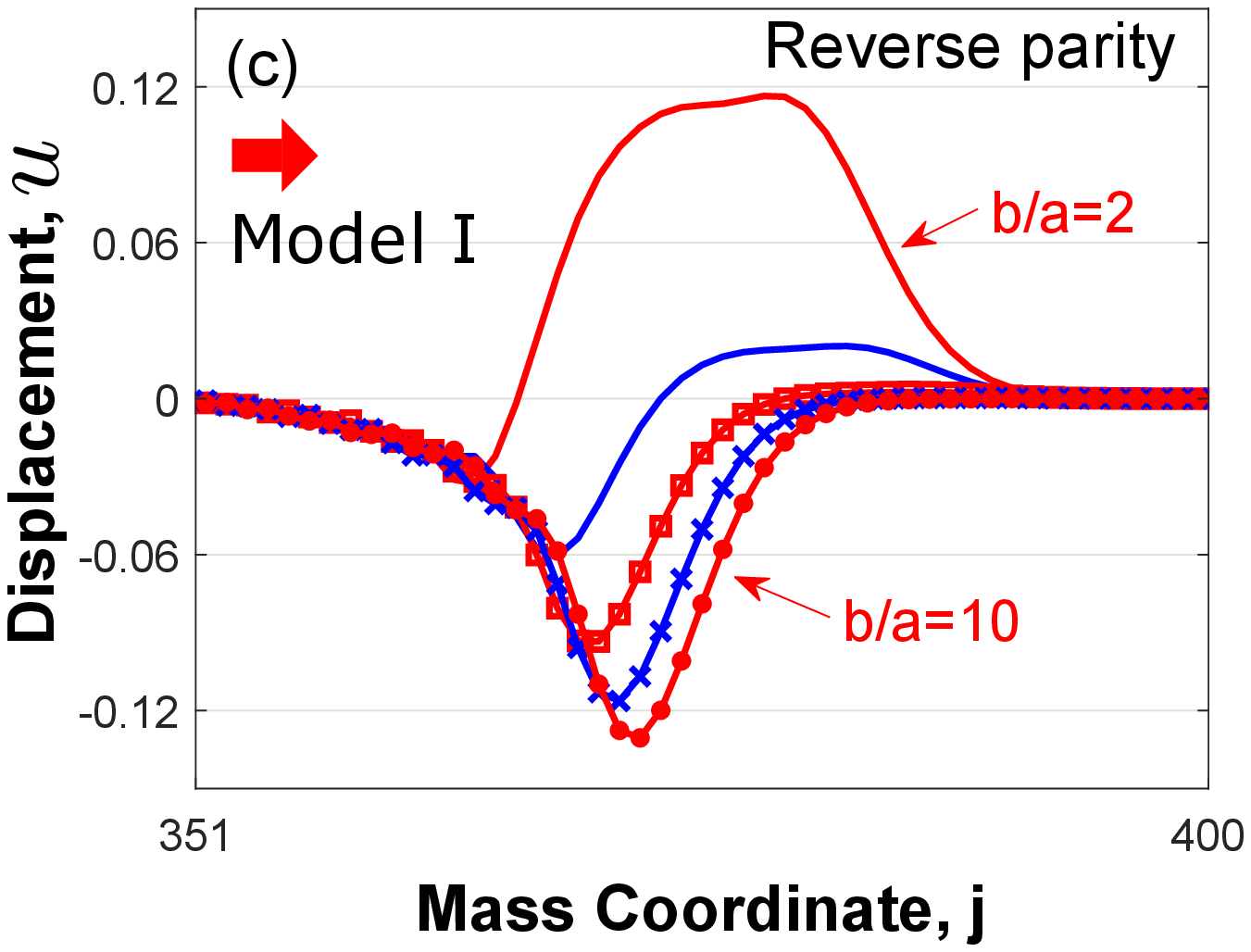}}
\subfigure{
	\includegraphics[width=3.25in]{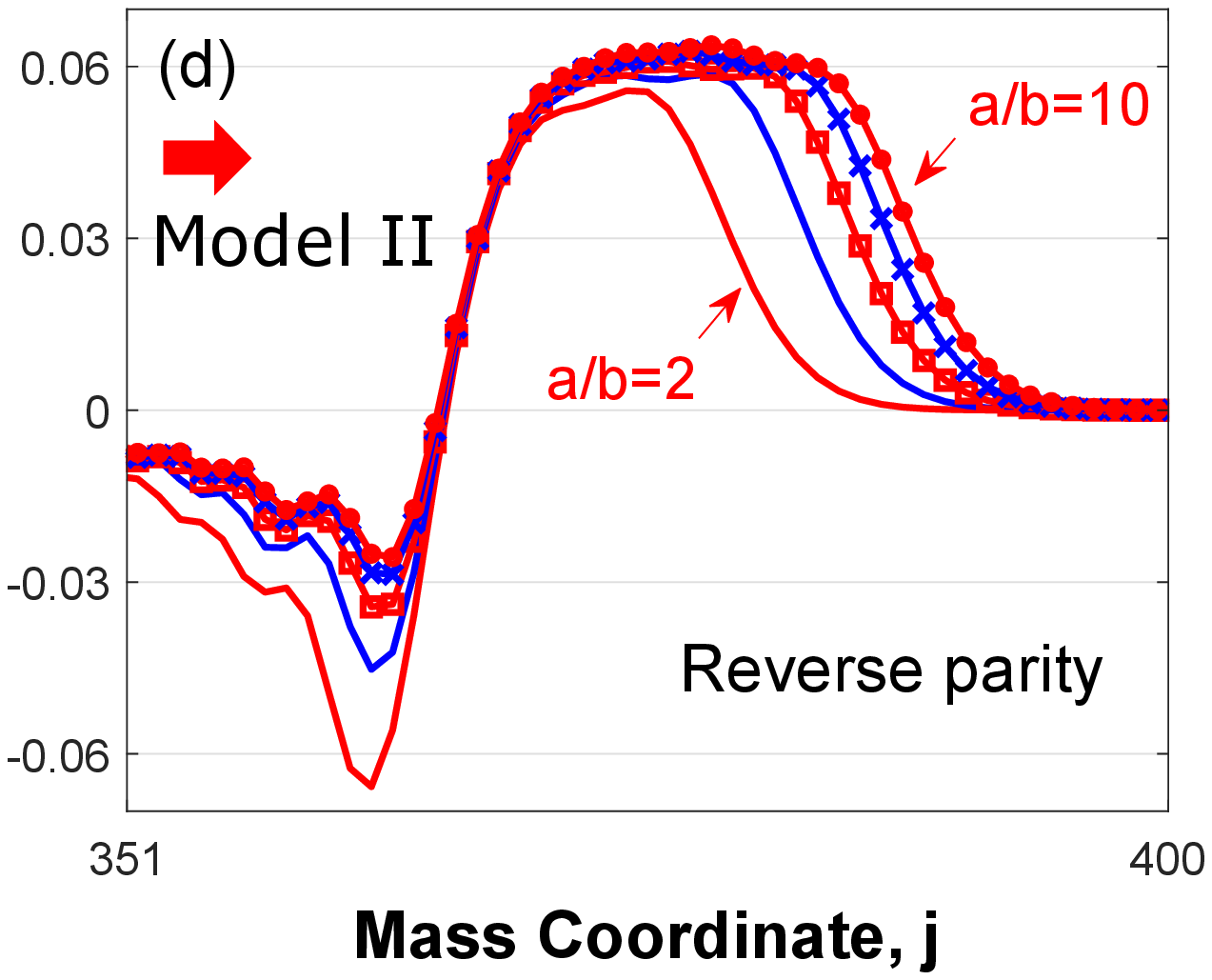}}
\caption{Transmission amplitudes for the two Models in Fig.\ \ref{fig5}:   Model I   on the left, (a) and (c), and   Model II   on the right, (b) and (d).  The black curve is the incident compression-tension  pulse; the red and blue {curves} are the transmitted waves. Incidence from the left (labeled by the blue arrow) for   Model I  is in (a) and from the right (labeled by red arrow) in (c), in both cases for  $b/a = 2,4,6,8,10$, demonstrating very significant non-reciprocal transmission; Similar phenomena are evident in (b) and (d), which consider  $a/b = 2,4,6,8,10$ for  Model II. }
\label{fig6}
\end{center}
\end{figure}%%%%%%%%%%%%%%%%%%%%%%%%%%%%%%%%%%%%%%%%%%%%%%%%%%%%%%%%%%%%%%%%%%%%%%

{\subsection{Optimal  Non-reciprocity} \label{sec4.15}}

Numerical simulations were performed to test for  non-reciprocal transmission. All results presented in this Section are obtained using the  parameters of Table \ref{table:parameters} with $100$ bilinear springs. 

We start with Model I in Fig.\ \ref{fig5}(a) for an incident compression-tension (CT) pulse. We fix the values of $a,c,d$ listed in Table \ref{table2} and vary the value of $b$ only. 
Figures\ \ref{fig6}(a) and \ref{fig6}(c) show the transmitted pulse amplitudes for different values of $b/a$. The incident pulse  from the left first shows an effective increase in the distance between compressive and tensile zones in the section on the left. Then the slow catch-up decreases the pulse amplitude in the section on the right. We therefore expect a compression-tension pulse with diminished  amplitude after transmission as shown in Fig.\ \ref{fig6}(a). Conversely,  a pulse propagating from the right first has an effective catch-up in the section on the right, which results in a change of  pulse type. In the section on the left, the second catch-up process evolves slowly. We therefore expect a TC pulse if no pulse type change occurs, e.g., cases $b/a = 6, 8, 10$ in Fig.\ \ref{fig6}(c), or the existence of both pulse types such as case $b/a = 4$, or a CT pulse with the smaller amplitude if the second change of pulse type happens like case $b/a = 2$.  Most importantly,  the amplitudes of the transmitted waves from the opposite directions are of  different orders of magnitude,  as evident from Figs.\ \ref{fig6}(a) and \ref{fig6}(c),  demonstrating significant non-reciprocal transmission.

\begin{figure}[h]%%%%%%%%%%%%%%%%%%%%%%%%%%%%%%%%%%%%%%%%%%%%%%%%%%%%%%%%%%%%%%%%%%%%%%
\begin{center}
\subfigure{
	\includegraphics[width=3.25in]{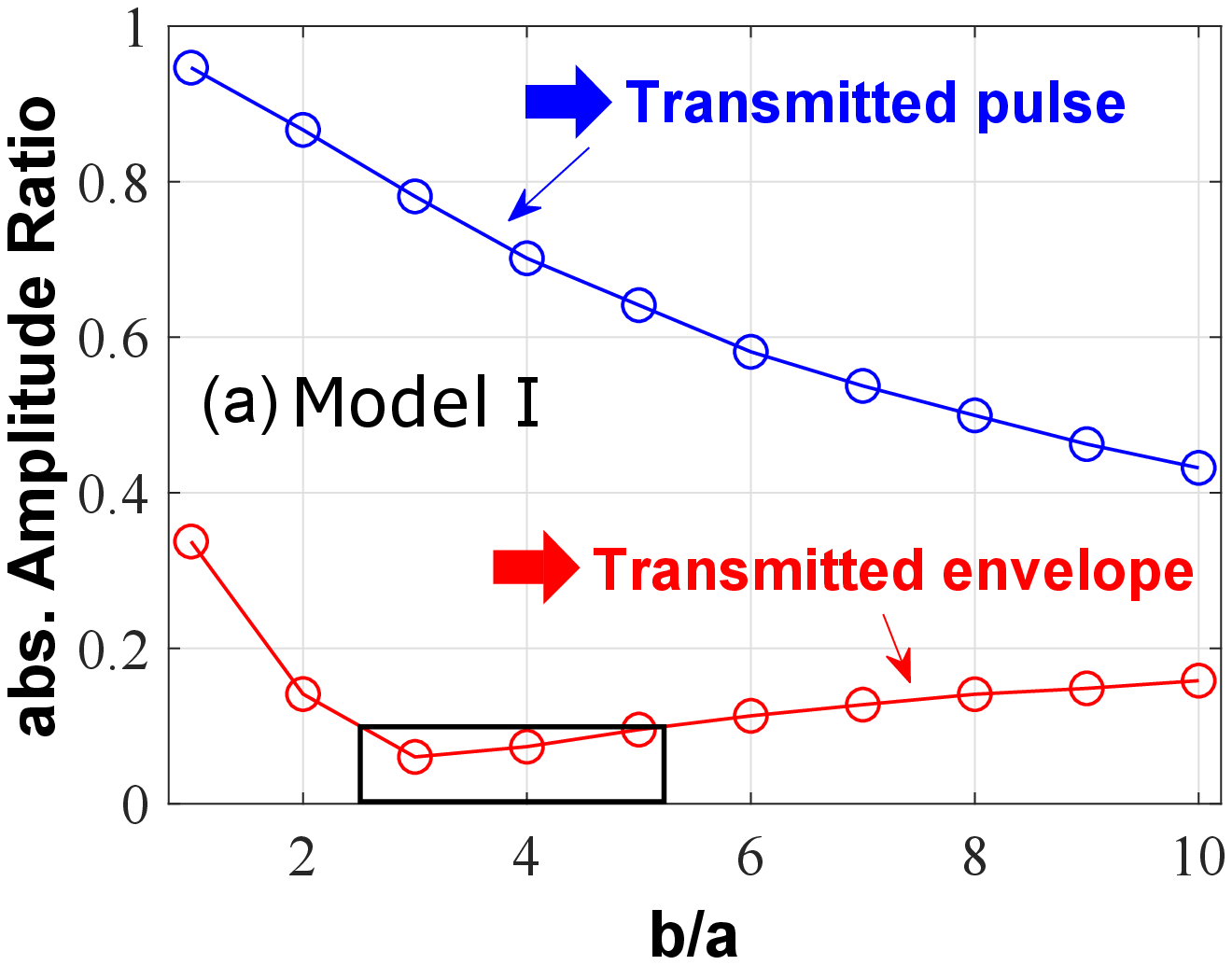}}
\subfigure{
	\includegraphics[width=3.25in]{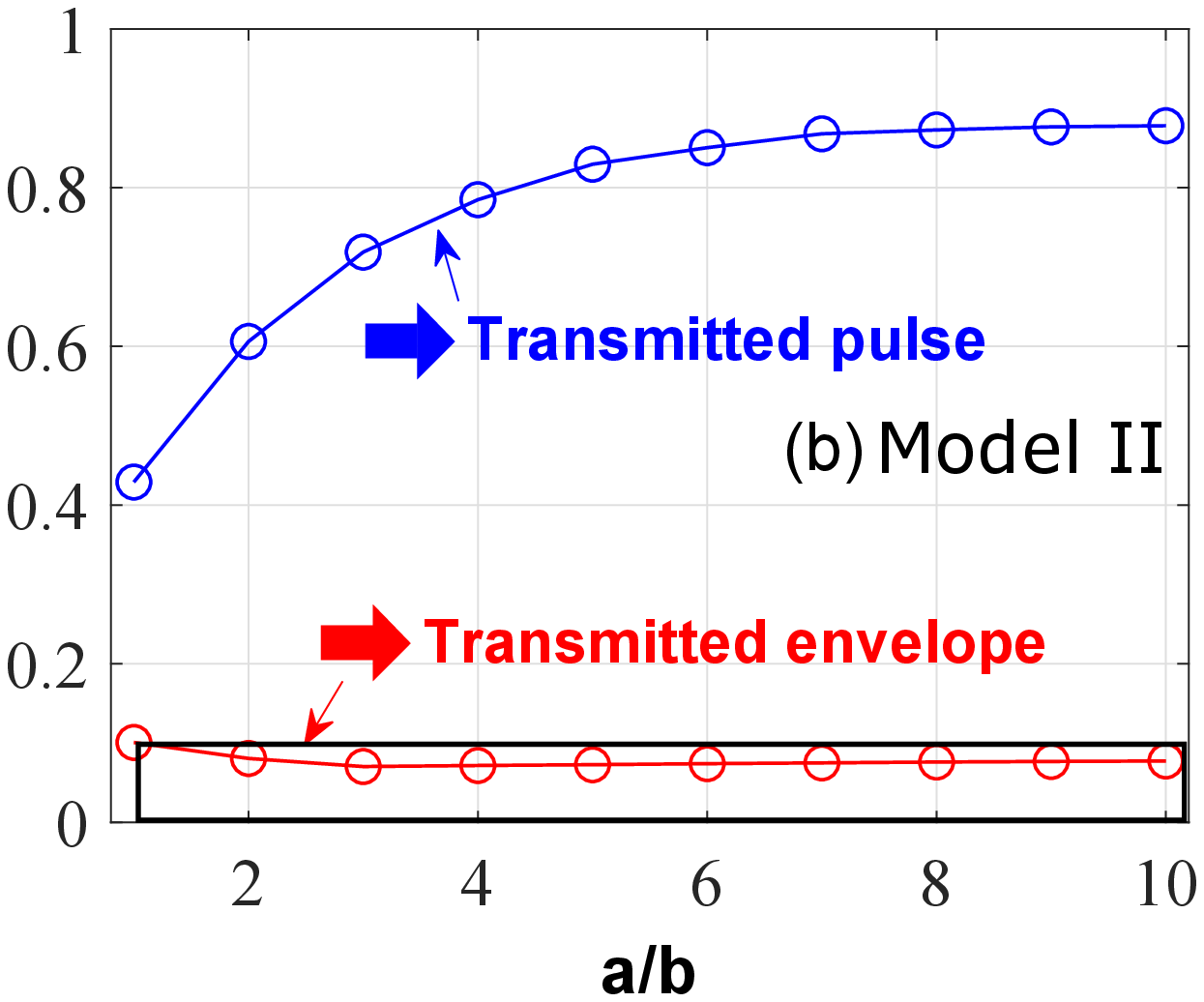}}
\caption{Transmission ratios for the two Models in Fig.\ \ref{fig5}. 
The transmission is quantified by measuring the amplitude ratios of the transmitted and incident pulses based on the data in Fig.\ \ref{fig6}. Pulses maintain the same type (CT) before and after transmission for incidence from the left; the transmitted envelopes consist of the CT and TC pulse for incidence from the right and we pick one of the greater absolute value for calculation. 
We consider that almost-zero transmission occurs when the absolute amplitude ratio is less than $10 \%$. The boxes indicate the range of $b/a$ and $a/b$ values that satisfy almost-zero transmission. 
%Both Models  realize almost one-way propagation because the pulse propagation from the right results in zero transmission while the transmitted pulse from the left remains (about one order of magnitude difference between their amplitudes). 
Model II has a broader parameter range for one-way propagation.}
\label{fig7}
\end{center}
\end{figure}%%%%%%%%%%%%%%%%%%%%%%%%%%%%%%%%%%%%%%%%%%%%%%%%%%%%%%%%%%%%%%%%%%%%%%
Transmission from opposite incidence directions is    quantified by calculating the amplitude ratios of the transmitted and incident pulses based on the data in Fig.\ \ref{fig6}.  These demonstrate significant wave non-reciprocity.  Figure \ref{fig7}(a) shows results for all $b/a$ cases. 
%this includes two plots for the propagation from the right because the transmitted envelope consists of two types of pulses. 
{Since the transmitted envelope consists of two types of pulses, we only take one of the greater absolute value into consideration.} 
%Zero transmission is of particular interest due to the possibility of one-way propagation. 
Here, we consider fully non-reciprocal or one-way propagation  to be approximately achieved when the absolute amplitude ratio in Fig.\ \ref{fig7}(a) is less than $10 \%$. To find the value of $b/a$ that satisfies this almost-zero transmission condition, we introduce a box to Fig.\ \ref{fig7}(a). The upper and lower edges of the box respectively indicates the $10 \%$ and $0$ amplitude ratios, and left and right edges give us the boundaries of the $b/a$ range. Since the pulse propagation from the right results in almost zero transmission but the transmitted pulse from the left remains significant in amplitude, and the difference between the amplitude ratios for the opposite propagation direction is about one order of magnitude, we conclude that essentially one-way propagation can be realized using Model I.
	
Next, Model II in Fig.\ \ref{fig5}(b) is considered, with only the value of $a$ varied and the ratio $a/b$ is used as the indicator for different tests;  the values of $b,c,d$ are listed in Table \ref{table2}.  The compression-tension pulse goes through the exactly same process as we described in the first Model. Figures \ref{fig6}(b) and \ref{fig6}(d) show the transmitted pulse amplitudes from the opposite directions for various values of $a/b$, demonstrating evident non-reciprocal transmission. In Fig.\ \ref{fig7}(b), we find a broader parameter range that gives almost-zero transmission in  Model II, as compared with Model I. As a result, one-way propagation can be realized in Model II over a wide range of parameters, allowing  tuning of the transmitted pulse.
\begin{figure}[h]%%%%%%%%%%%%%%%%%%%%%%%%%%%%%%%%%%%%%%%%%%%%%%%%%%%%%%%%%%%%%%%%%%%%%%
\begin{center}
\subfigure{
	\includegraphics[width=3.25in]{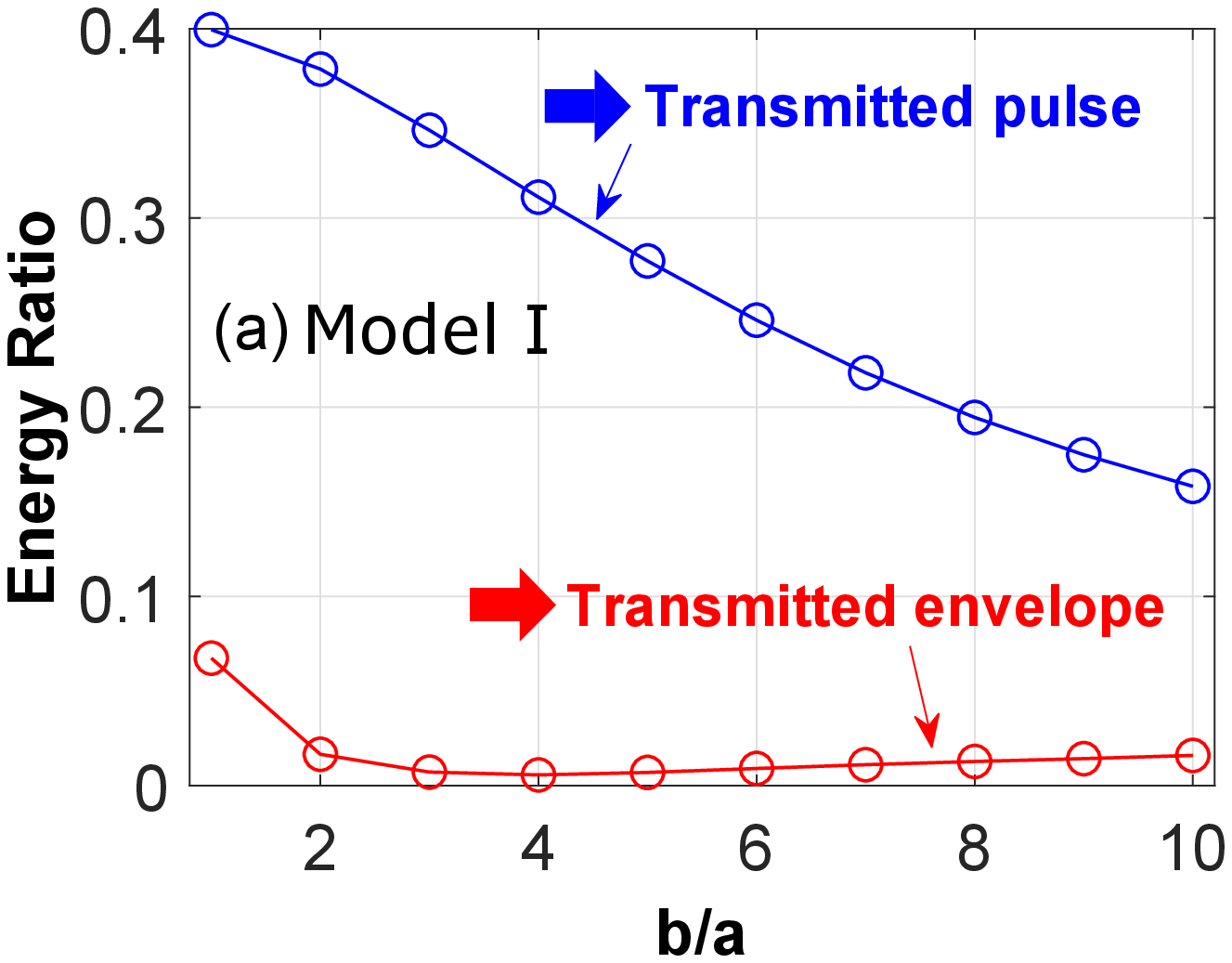}}
\subfigure{
	\includegraphics[width=3.25in]{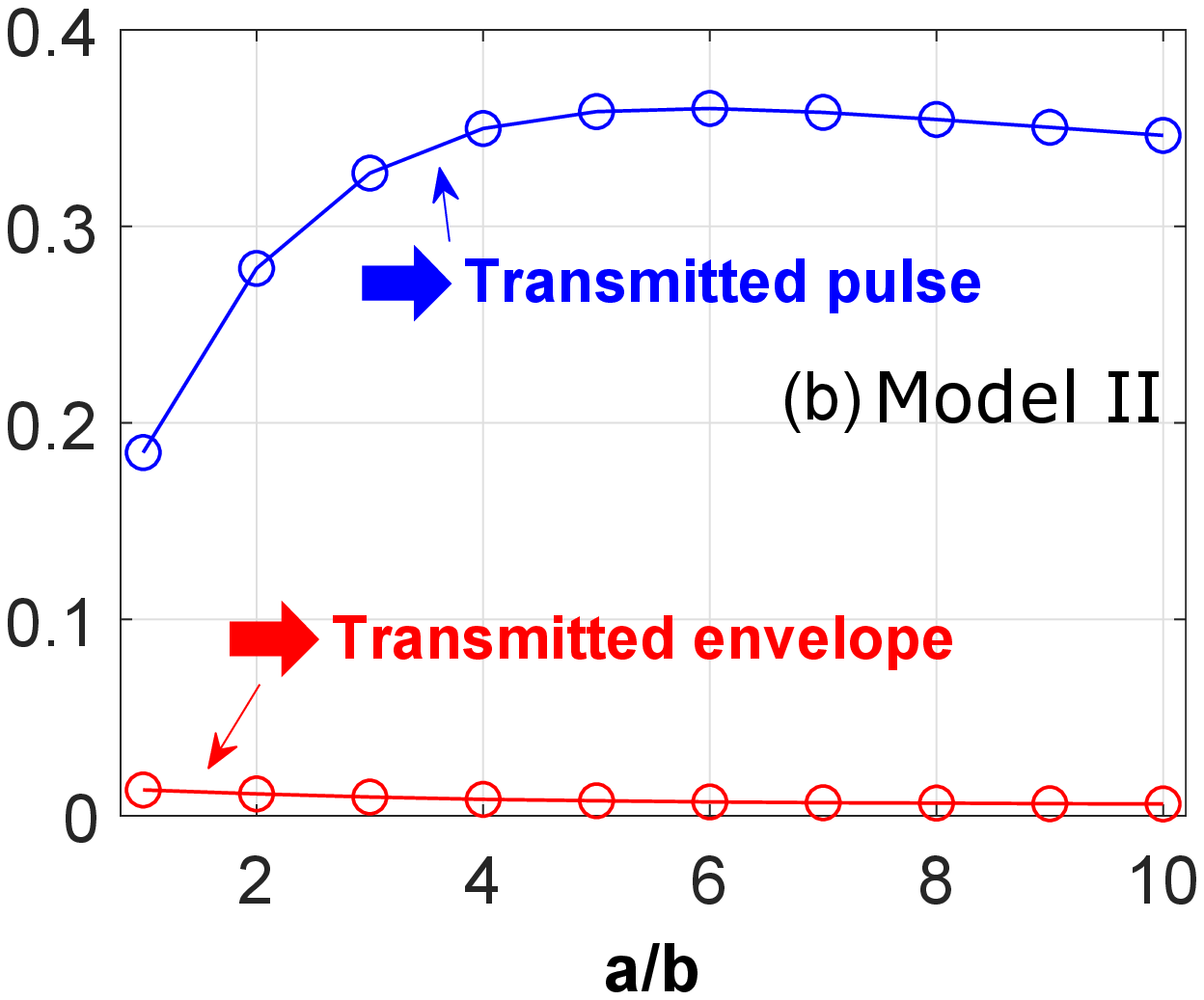}}
{\caption{Transmitted energy ratios for the two Models in Fig.\ \ref{fig5}.  
%Both energy and absolute amplitude ratios show the similar tendencies against corresponding parameters as compared with . 
The  values of $b/a$ and $a/b$ which satisfy the almost-zero transmission condition in Fig.\ \ref{fig7} also produce  significant non-reciprocal  energy transmission.}
\label{fig8}}
\end{center}
\end{figure}%%%%%%%%%%%%%%%%%%%%%%%%%%%%%%%%%%%%%%%%%%%%%%%%%%%%%%%%%%%%%%%%%%%%%%

{Finally, we consider the non-reciprocal effect from the  perspective of total incident and transmitted energy,  kinetic plus   potential.
We calculate the input energy after the external forcing is fully applied; the transmitted energy is measured after the transmitted pulse has fully entered the linear part of the chain system as shown in Fig.\ \ref{fig6}. The velocities of the masses yields the kinetic energy and the relative displacements between  adjacent masses defines potential energy.
We focus on the non-dimensional ratio of transmitted to incident energy. Figure.\ \ref{fig8}(a) shows the ratio for various $b/a$ values in Model I. Comparison with the results in Fig.\ \ref{fig7}(a) shows that both absolute amplitude ratios and energy ratios show similar tendencies as functions of $b/a$. This phenomenon is also found in Fig.\ \ref{fig8}(b) which depicts all $a/b$ cases in Model II. Therefore, $b/a$ and $a/b$ values which satisfy the almost-zero transmission condition from the amplitude perspective  also produce  significantly low transmitted energy. Furthermore, we  find that a considerable  amount of the  incident energy is lost in transmission, over half as Fig.\ \ref{fig8} shows.  The remaining energy is mainly reflected   with a small amount lost  from damping.}

{Only the compression-tension (CT) incident pulse has been considered in this Section.  Non-reciprocal wave behavior is also found for TC input pulses, with   details given in Appendix \ref{appD}. }

%%%%%%%%%%%%%%%%%%%%%%%%%%%%%%%%%%%%%%%%%%%%%%%%%%%%%%%%%%%%%%%%%%%%%%
{\subsection{The Effect of  Incident Frequency on Non-reciprocity} \label{sec4.2}}

{%In previous Sections, we use a constant external forcing frequency ($\omega = 0.5$) to test the fundamental spatial modulations and non-reciprocal models. 
Here, we consider non-reciprocal transmission for a given model optimized for one value of input frequency $\omega$ and examine how it behaves for different incident frequencies. }
\begin{figure}[h!]%%%%%%%%%%%%%%%%%%%%%%%%%%%%%%%%%%%%%%%%%%%%%%%%%%%%%%%%%%%%%%%%%%%%%%
\begin{center}
\subfigure[Transmitted amplitude]{
	\includegraphics[width=3.25in]{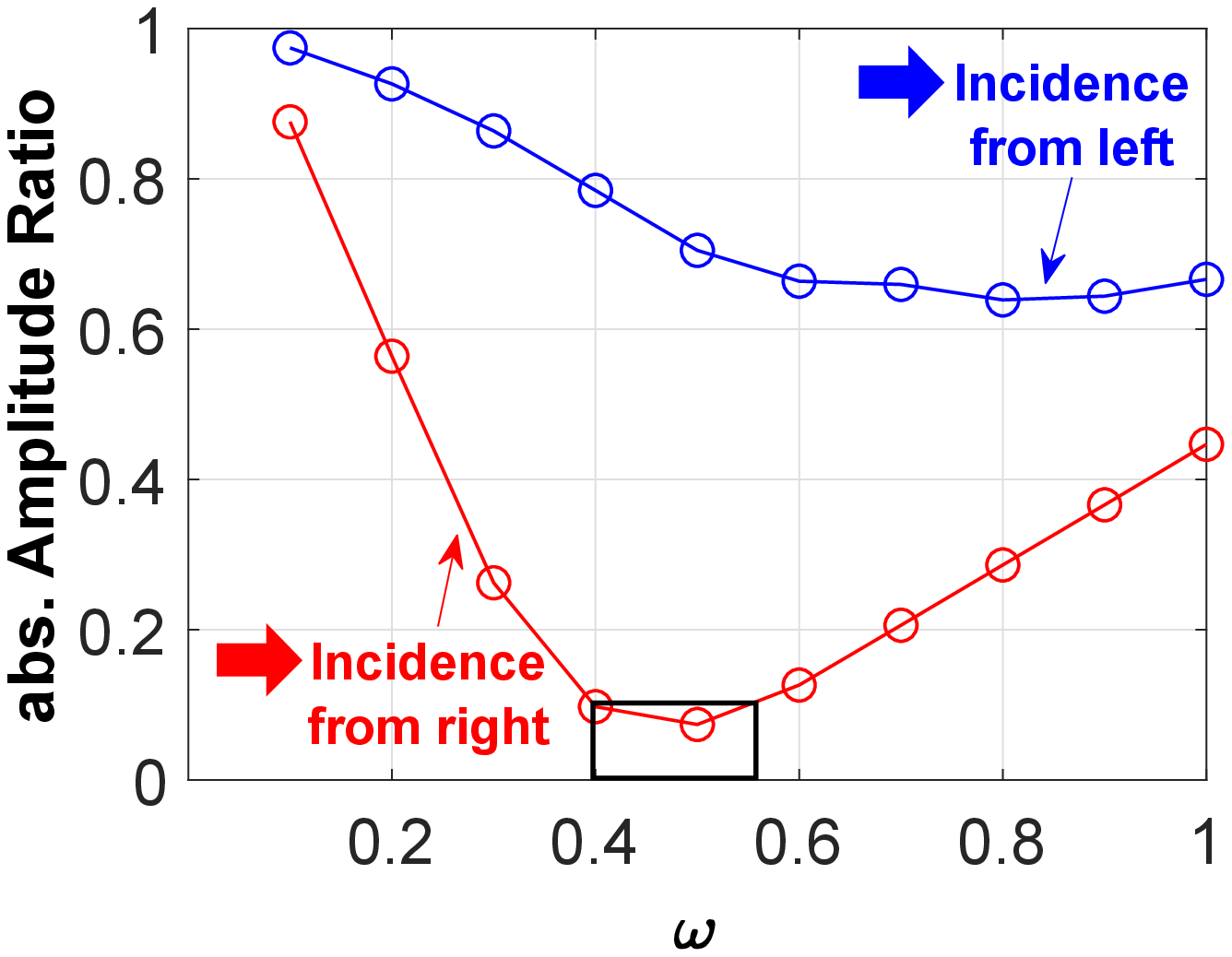}}
\subfigure[Transmitted energy]{
	\includegraphics[width=3.25in]{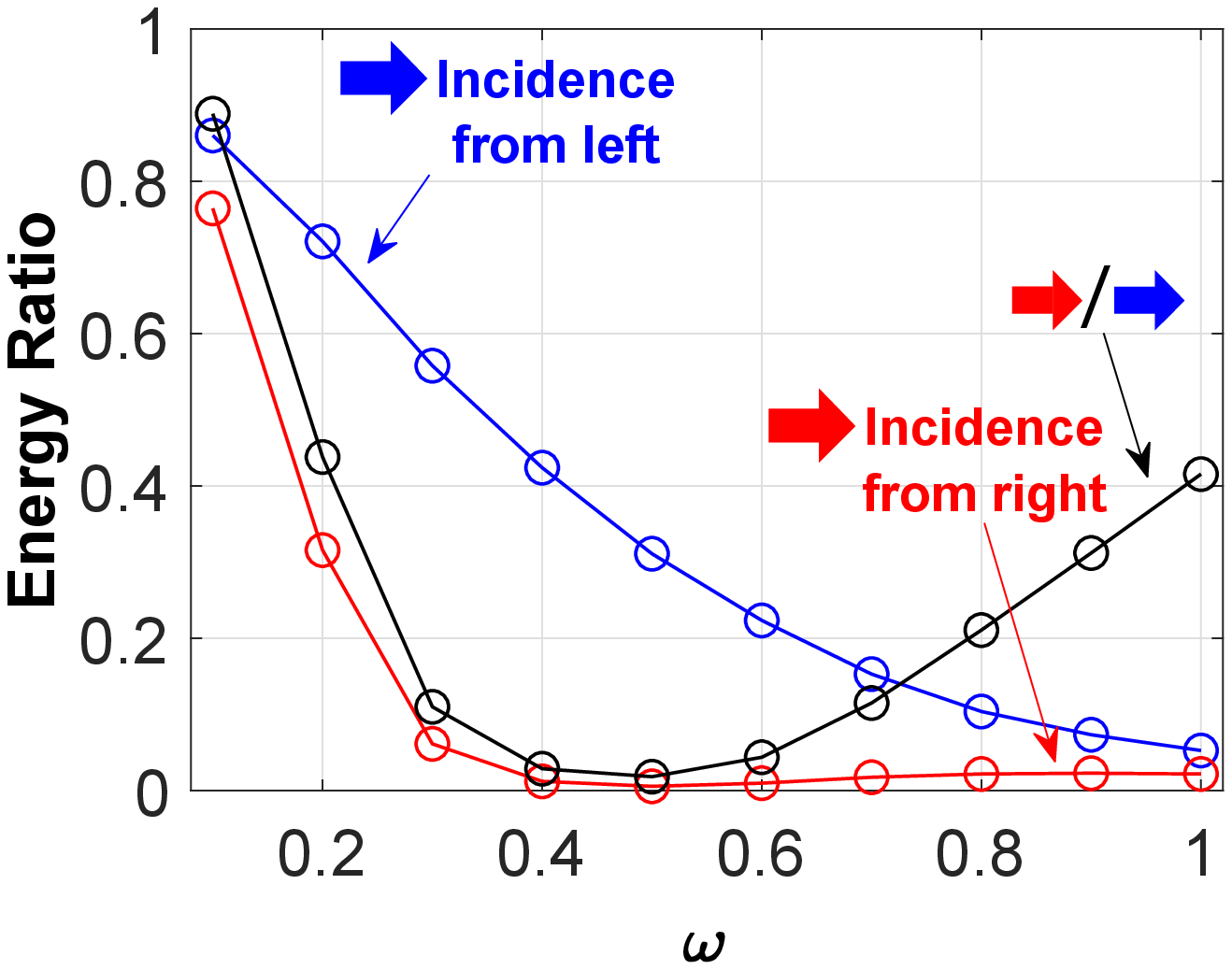}}
{\caption{Transmission as a function of frequency for  Model I with  $b/a = 4$ from amplitude and energy perspectives. The box in (a) indicates the frequency range for almost-zero transmission according to its definition (ratio $< 10 \%$). 
The black curve in (b) is the ratio of the transmitted energy for incidence from the left vs.\ incidence from the right.}  
%The results indicate approximate one-way propagation Only a small frequency range around $\omega = 0.5$ is able to realize .}
\label{fig9}}
\end{center}
\end{figure}%%%%%%%%%%%%%%%%%%%%%%%%%%%%%%%%%%%%%%%%%%%%%%%%%%%%%%%%%%%%%%%%%%%%%%

{Significant non-reciprocity was found in  Sect.\ \ref{sec4} for the case  $b/a = 4$, based on the results  in Fig.\ \ref{fig7}(a) and \ref{fig8}(a). We therefore focus on  Model I with   fixed $b/a = 4$ and examine what happens as $\omega$ is varied.  Figure.\ \ref{fig9}(a) shows that the  transmitted amplitude is strongly dependent on the incident frequency. The almost-zero transmission condition (absolute amplitude ratio $< 10 \%$), is satisfied over a relatively small frequency range around $\omega = 0.5$, as the box in Fig.\ \ref{fig9}(a) shows.  The behavior of the right-to-left energy ratio,  shown in Figure.\ \ref{fig9}(b), is consistent  with this conclusion.
These limited parameter studies suggest that the bilinear non-reciprocity effect is a relatively narrow band phenomenon.}

{However, for a given input forcing frequency $\omega$, there is always  a model (a set of parameters) giving significant non-reciprocal transmission. We consider non-dimensional frequency from $\omega = 0.3$ to $0.8$ for  Model I and  vary the parameter $b/a$ to obtain optimal non-reciprocity (the rest of parameters remain the same as Table\ \ref{table2}). Table\ \ref{table_omega_ii} shows the parameter $b/a$ and corresponding transmission ratios when significant non-reciprocal propagation occurs. We  note that all absolute amplitude ratios in this table are  less than $15 \%$ and all energy ratios are less than $1  \%$.} 

\begin{table}[h]%%%%%%%%%%%%%%%%%%%%%%%%%%%%%%%%%%%%%%%%%%%%%%%%%%%%%%%%%%%%%%%%%%%%%%
{\caption{The model parameter $b/a$ and corresponding transmission ratios when non-reciprocal wave motion occurs for different values of the incident pulse frequency.  $A_L$ and $E_L$ are the  amplitude and energy magnitudes for incidence from the left, relative to incident values.  
$A_{R/L}$ and $E_{R/L}$ represent  ratios for incidence from the right vs.\ incidence from the left.
}
\label{table_omega_ii}}
\begin{center}
\begin{tabular}{c ccccc }
%& & & & & \\ % put some space after the caption
\hline\hline
 $\omega$  &   $b/a$   &  $A_L$ & $A_{R/L}$ &  $E_L$ & $E_{R/L}$ \\
\hline
0.3 &  10 & 0.660 & 0.126 & 0.407 & 0.033  \\
0.4 &  6 & 0.672 & 0.099 & 0.356 & 0.024 \\
0.6 &  2 & 0.825 & 0.088 & 0.311 & 0.025  \\
0.7 &  1.5 & 0.876 & 0.110 & 0.207 & 0.051 \\
0.8 &  1.2 & 0.757 & 0.146 & 0.147 & 0.095  \\ 
\hline\hline
\end{tabular}
\end{center}
\end{table}%%%%%%%%%%%%%%%%%%%%%%%%%%%%%%%%%%%%%%%%%%%%%%%%%%%%%%%%%%%%%%%%%%%%%%

%{From Table \ref{table_omega_ii}, we notice that the transmitted energy is quite low for forcing frequency $\omega$ of higher value (e.g. for $\omega = 0.8$, the transmitted energy is less than $15 \%$). Thus, it is of less interest to discuss cases of relatively lower transmitted energy. Hence, lower frequency $\omega$ is preferable for demonstrations and applications.  We assume that almost-zero transmission occurs when the  amplitude ratio is smaller than $10 \%$ and the energy ratio smaller than $1.5 \%$.}

%%%%%%%%%%%%%%%%%%%%%%%%%%%%%%%%%%%%%%%%%%%%%%%%%%%%%%%%%%%%%%%%%%%%%%
{\subsection{ Propagation of Different Types of Incident Wave}
\label{appE}}

{Our design of non-reciprocal bilinear system is based on a single cycle of a  CT or TC pulse. Here we  explore more general scenarios where the incident wave may consist of several cycles of a single pulse type or  is a combination of both pulse types. We consider  Model I with $b/a = 4$ for testing (see Fig.\ \ref{fig5}(a), with the other parameters the same as in Table \ref{table2}) .}

\begin{figure}[h] %%%%%%%%%%%%%%%%%%%%%%%%%%%%%%%%%%%%%%%%%%%%%%%%%%%%%%%%%%%%%%%%%%%%%%
%[hbt!]
\begin{center}
\subfigure{
	\includegraphics[width=0.45\textwidth]{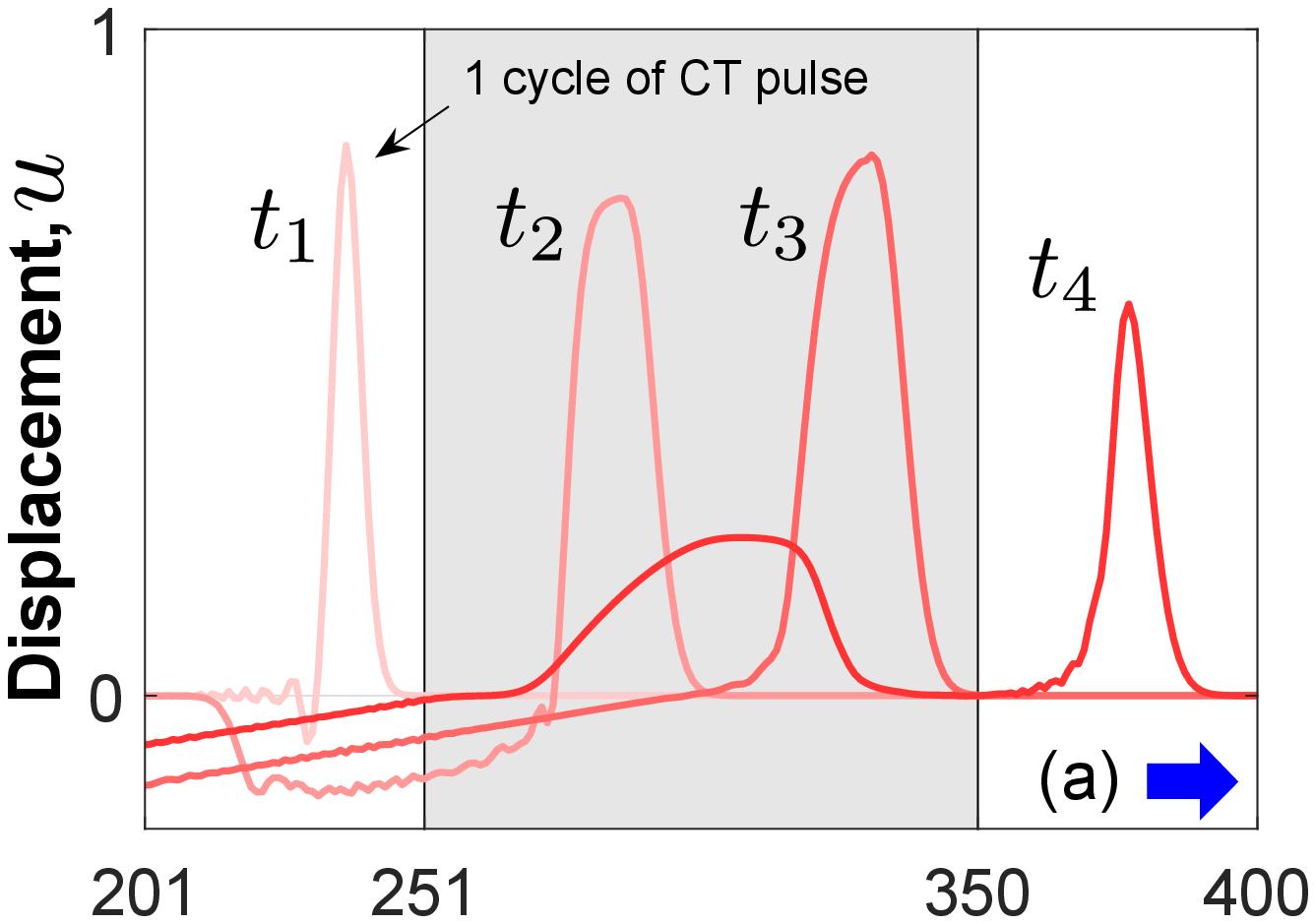}}
\subfigure{
	\includegraphics[width=0.45\textwidth]{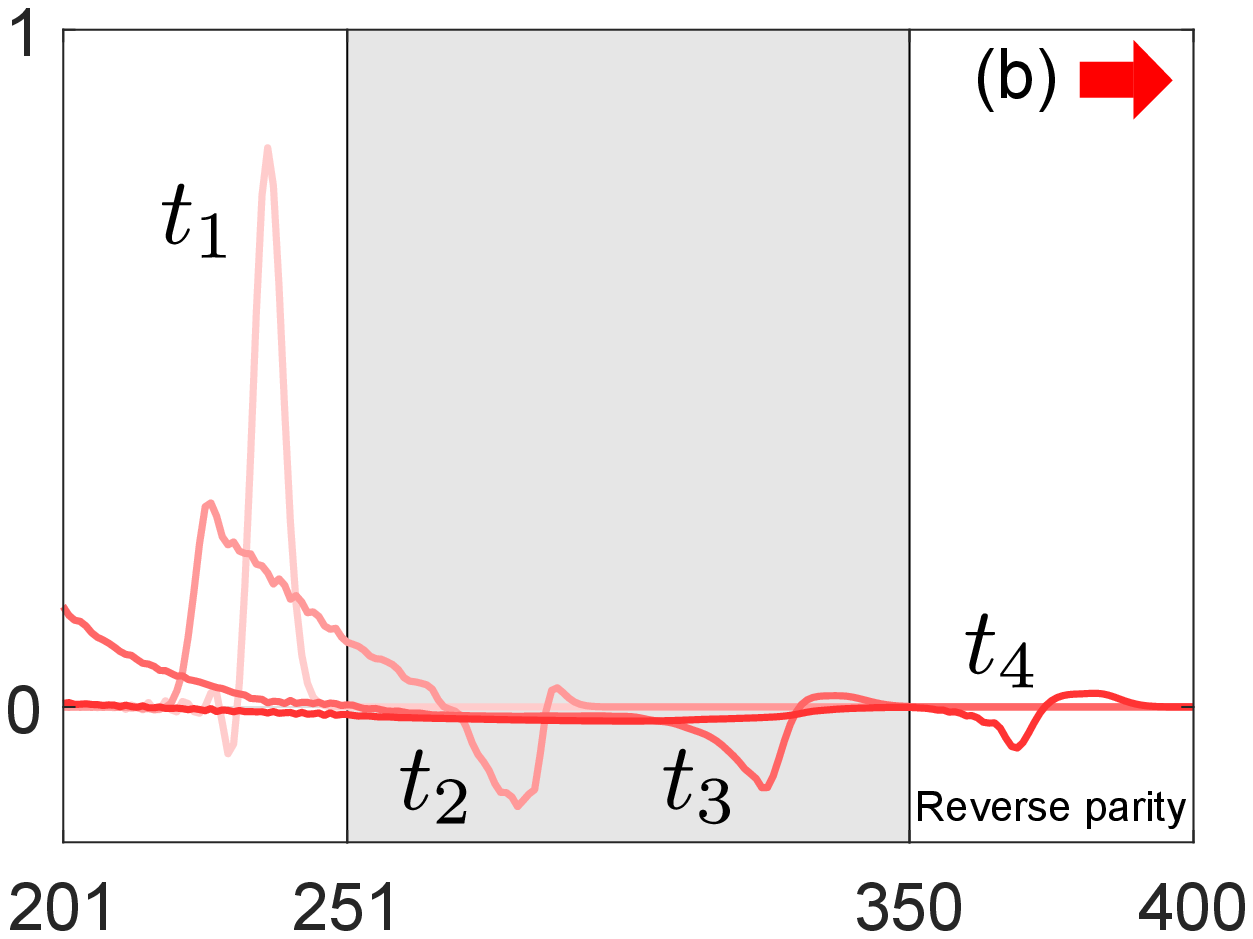}}
	\subfigure{
	\includegraphics[width=0.45\textwidth]{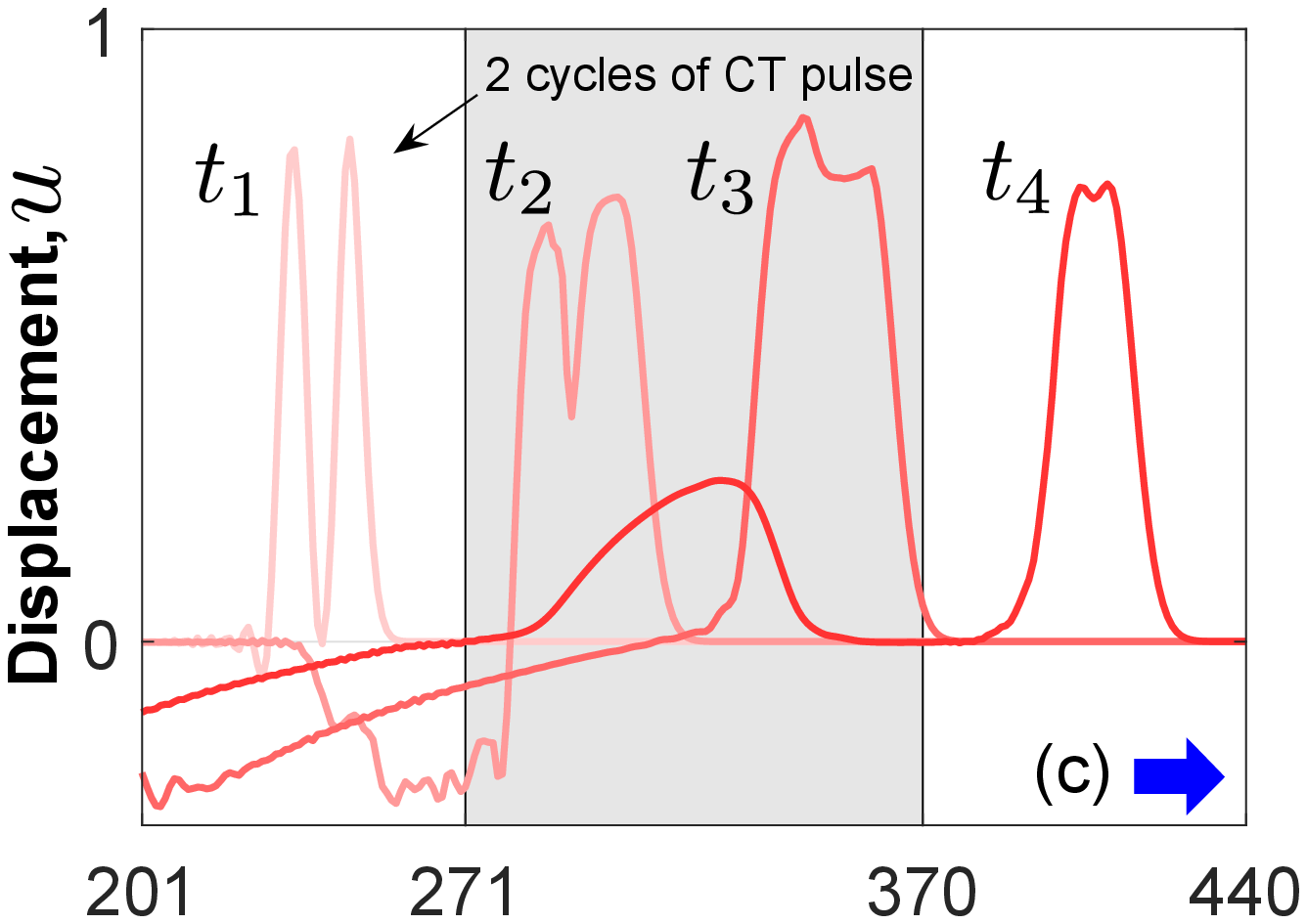}}
\subfigure{
	\includegraphics[width=0.45\textwidth]{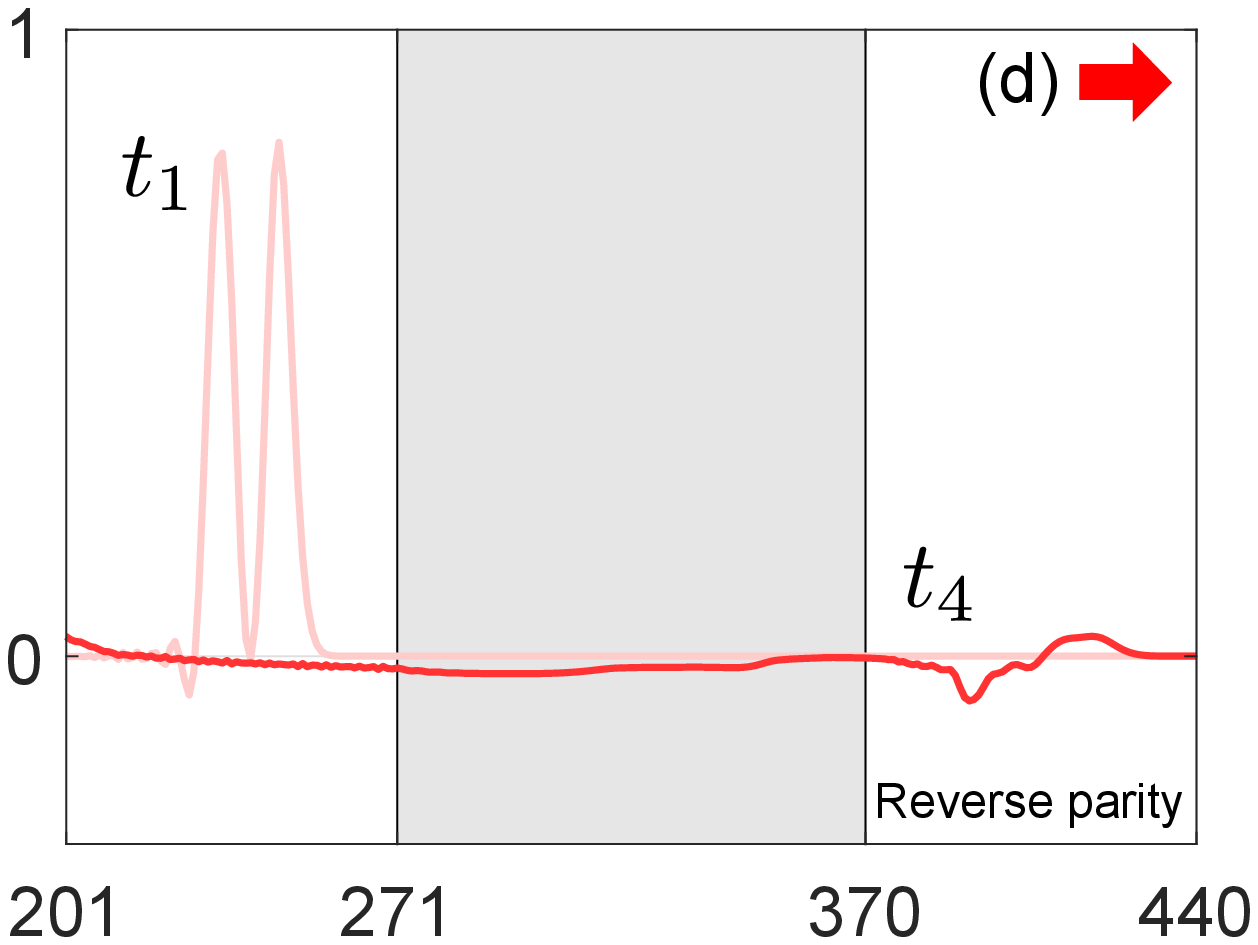}}
\subfigure{
	\includegraphics[width=0.45\textwidth]{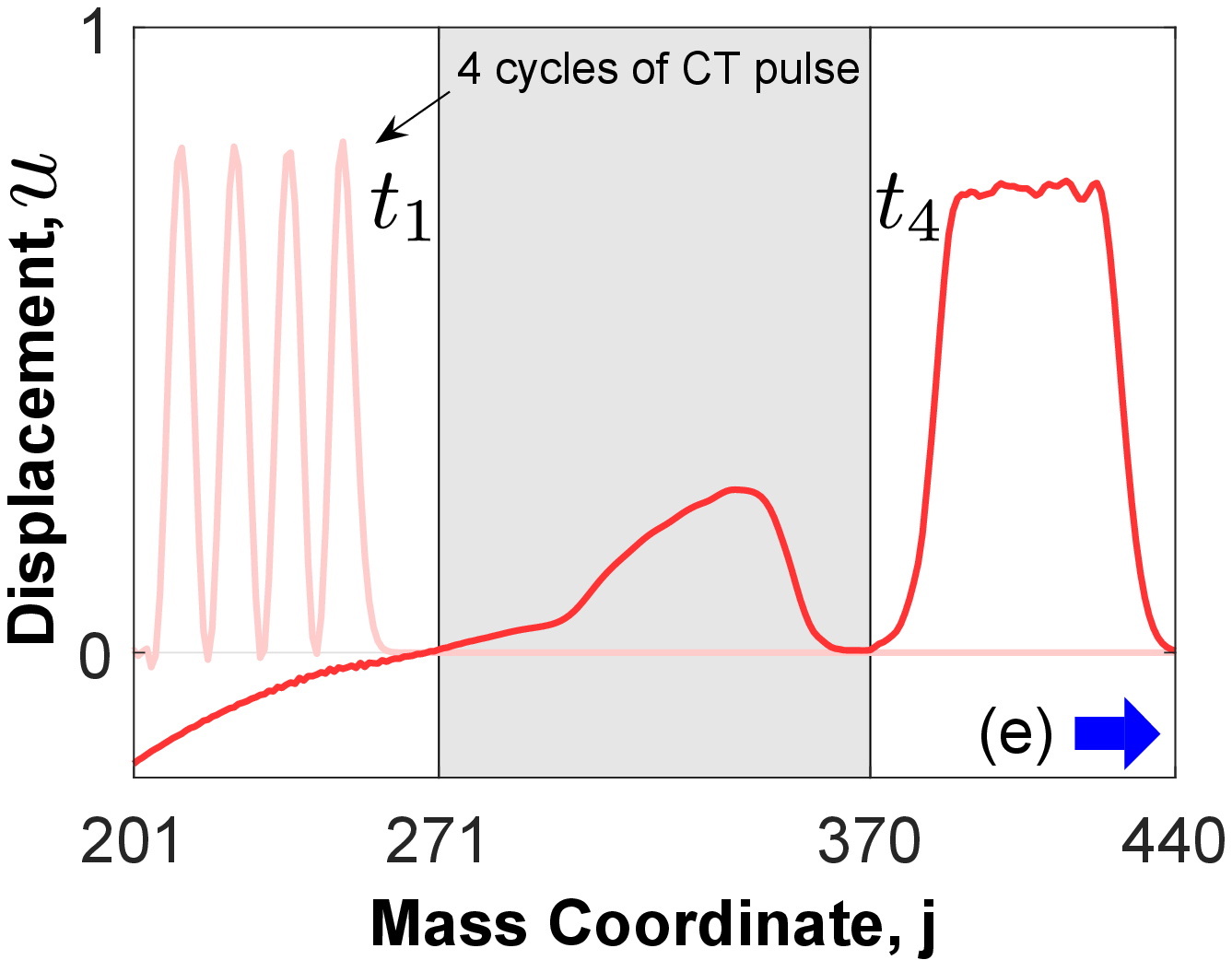}}
\subfigure{
	\includegraphics[width=0.45\textwidth]{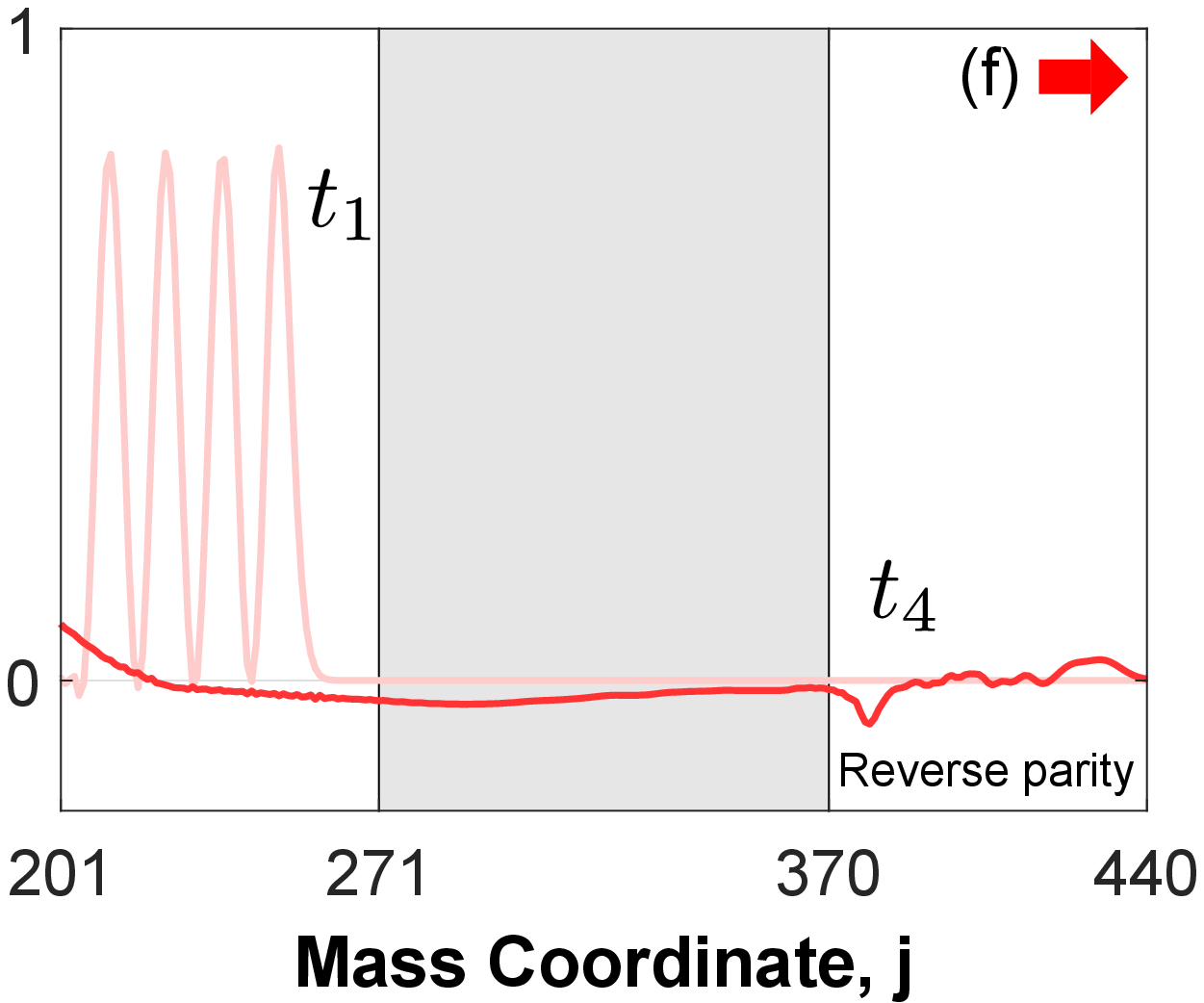}}
{\caption{Dynamic properties of   Model I with $b/a = 4$  when the incident wave consists of  multiple  cycles of a CT pulse. The plots on the left (right) correspond to incidence from the left (right).}
\label{fig15}}
\end{center}
\end{figure}

{Significant non-reciprocal wave motion is still observed if the incident wave    consists of several cycles  of a single pulse type, CT or TC. Figure \ref{fig15} shows the dynamic response of   Model I for an incident wave consisting of $n$ cycles of a CT pulse, for $n=1,\, 2$ and $4$, corresponding to the forcing (see Eq.\ \eqref{dimensionless_force})
\beq{-3}
f =  f_0 \, \mathcal{H}(t) \, \mathcal{H}(\frac{2n \pi}{\omega} - t) \, \sin \omega \, t \, .
\eeq
In each case of the  wave with multiple cycles incident from the left it is observed that the transmitted wave  is a single cycle but extended in space and time. This characteristic shape is due to an approximate zero deformation zone in the transmitted wave. Conversely, a multiple cycle  wave incident from the right produces a very low amplitude transmitted wave.}
%%%%%%%%%%%%%%%%%%%%%%%%%%%%%%%%%%%%%%%%%%%%%%%%%%%%%%%%%%%%%%%%%%%%%%

\begin{figure} %%%%%%%%%%%%%%%%%%%%%%%%%%%%%%%%%%%%%%%%%%%%%%%%%%%%%%%%%%%%%%%%%%%%%%
%[hbt!]
\begin{center}
\subfigure{
	\includegraphics[width=0.45\textwidth]{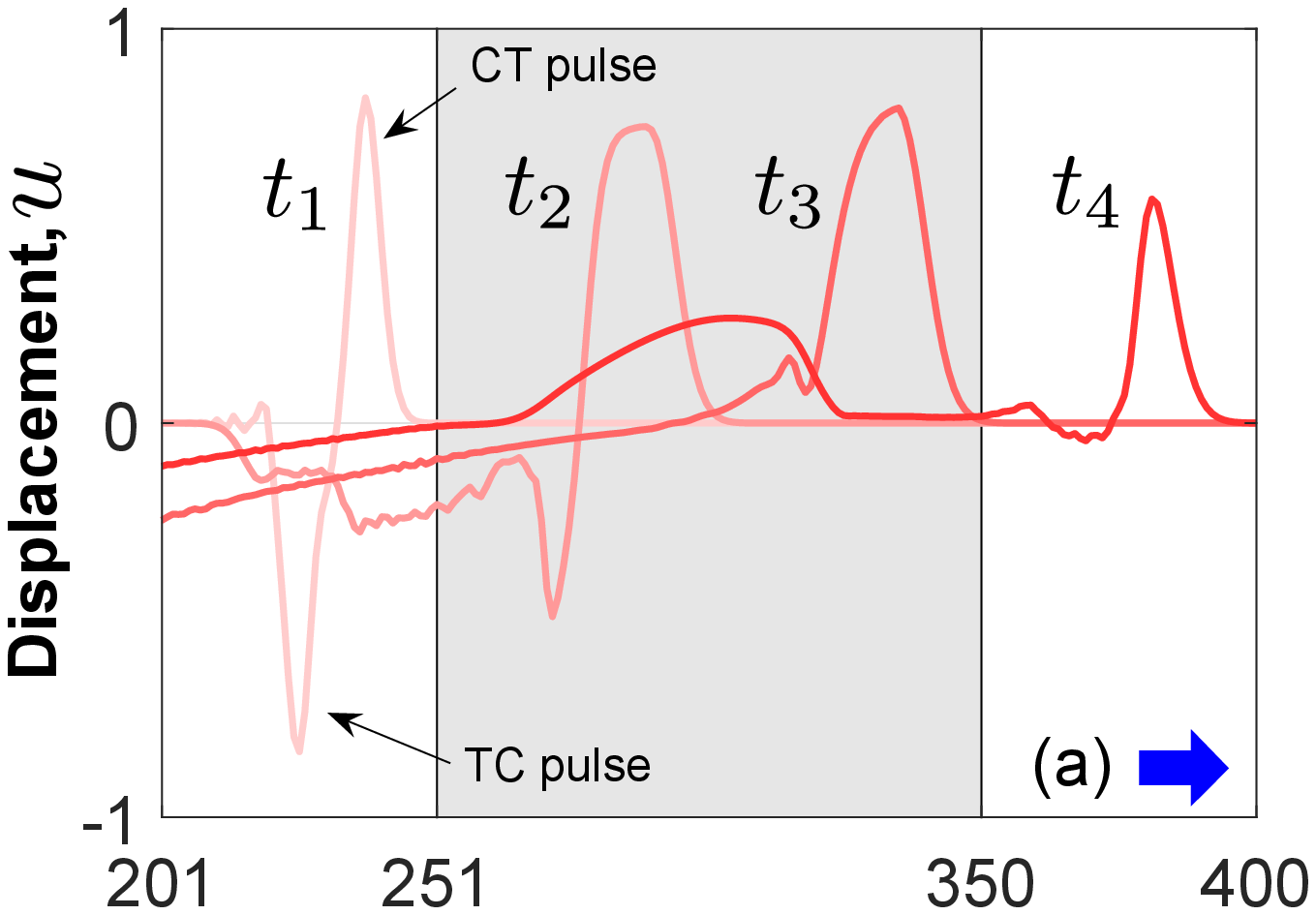}}
\subfigure{
	\includegraphics[width=0.45\textwidth]{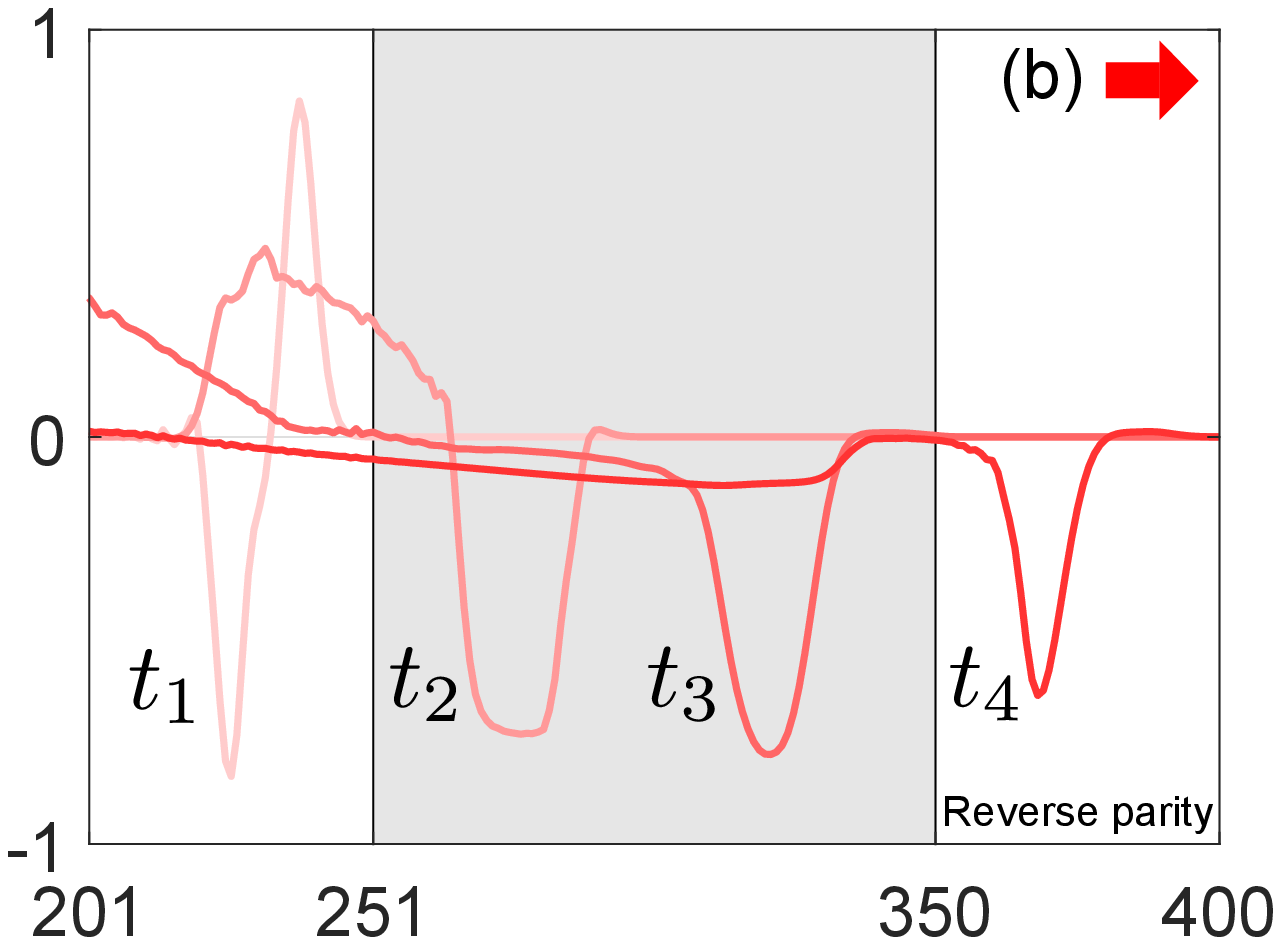}}
\subfigure{
	\includegraphics[width=0.45\textwidth]{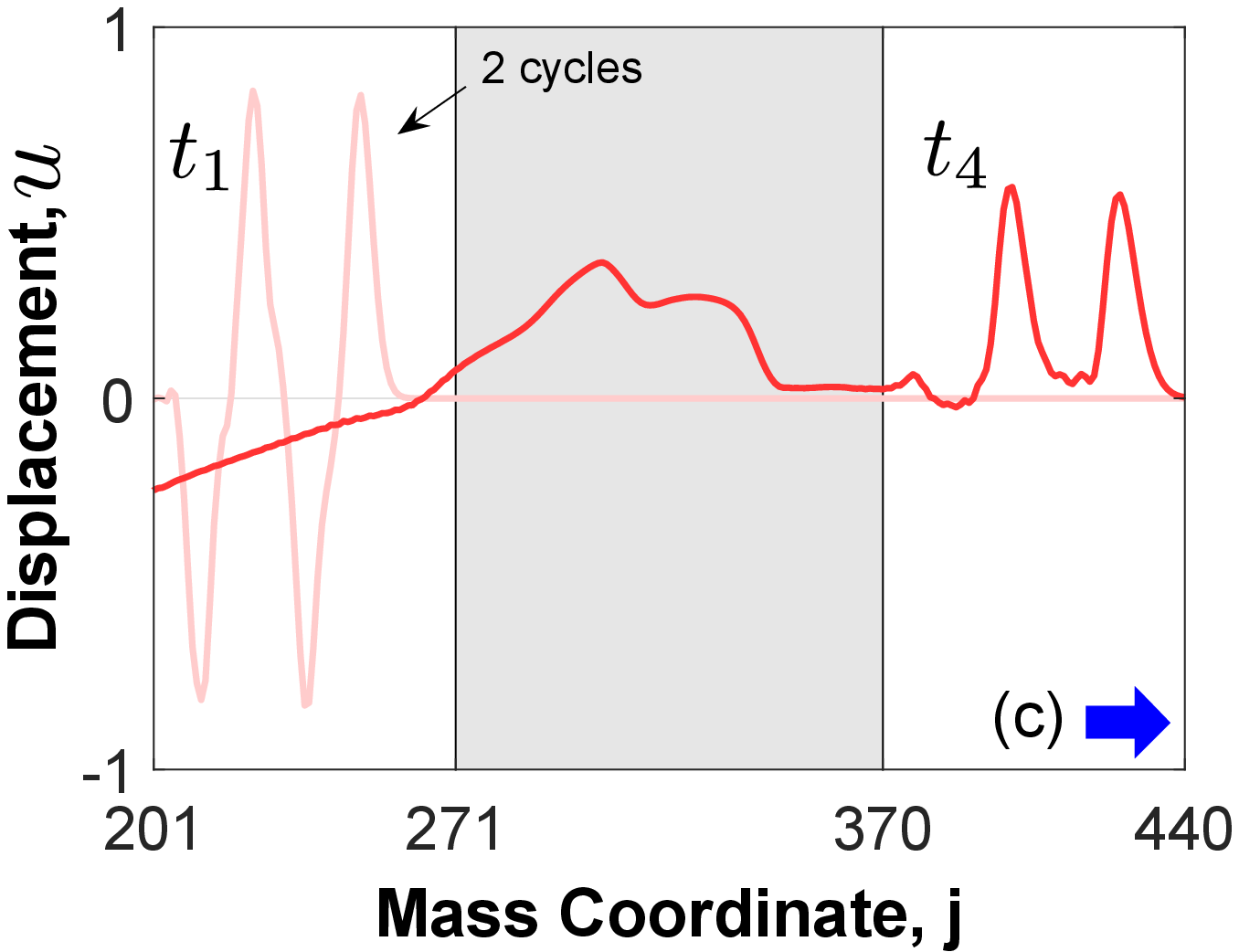}}
\subfigure{
	\includegraphics[width=0.45\textwidth]{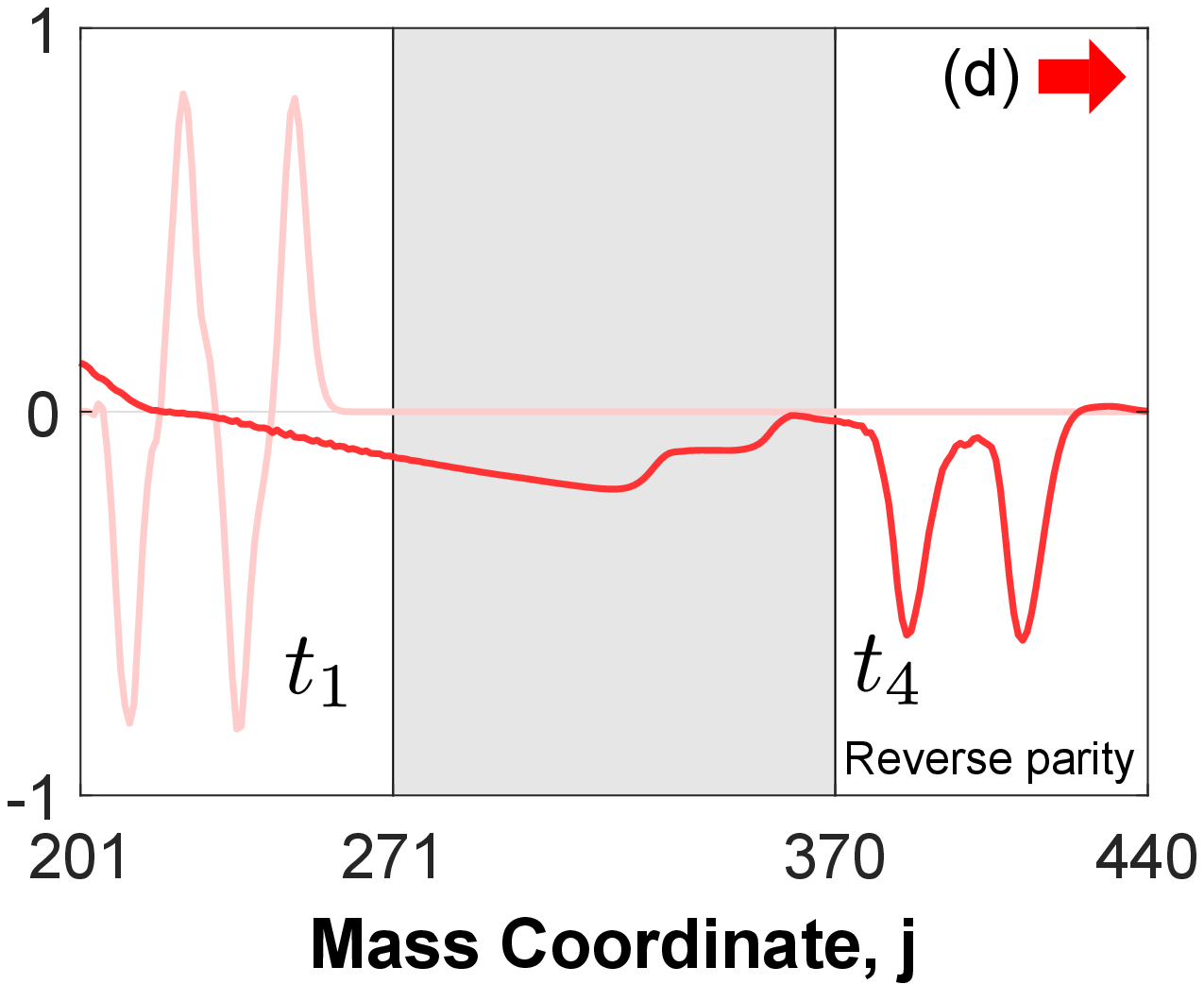}}
{\caption{Dynamic properties of  Model I for an incident wave consisting of both pulse types. The incident wave in (a) and (b) corresponds to the forcing Eq.\ \eqref{-4}. }
\label{fig14}}
\end{center}
\end{figure}

{Finally, we consider an incident wave that contains both CT and TC pulse types.  The forcing 
\beq{-4}
f =  f_0 \, \mathcal{H}(t) \, \mathcal{H}(\frac{2 \pi}{\omega} - t) \, \sin \omega \, t \, 
- f_0 \, \mathcal{H}(t-\frac{2 \pi}{\omega}) \, \mathcal{H}(\frac{4 \pi}{\omega} - t) \, \sin \omega \, t \,  
\eeq
produces the incident wave in Figs.\ \ref{fig14}(a) and (b), which is a CT pulse followed by a TC pulse.   The incident wave in  Figs.\ \ref{fig14}(c) and (d) is two cycles of the CT/TC pulse. 
In both cases  Figure \ref{fig14} shows that there is always a non-zero transmitted pulse of either CT or TC type, with the pulse type dependent on the incidence direction.  Thus, incidence from the left (right) produces a transmitted wave that is purely of CT (TC) type.  
 This phenomenon can be understood by recalling Fig.\ \ref{fig6} for CT incidence and Fig.\ \ref{fig13}  in Appendix \ref{appD} for TC incidence. 
When a wave consisting of both pulse types propagates from one side, one of them results in zero transmission (e.g. TC input pulse propagates form left in Model I, as Fig.\ \ref{fig13}(a) shows) and the other a transmitted pulse with type unchanged and amplitude slightly decreased (e.g. CT input pulse propagates form left in Model I, as Fig.\ \ref{fig6}(a) shows). In sum, the transmission  is purely CT or TC for incidence from the left or right, respectively. 
This filtering effect  is strongly-non-reciprocal.}
%%%%%%%%%%%%%%%%%%%%%%%%%%%%%%%%%%%%%%%%%%%%%%%%%%%%%%%%%%%%%%%%%%%%%%

%%%%%%%%%%%%%%%%%%%%%%%%%%%%%%%%%%%%%%%%%%%%%%%%%%%%%%%%%%%%%%%%%%%%%%
\section{Conclusion} \label{sec6}
We have demonstrated a passive 1D spring-mass-damper chain structure that  breaks elastic wave reciprocity. Amplitude-independent  non-reciprocity is a result of introducing bilinear springs and carefully designing the spatially inhomogeneous and asymmetric spatial modulations of the bilinear stiffness. A compression-tension pulse is used to quantify non-reciprocal transmission.  Incident pulses from opposite directions result in significantly different transmitted pulses; almost-zero transmission with significant transmission from the other direction has been demonstrated.  The results shown here indicate that a simple 1D bilinear chain system can produce non-reciprocal wave dynamics, and is therefore suited for   a variety of %{broad-band} 
elastic wave control mechanisms, such as one-way propagation, pulse type inversion, pulse type filtering, etc. Moreover, these simple models can  work as the building blocks for some more complex, higher dimensional non-reciprocal structure designs.

%%%%%%%%%%%%%%%%%%%%%%%%%%%%%%%%%%%%%%%%%%%%%%%%%%%%%%%%%%%%%%%%%%%%%%
\begin{acknowledgment}
This work is supported by the  NSF EFRI program under award No. 1641078.
\end{acknowledgment}

%%%%%%%%%%%%%%%%%%%%%%%%%%%%%%%%%%%%%%%%%%%%%%%%%%%%%%%%%%%%%%%%%%%%%%
%\appendix
\section*{Appendix} \medskip

\begin{appendices}
\section{ Stiffness and Damping Coefficients in the PML}  \label{appA}

A perfectly matched layer (PML) is attached to the test chain at each end  to eliminate reflections, see Fig.\ \ref{fig1}. Suppose that the index $j$ starts from 1 at the beginning of the PML on the left. Each PML is a linear spring-mass-damper chain in which the damping coefficients are ''ramped-up'' to avoid internal reflections. The location-dependent parameters of the left and right PMLs are symmetric about the central test chain. Here we concentrate on the PML on the left for which the dimensionless stiffness is $\kappa_j = 1$. The PML index takes the values $1 \leq j \leq N_{pml}$, with dimensionless damping coefficient    
\beq{1=1}
\zeta_j = \zeta_{max} \, \Big ( \frac{N_{pml}+1 -j}{N_{pml}} \Big ) ^3 \, ,
\eeq
where $\zeta_{max}$ is the maximum damping coefficient at the PML end. All the numerical experiments in this paper take $\zeta_{max} = 10$.

\begin{figure}[hbt!]%%%%%%%%%%%%%%%%%%%%%%%%%%%%%%%%%%%%%%%%%%%%%%%%%%%%%%%%%%%%%%%%%%%%%%
\begin{center}
\subfigure[Without damping.]{
\includegraphics[width=2.2in]{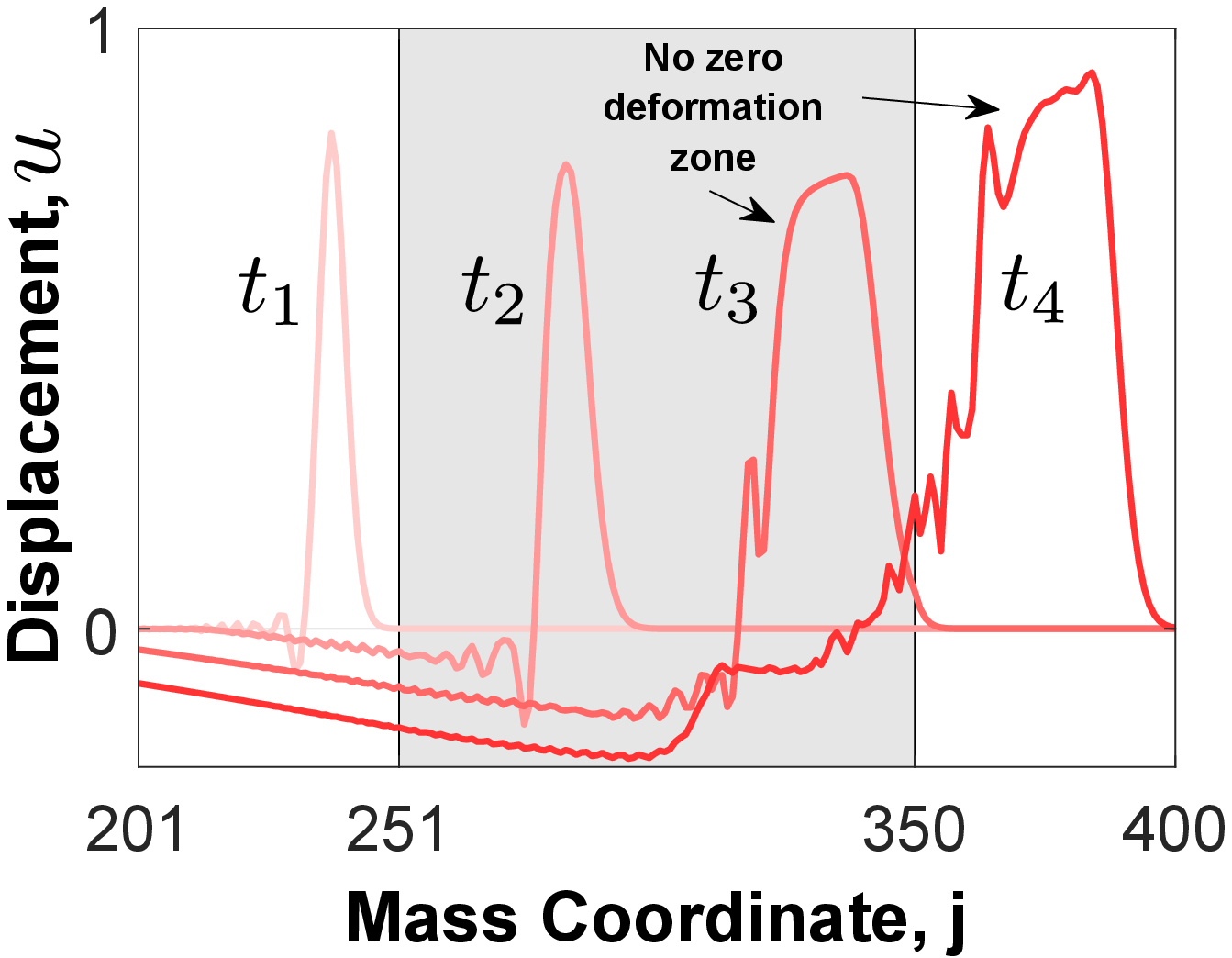}}
\subfigure[With damping, $\zeta_0 = 0.05$.]{
\includegraphics[width=2.2in]{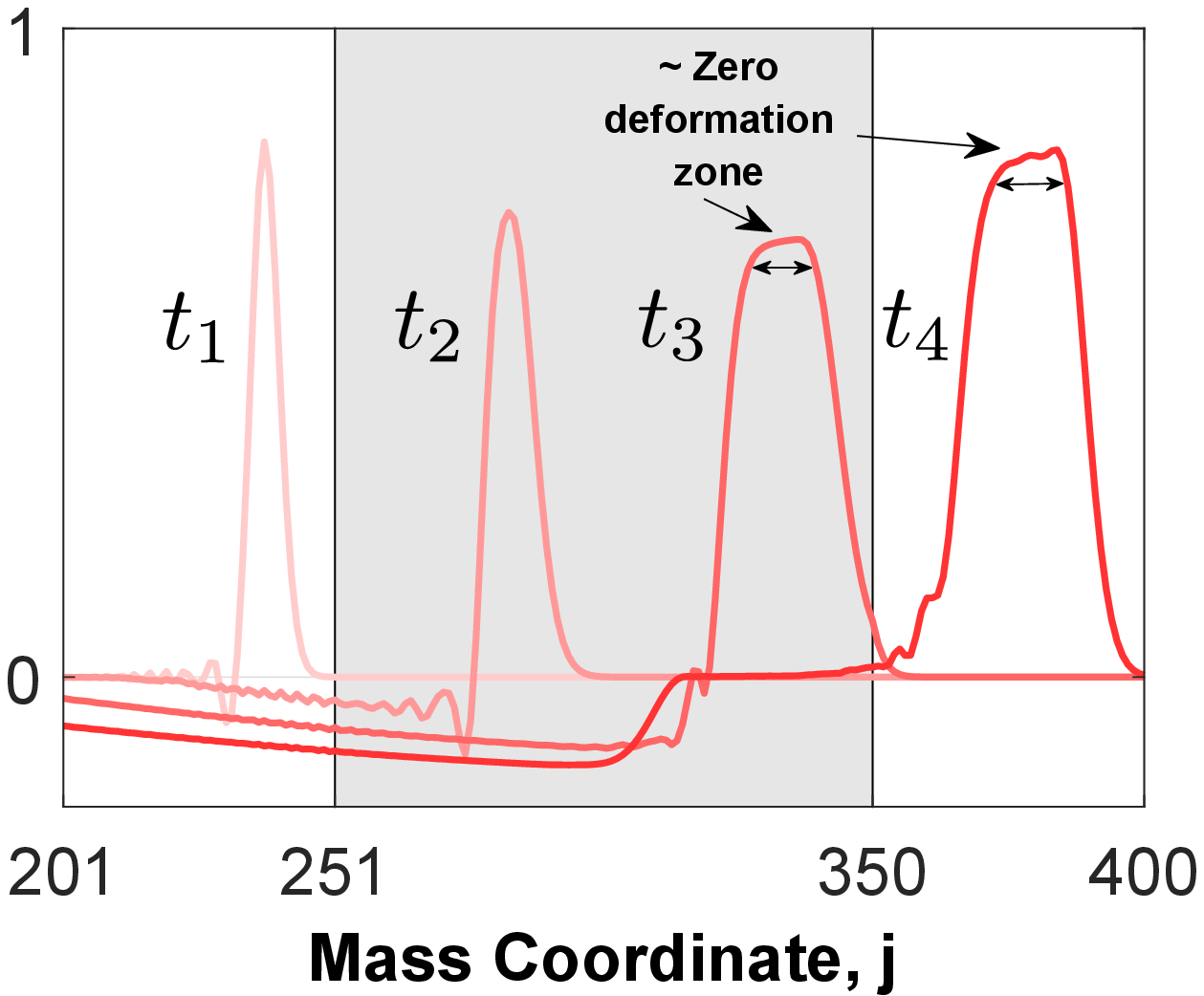}}
\subfigure[With damping, $\zeta_0 = 0.1$.]{
\includegraphics[width=2.2in]{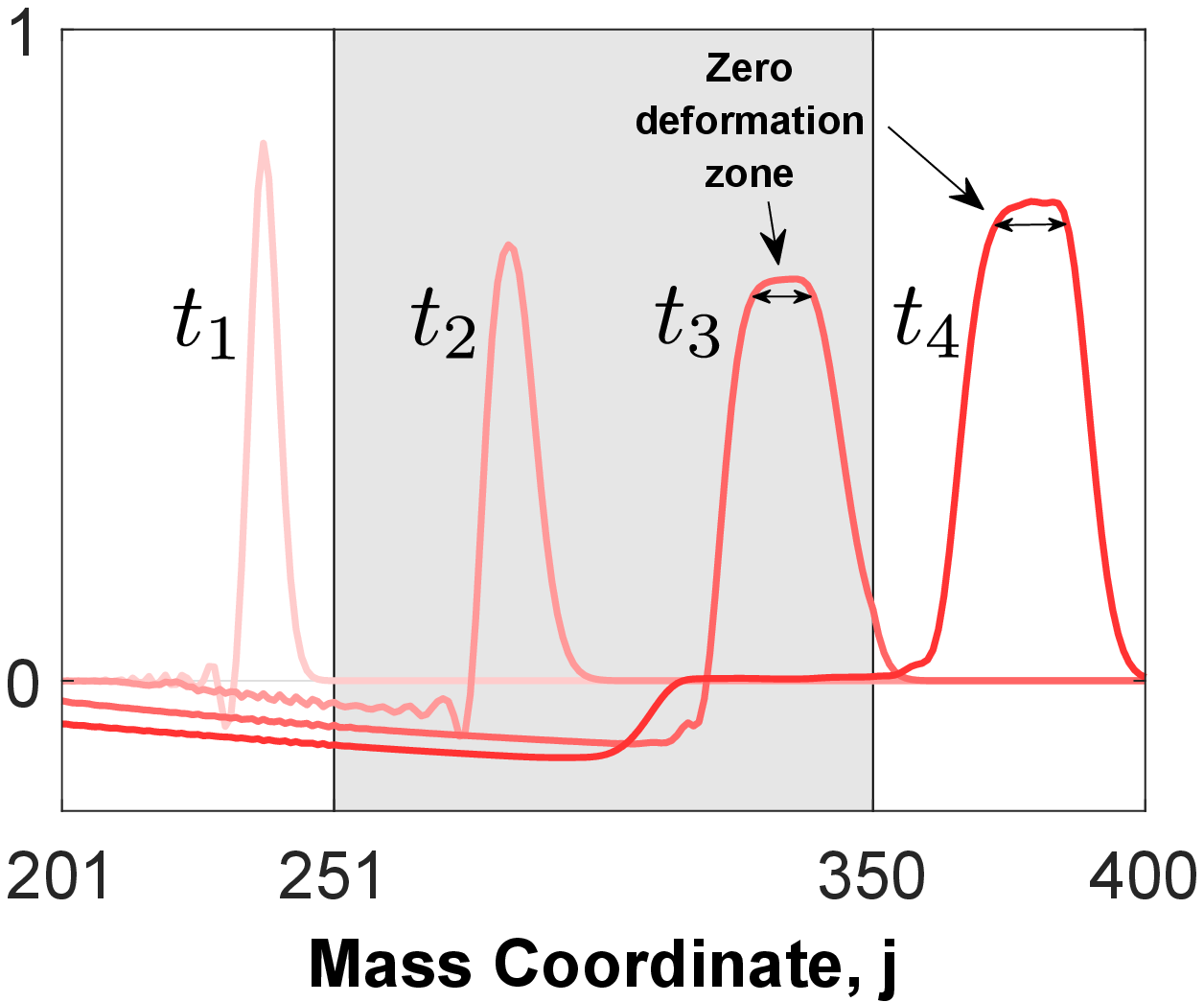}}
{\caption{A comparison of the response with and without damping included in the bilinear part. (a) depicts the displacement fields at different instants for the modulation of Fig.\ \ref{fig3}(a) without damping; (b) and (c) with constant weak damping ($\zeta_0 = 0.05$ and $0.1$, respectively). }
\label{fig10}}
\end{center}
\end{figure}

{\section{The Effect of Damping in the Bilinear Chain}  \label{appA_1}}
{We include damping in the bilinear section in order to understand how any realistic damping effects  wave propagation. Figure \ref{fig10} compares cases with and without   damping included in the bilinear chain. We   find that the weak dampers in the bilinear chain do not significantly affect the propagation results in terms  of the pulse shape and size. 
Shock-like wave structures can be observed in Fig.\ \ref{fig10}(a). Damping has a strong smoothing effect as   can be seen in 
Figs.\ \ref{fig10}(b) and   \ref{fig10}(c). In particular, Fig.\ \ref{fig10}(c) shows a clear pulse shape with the characteristic feature of a zero deformation zone.  For this reason we choose to  set dimensionless damping coefficient $\zeta_0 = 0.1$ for  
numerical simulations in this paper.}
%However, damping can effectively smooth the shock-like wave structures as is evident in Fig.\ \ref{fig10}(a).} 
%%%%%%%%%%%%%%%%%%%%%%%%%%%%%%%%%%%%%%%%%%%%%%%%%%%%%%%%%%%%%%%%%%%%%%

\begin{figure}[hbt!]%%%%%%%%%%%%%%%%%%%%%%%%%%%%%%%%%%%%%%%%%%%%%%%%%%%%%%%%%%%%%%%%%%%%%%
\begin{center}
\subfigure[Slope = 0.5/100.]{
\includegraphics[width=3.25in]{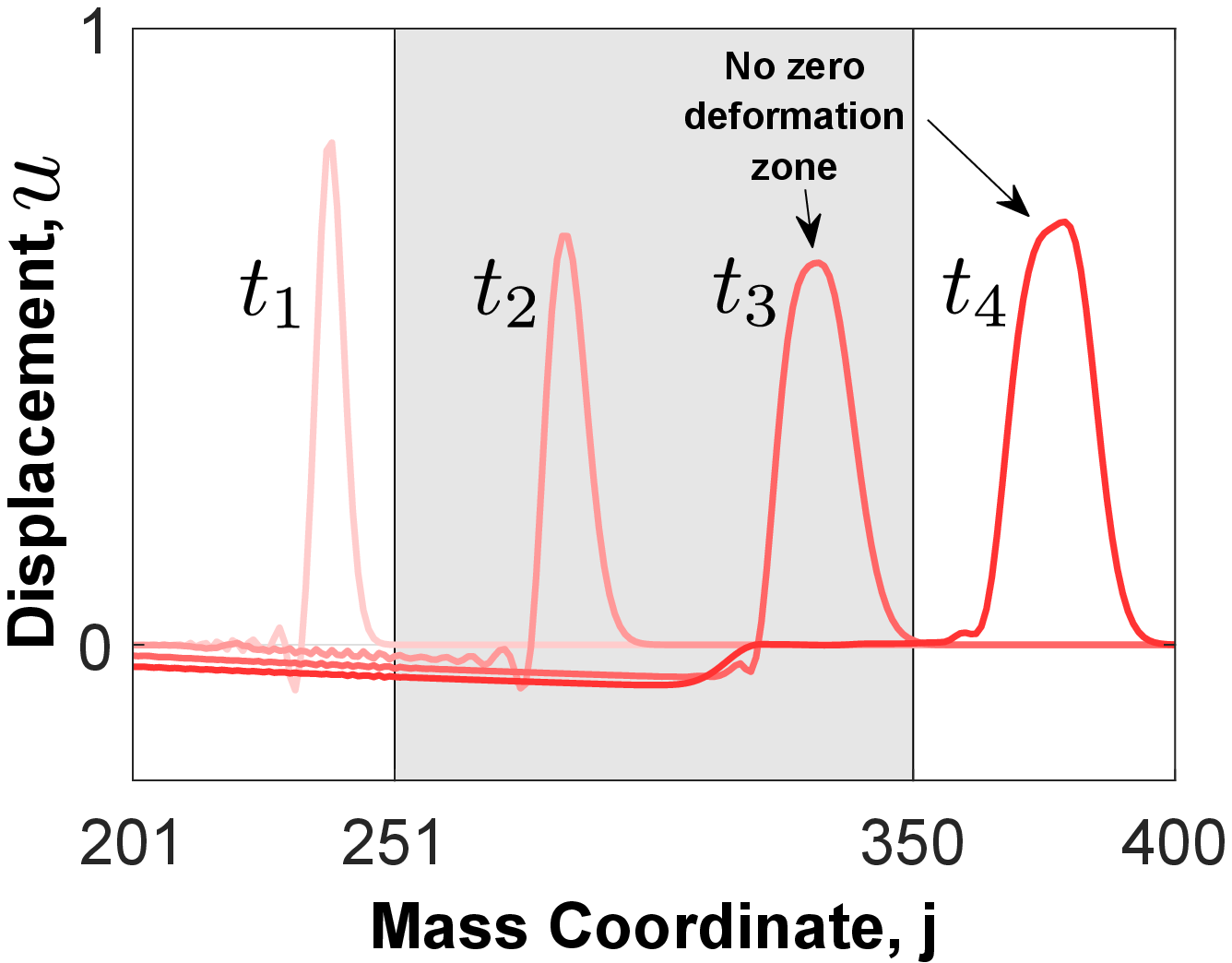}}
\subfigure[Slope = 2/100.]{
\includegraphics[width=3.25in]{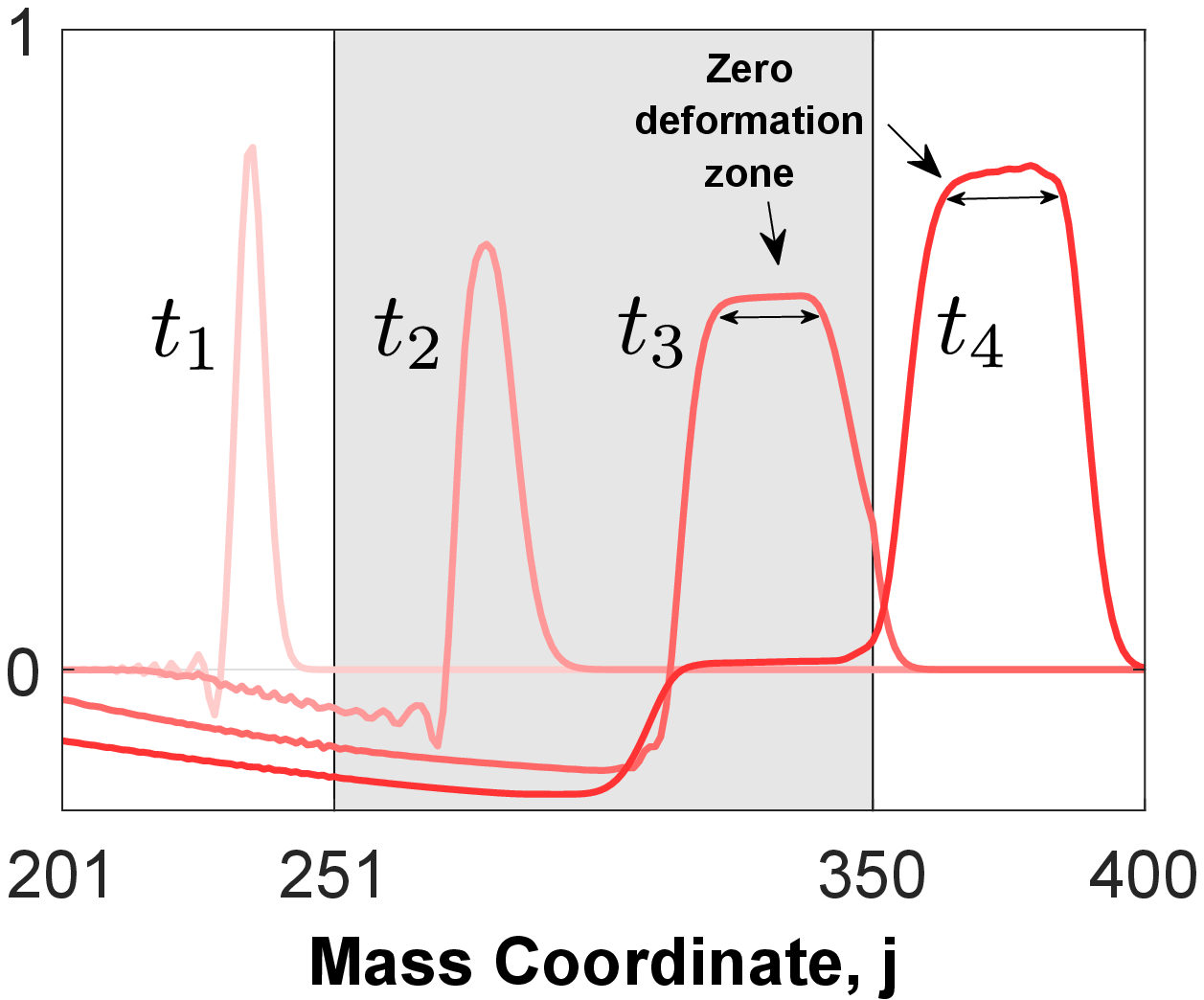}}
{\caption{The effect of stiffness modulation slope in Fig.\ \ref{fig3}(a) on the length of the zero deformation zone. A small zero deformation  is observed in (a). However (b) indicates that the greater slope leads to a significantly longer zero deformation regime.}
\label{fig11}}
\end{center}
\end{figure}%%%%%%%%%%%%%%%%%%%%%%%%%%%%%%%%%%%%%%%%%%%%%%%%%%%%%%%%%%%%%%%%%%%%%%
%%%%%%%%%%%%%%%%%%%%%%%%%%%%%%%%%%%%%%%%%%%%%%%%%%%%%%%%%%%%%%%%%%%%%%
\section{Spatial Modulation of the Bilinear Stiffness}  \label{appB}
{The fundamental modulations in Sect.\ \ref{sec3} are designed based on the rule of spatial inhomogeneity and asymmetry, necessary but not sufficient for achieving wave non-reciprocity.  
Spatial inhomogeneity and asymmetry are obtained by a linear   modulation of the stiffness.  Figure \ref{fig11} shows how the slope of the stiffness curve effects transmission, in this case through the length of the zero deformation regime. The reason is that larger values of $\Delta_{j,c}$  increase the slope of stiffness curve, leading to greater compressive wave speeds, which in turn results  in    larger sizes of the  zero deformation zone.}

\begin{figure}[hbt!]%%%%%%%%%%%%%%%%%%%%%%%%%%%%%%%%%%%%%%%%%%%%%%%%%%%%%%%%%%%%%%%%%%%%%%
\begin{center}
\subfigure[$\,$ Displacement  for modulation of Fig.\ \ref{fig3}(a).]{
\includegraphics[width=3.25in]{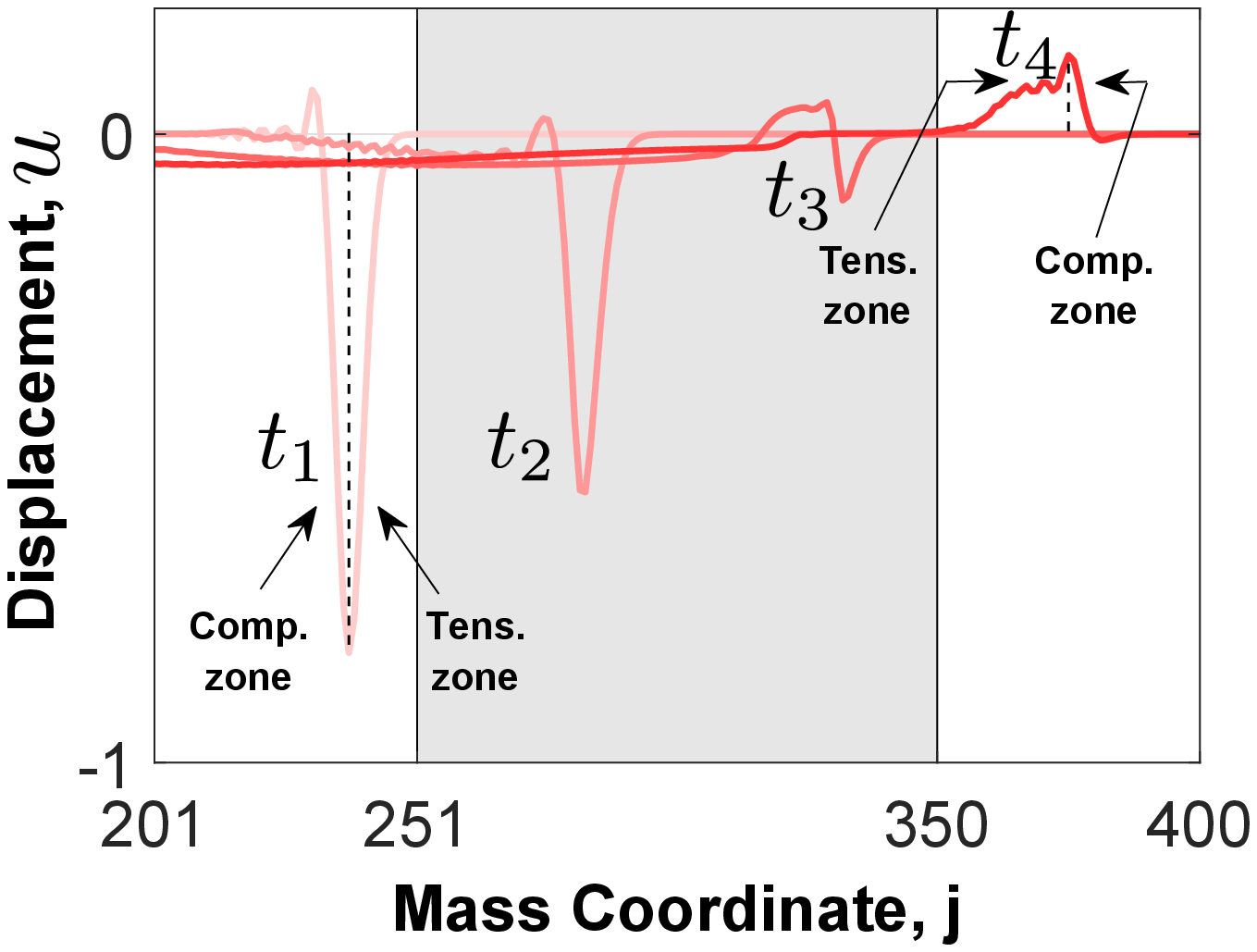}}
\subfigure[$\,$ Displacement  for modulation of Fig.\ \ref{fig3}(b).]{
\includegraphics[width=3.25in]{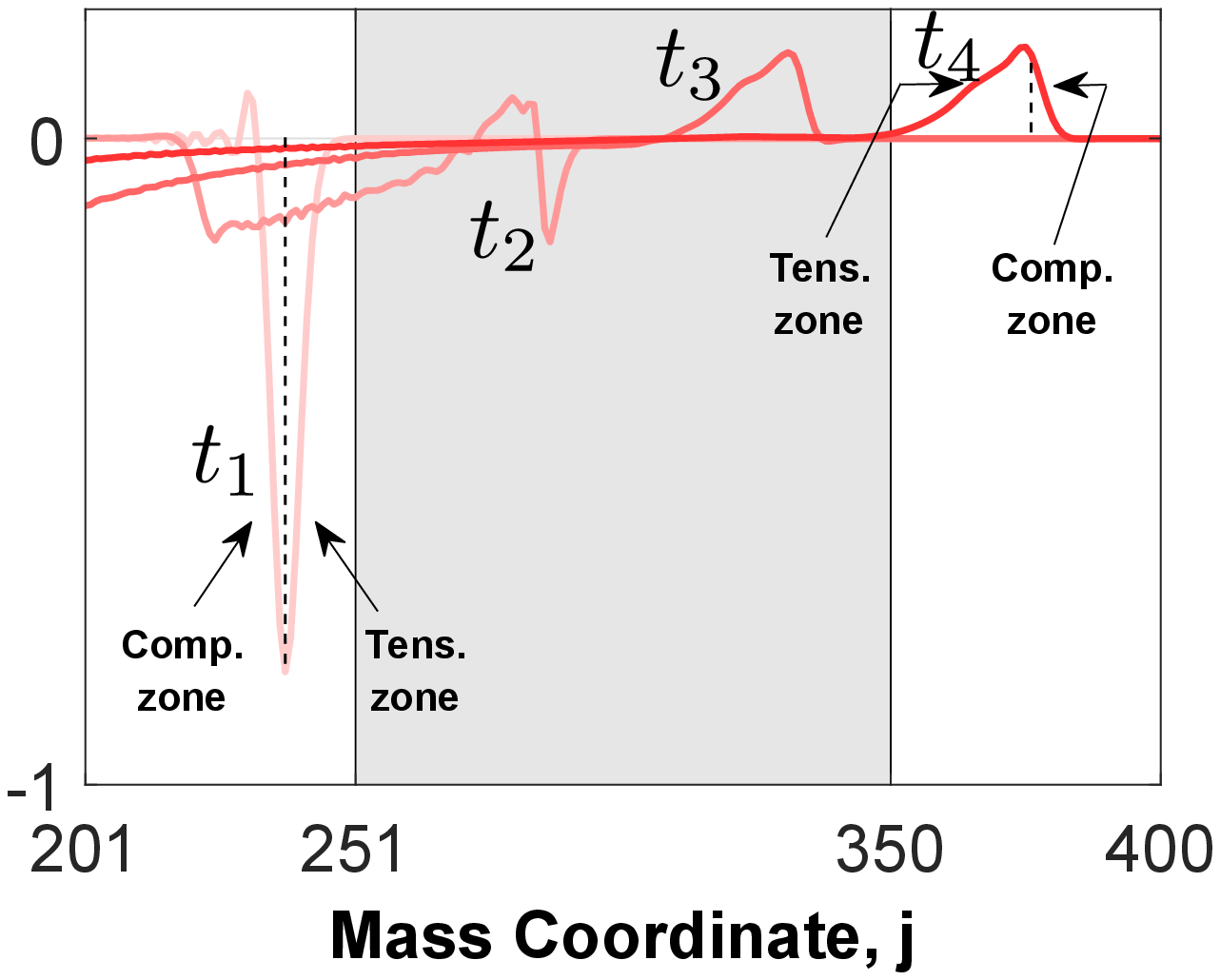}}
\subfigure[$\,$ Displacement  for modulation of Fig.\ \ref{fig4}(a).]{
\includegraphics[width=3.25in]{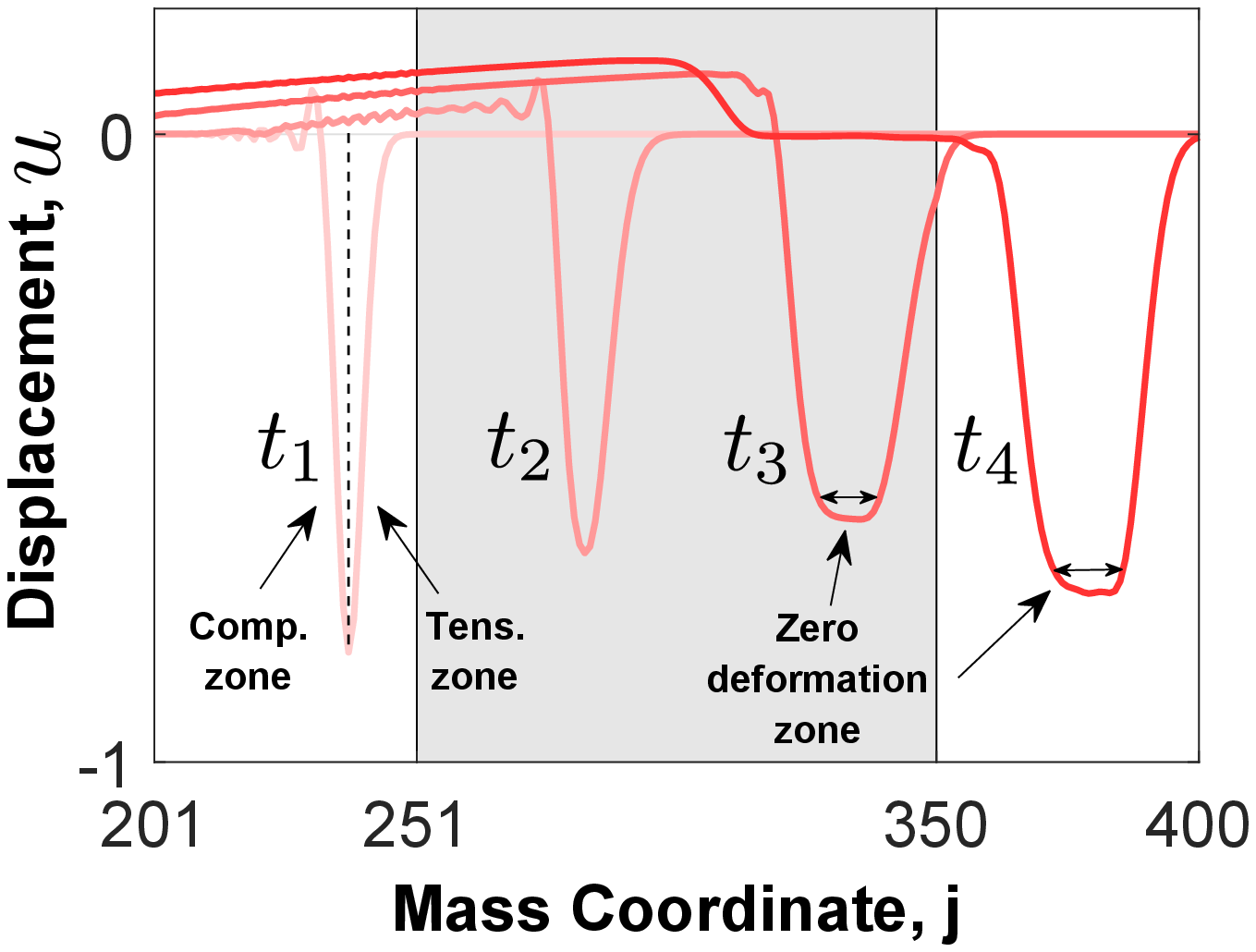}}
\subfigure[$\,$ Displacement  for modulation of Fig.\ \ref{fig4}(b).]{
\includegraphics[width=3.25in]{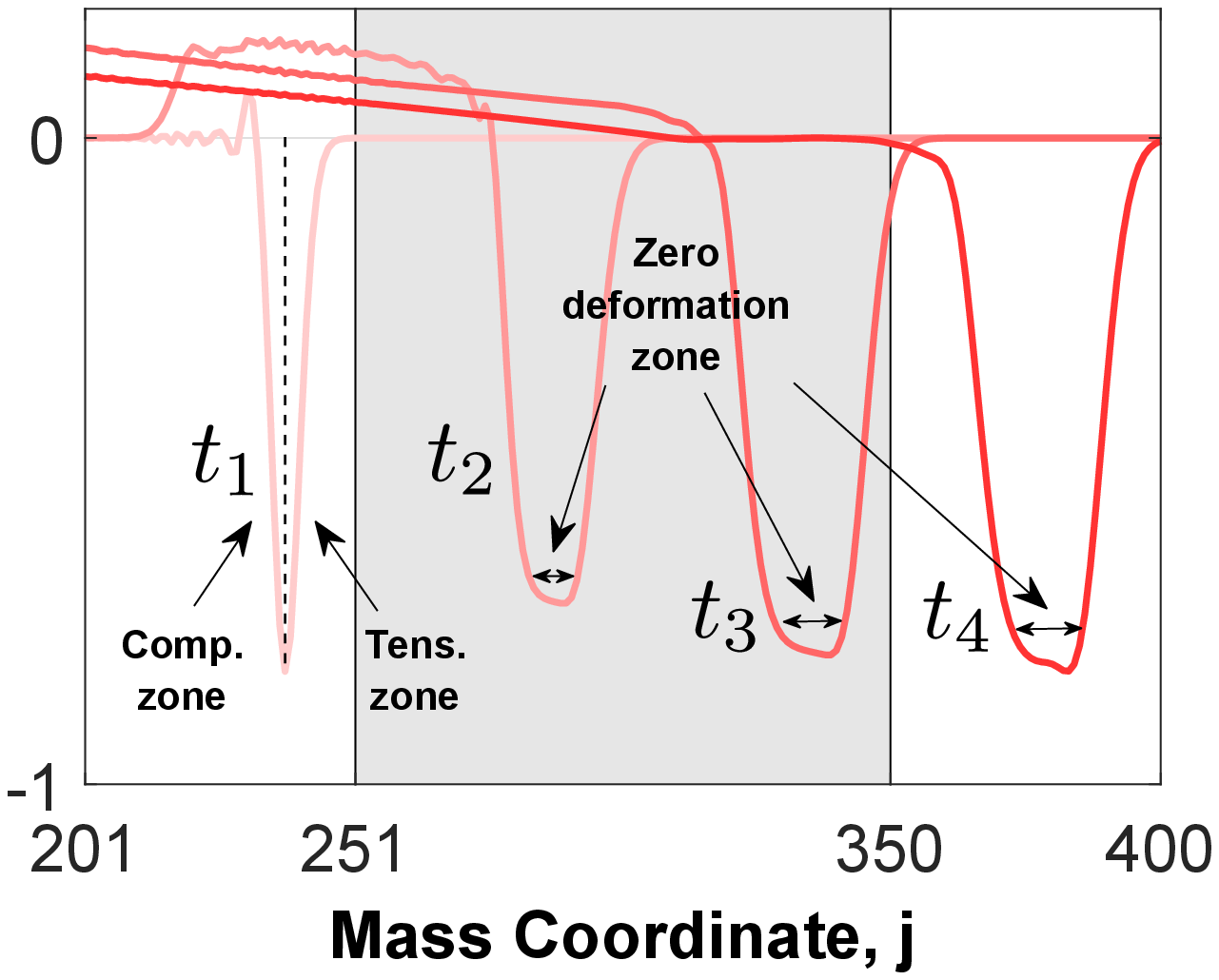}}
\caption{Dynamic properties of the bilinear spring chain. (a) and (b) shows the displacement fields of the bilinear chain at four different moments for modulations given by  Figs.\ \ref{fig3}(a) and \ref{fig3}(b), respectively. (c) and (d) shows the displacement fields for modulations shown in Figs.\ \ref{fig3}(a) and \ref{fig3}(b) respectively.}
\label{fig12}
\end{center}
\end{figure}

A tension-compression (TC) pulse is used for testing the different configurations, generated using the negative sign in the external loading Eq.\ \eqref{dimensionless_force}.  Assuming  that the stiffness of the bilinear spring is greater in compression than in tension, we set $\Delta_{j,c} > \Delta_{j,t} = 0$ and $\Delta_{j,c}$ changes linearly over location according to  Eqs.\ \eqref{delta_incr} and \eqref{delta_decr}. Figures \ref{fig12}(a) and  \ref{fig12}(b) depict the displacement fields along the test chain at four different moments for the stiffness modulations of Fig.\ \ref{fig3}(a) and \ref{fig3}(b), respectively. Conversely,  when the bilinear spring has a greater stiffness in tension than compression: $\Delta_{j,t} > \Delta_{j,c} = 0$ where $\Delta_{j,t}$ changes linearly according to  Eqs.\ \eqref{delta_incr_2} and \eqref{delta_decr_2},  the displacement fields for the modulation of Figs.\ \ref{fig4}(a) and \ref{fig4}(b), are shown in  Figs.\ \ref{fig12}(c) and \ref{fig12}(d), respectively. 

%%%%%%%%%%%%%%%%%%%%%%%%%%%%%%%%%%%%%%%%%%%%%%%%%%%%%%%%%%%%%%%%%%%%%%

%%%%%%%%%%%%%%%%%%%%%%%%%%%%%%%%%%%%%%%%%%%%%%%%%%%%%%%%%%%%%%%%%%%%%%
\section{Dimensionless Stiffness for Incidence from the Right}  \label{appC}

Two asymmetric stiffness-location modulation models for  breaking wave reciprocity are considered, as shown in Fig.\ \ref{fig5}. The designs are based on the fundamental asymmetric nonlinear configurations of Section \ref{sec3}. For pulse propagation from the right (as the red arrow indicates), the index $j$ starts from 1 at the right end of the test chain in Fig.\ \ref{fig1} (reverse parity), and takes the value $N_l +1 < j \leq N_l+\frac{d}{c+d} \, N_{bl}$ for the right section of the bilinear part and $N_l+\frac{d}{c+d} \, N_{bl} < j \leq N_l+N_{bl}$ for the left. The dimensionless stiffness increments $\Delta_{j,c}$ and $\Delta_{j,t}$ are
\beq{1=2}
\Delta_{j,c}= 
\begin{cases}
	0 \, , & \text{right section} \, , \\
	a \, \, \frac{j \, - \, N_{l} \, - \, \frac{d}{c+d} \, N_{bl} \, - \, 1}{\frac{c}{c+d} \, N_{bl} \, - \, 1} \, , & \text{left section} \, ,
    \end{cases}
\qquad
\Delta_{j,t}= 
\begin{cases}
	b \, \, \frac{N_l \, + \, \frac{d}{c+d} \, N_{bl} \, + \, 1 \, - \, j}{\frac{d}{c+d} \, N_{bl} \, - \, 1} \, , & \text{right section} \, , \\
	0 \, , & \text{left section} \, .
    \end{cases}
\eeq

%%%%%%%%%%%%%%%%%%%%%%%%%%%%%%%%%%%%%%%%%%%%%%%%%%%%%%%%%%%%%%%%%%%%%%
{\section{ Propagation  of a Tension-compression (TC) Input Pulse}  \label{appD}}

\begin{figure}[h] %%%%%%%%%%%%%%%%%%%%%%%%%%%%%%%%%%%%%%%%%%%%%%%%%%%%%%%%%%%%%%%%%%%%%%
%[hbt!]
\begin{center}
\subfigure{
	\includegraphics[width=0.45\textwidth]{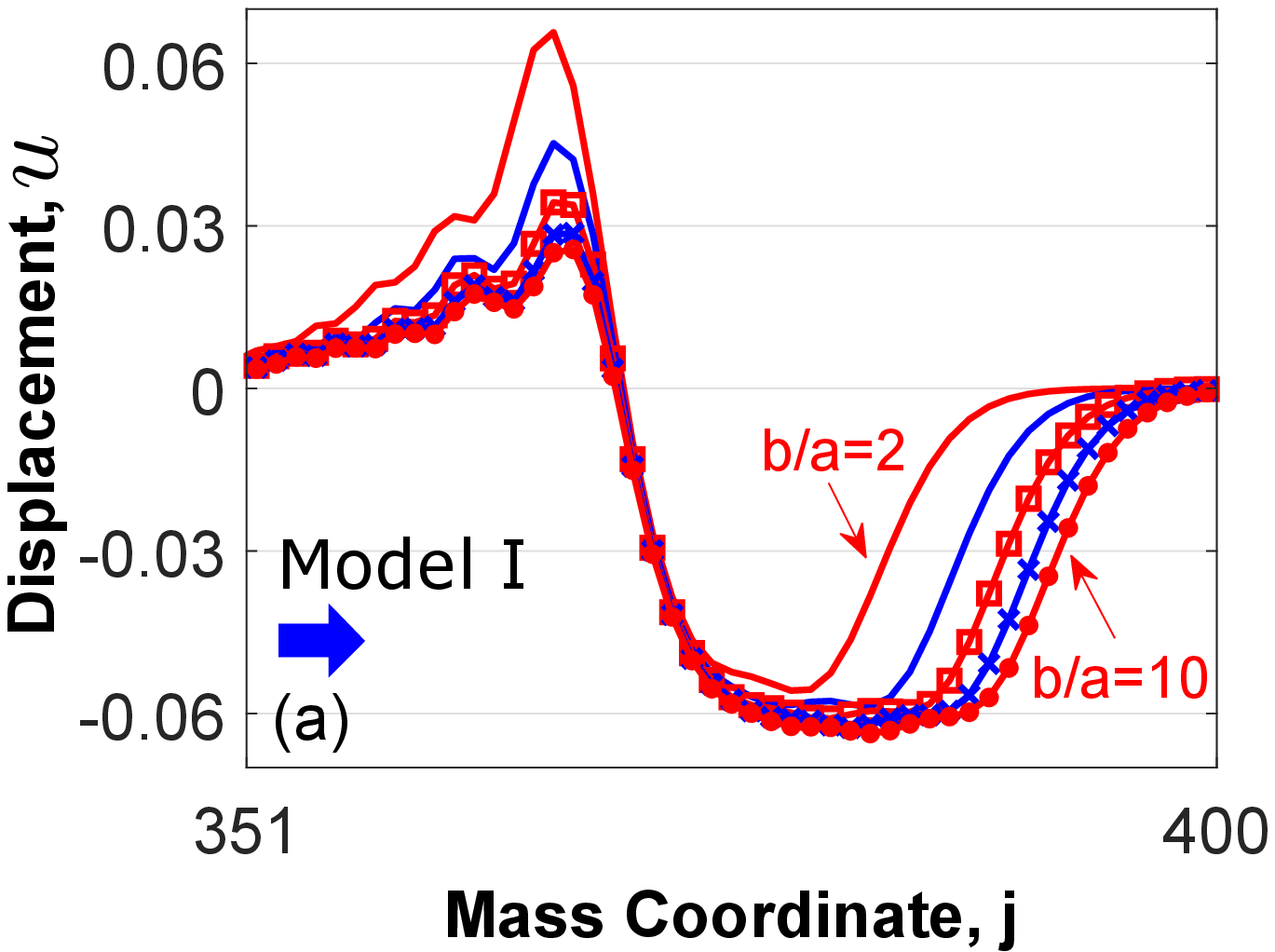}}
\subfigure{
	\includegraphics[width=0.45\textwidth]{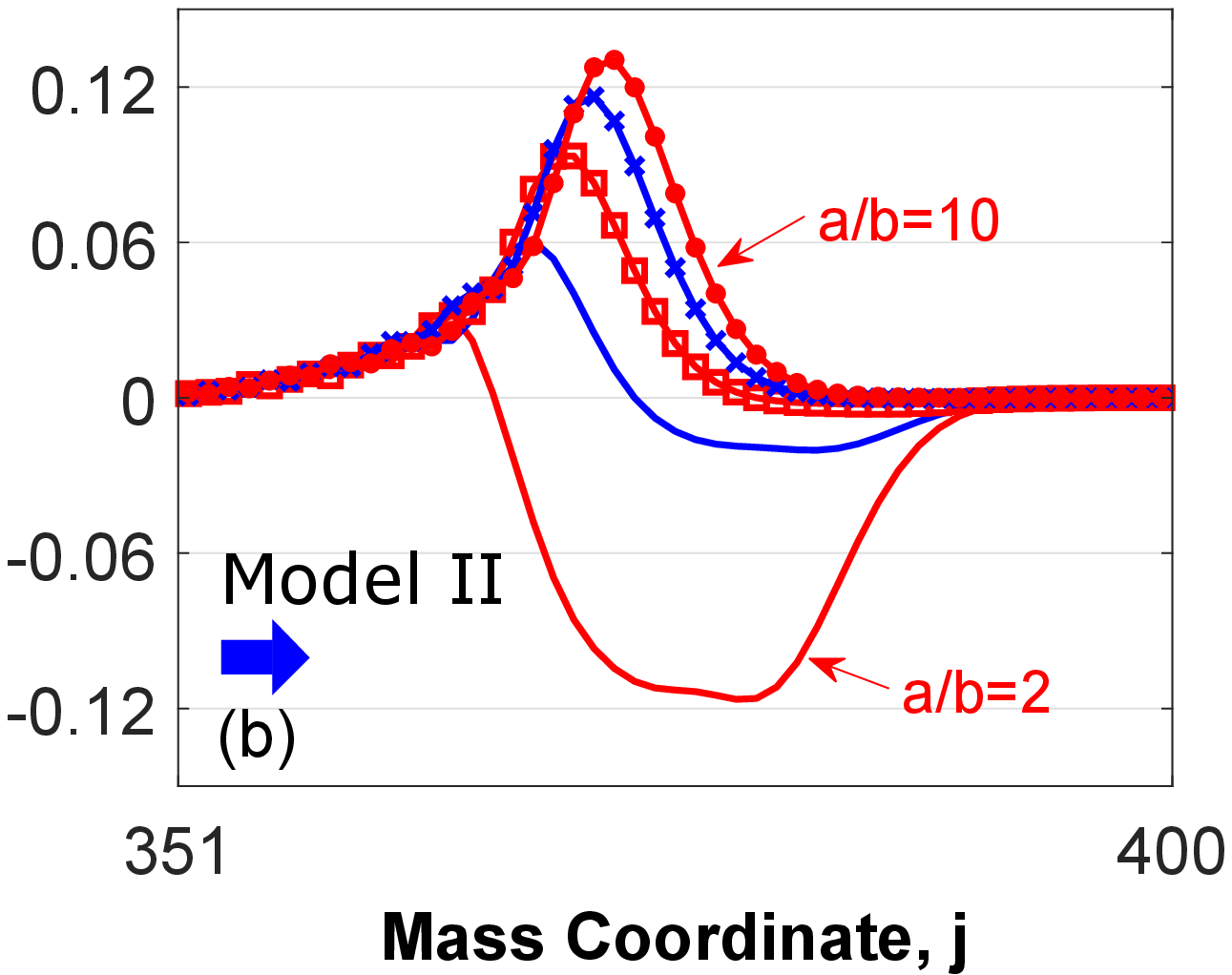}}
\subfigure{
	\includegraphics[width=0.45\textwidth]{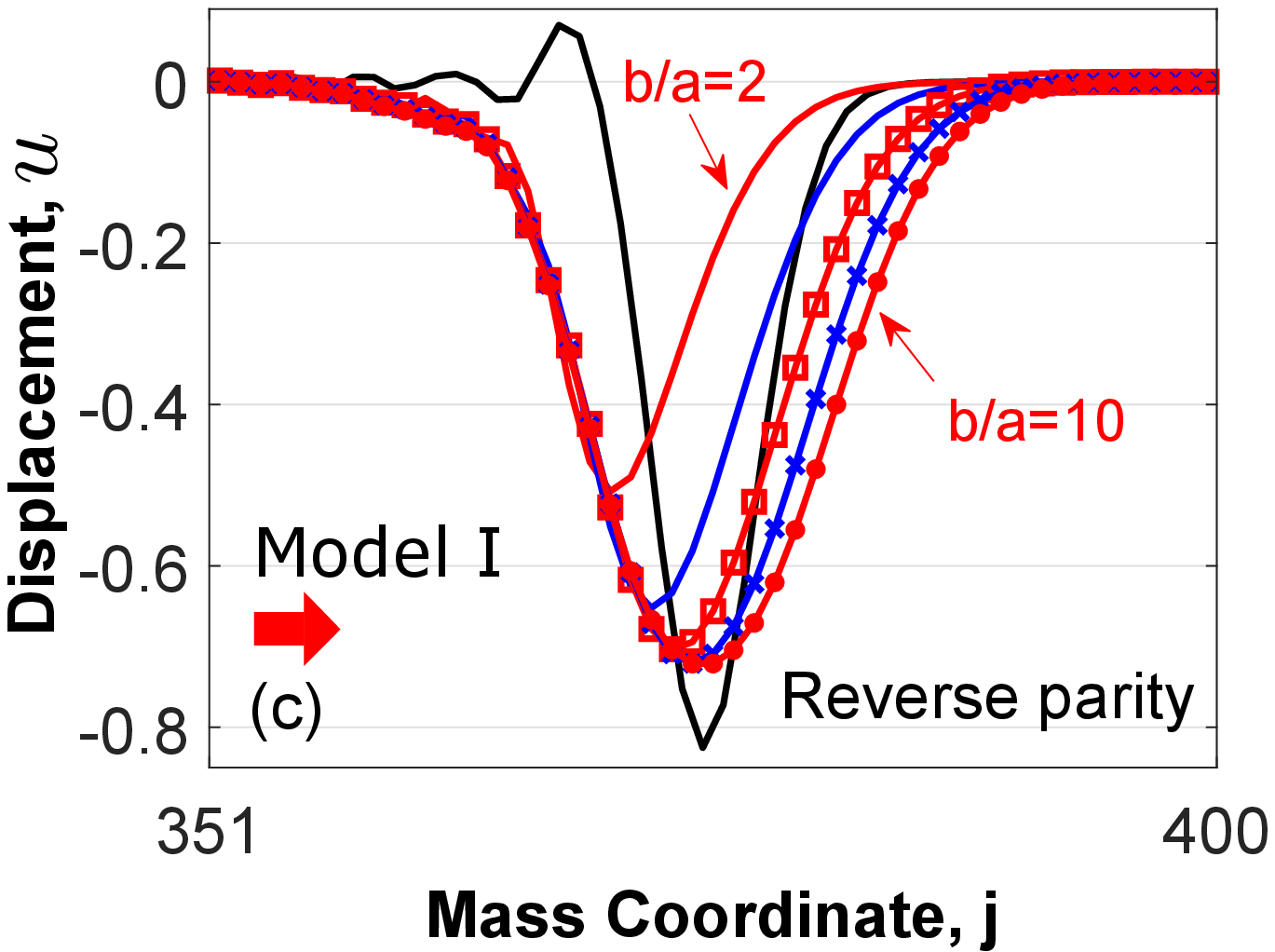}}
\subfigure{
	\includegraphics[width=0.45\textwidth]{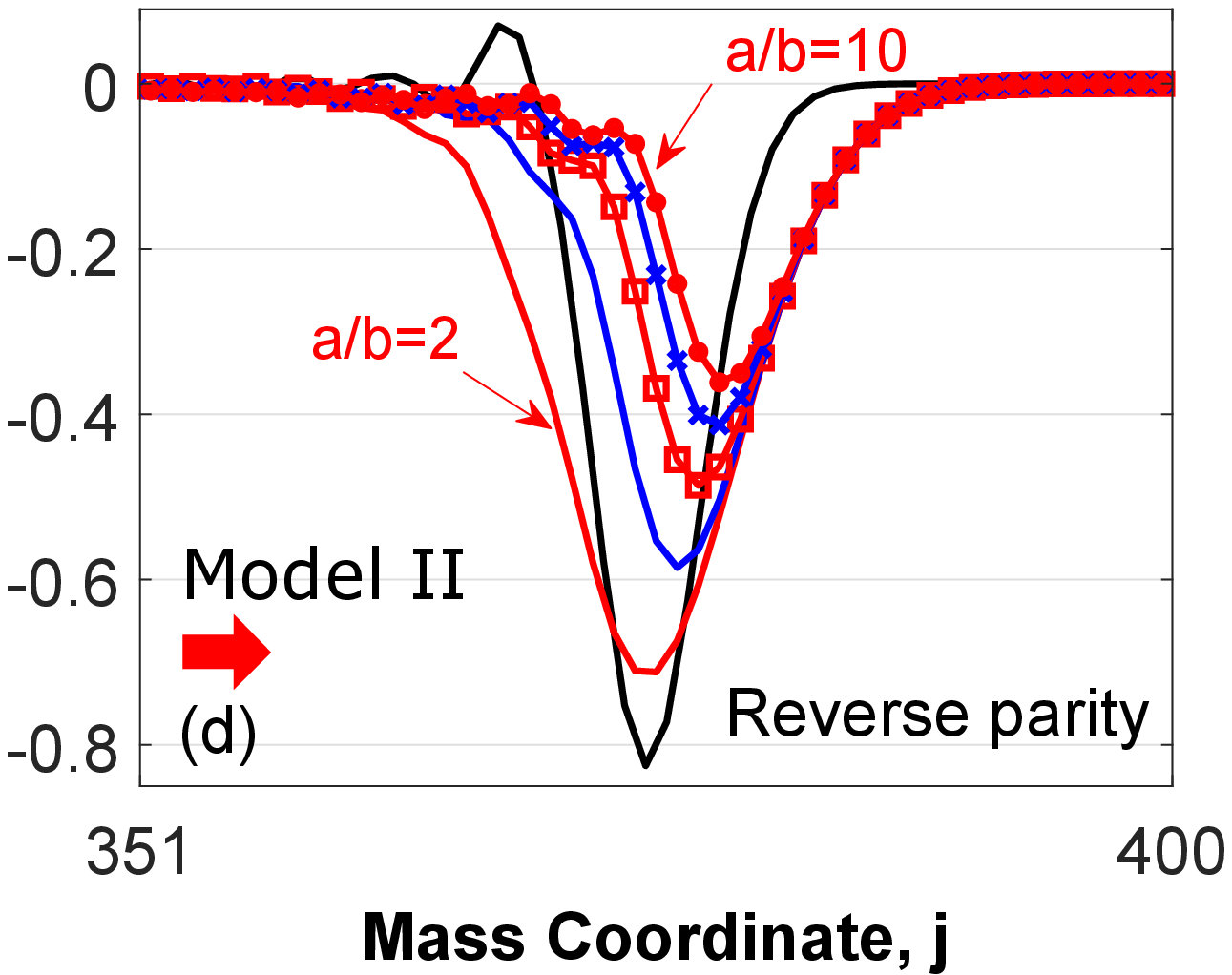}}
{\caption{Transmission amplitudes for the two Models in Fig.\ 5 with a TC input pulse: Model I   on the left, (a) and (c), and   Model II   on the right, (b) and (d).  The black curve is the incident TC pulse; the red and blue curves are the transmitted waves. Incidence from the left (labeled by the blue arrow) for Model I is in (a) and from the right (labeled by red arrow) in (c), in both cases for $b/a = 2,4,6,8,10$, demonstrating very significant non-reciprocal transmission; Similar phenomena are evident in (b) and (d), which consider $a/b = 2,4,6,8,10$ for Model II. }
\label{fig13}}
\end{center}
\end{figure}

{A compression-tension (CT) incident pulse is considered in   Sect.\ \ref{sec4} for  demonstrating non-reciprocal wave effects. In the same, way, it is possible to realize non-reciprocal wave behavior  when the input pulse is TC, as shown in Fig.\ \ref{fig13}. Figures \ref{fig13}(a) and \ref{fig13}(c) show the transmitted pulse amplitudes for different $b/a$ values. These results are equivalent to sign inversion of the data presented  in Figs.\ \ref{fig6}(d) and \ref{fig6}(b), which are the results for  CT pulse incidence. The same phenomenon can be found in for Model II. 
Table \ref{table4} summarizes the relationship between the simulation results using the two incident pulse types.}
%%%%%%%%%%%%%%%%%%%%%%%%%%%%%%%%%%%%%%%%%%%%%%%%%%%%%%%%%%%%%%%%%%%%%%
\begin{table}[h]%%%%%%%%%%%%%%%%%%%%%%%%%%%%%%%%%%%%%%%%%%%%%%%%%%%%%%%%%%%%%%%%%%%%%%
{\caption{The relation between the simulation results of the cases using CT pulse in Fig.\ \ref{fig6} and TC pulse in Fig.\ \ref{fig13}. $L$ and $R$ represents the case of incidence from the left and right, respectively. The negative sign in the second row denotes sign inversion of the displacements.}
\label{table4}}
\begin{center}
\begin{tabular}{c cccc}
%& & & & \\ % put some space after the caption
\hline\hline
\, & Model I ($L$) & Model I ($R$) & Model II ($L$) & Model II ($R$) \\
\hline
CT pulse \quad{} & Fig.\ \ref{fig6}(a) & Fig.\ \ref{fig6}(c) & Fig.\ \ref{fig6}(b) & Fig.\ \ref{fig6}(d)  
\\
TC pulse \quad{} & Fig.\ \ref{fig13}(a) = - Fig.\ \ref{fig6}(d)  & Fig.\ \ref{fig13}(c) = - Fig.\ \ref{fig6}(b) & Fig.\ \ref{fig13}(b) = - Fig.\ \ref{fig6}(c) & Fig.\ \ref{fig13}(d) = - Fig.\ \ref{fig6}(a) \\ 
\hline\hline
\end{tabular}
\end{center}
\end{table}%%%%%%%%%%%%%%%%%%%%%%%%%%%%%%%%%%%%%%%%%%%%%%%%%%%%%%%%%%%%%%%%%%%%%%

\end{appendices}

%%%%%%%%%%%%%%%%%%%%%%%%%%%%%%%%%%%%%%%%%%%%%%%%%%%%%%%%%%%%%%%%%%%%
%\bibliography{../../SHARED_BIBLIOGRAPHY/AN_BIG_BIB}
%\bibliographystyle{unsrt}
\bibliographystyle{asmems4}

\end{document}